\documentclass[notitlepage, nofootinbib, no reprint, amsmath, amssymb, aps, twocolumn, pra]{revtex4-2}

\usepackage{amssymb}
\usepackage{hyperref}
\usepackage{bm}
\usepackage{soul}
\usepackage{dsfont}
\usepackage{amsthm}
\usepackage{amsmath}
\usepackage{verbatim} 
\usepackage{qcircuit}
\usepackage{tabularx}
\usepackage{color,soul}
\usepackage{esint}
\usepackage{physics}
\usepackage{braket}
\usepackage{bbold}
\usepackage{amssymb}
\usepackage{float}
\usepackage[normalem]{ulem}

\usepackage{todonotes}

\def\f{f_{\alpha}}

\def\f2{\mathbb F_2^L}
\def\fll{\mathbb F_2^{2L}}
\def\scv{V}
\def\ent{S_{\min}^{\,\text{ent}}}

\def\hfex{h^{\text{ext}}}

\newcommand\mf[1]{\mathbf{#1}}

\usepackage{mathtools}

\DeclarePairedDelimiterX{\infdivx}[2]{(}{)}{%
  #1\;\delimsize\|\;#2%
}

\newtheorem{theorem}{Theorem}

\newtheorem{lemma}{Lemma}[section]

\theoremstyle{remark}
\newtheorem*{remark}{Remark}

\begin{document}

\author{Konstantin Antipin}
\email{kv.antipin@physics.msu.ru}
\affiliation{Faculty of Physics, M.V. Lomonosov Moscow State University, Moscow,  Russia}
\affiliation{Skolkovo Institute of Science and Technology, Moscow, Russia }

\author{Richik Sengupta}
\email{r.sengupta@skoltech.ru}
\affiliation{Artificial Intelligence Research Institute (AIRI), Moscow, Russia}
\affiliation{Skolkovo Institute of Science and Technology, Moscow, Russia}

\title{Entanglement of superpositions in zero-energy subspace: \\insights from stabilizer theory}

\begin{abstract}
Exact zero-energy eigenstates built from entangled dimers
have recently been found in a large class of spin
Hamiltonians.
While the individual states have a simple structure,
their superpositions can have very different entanglement
properties. In this work, we study these superpositions
for geometrically local spin-$1/2$ chains with periodic
boundary conditions. Using stabilizer theory and polynomial
methods, we characterize the subspace they span and find
the minimum entanglement entropy between two halves of
the chain. For Hamiltonians generated by a single local
Pauli interaction with noncommuting translates, this minimum
obeys a volume law at fixed interaction range.
The structure of the least-entangled superpositions
also suggests further exact eigenstate constructions,
including zero-energy states separable between the
two halves of the chain and states at nonzero energies.
\end{abstract}

\maketitle

\section{Introduction}
\label{sec:introduction}

The entanglement structure of highly excited eigenstates provides a useful
perspective on thermalization in isolated quantum many-body systems.
The eigenstate thermalization hypothesis~(ETH) relates eigenstate
expectation values of local observables to equilibrium statistical
mechanics~\cite{Deutsch1991,Srednicki1994}, while thermal eigenstates at
finite energy density typically exhibit volume-law entanglement.
Reviews of thermalization and its relation to entanglement can be found
in Refs.~\cite{DAlessio2016,Mori2018}.
Quantum many-body scars have attracted considerable interest as
exceptional nonthermal eigenstates embedded in otherwise thermalizing
spectra~\cite{Turner2018,Serbyn2021,Moudgalya2022}. Their existence
motivates the search for exact eigenstates with atypical correlations
and entanglement, and for systematic methods to characterize their
departure from thermal behavior.

Degenerate eigenspaces provide a natural setting for investigating these
questions. Any linear combination of exactly degenerate eigenstates is
again an eigenstate, but its entanglement can differ substantially from
that of its constituent states. In particular, coherent superpositions of
highly entangled states can have much lower entanglement. Consequently,
the entanglement of a chosen eigenbasis does not fully characterize the
corresponding eigenspace. This motivates studying the range of
entanglement attainable within the entire subspace, including its minimum
over normalized superpositions. Unlike the entanglement of individual
basis vectors, this minimum is independent of the basis chosen to
represent the subspace.

The importance of considering superpositions has been demonstrated for
both ground-state and excited-state manifolds. In
Ref.~\cite{Okatev2025}, entanglement in spin systems with degenerate
ground states was investigated by sampling random superpositions with
the Haar measure and analyzing the resulting entropy distributions.
For degenerate zero-energy eigenspaces, a subspace-disentangling
algorithm was developed in Ref.~\cite{Karle2021} to search for
least-entangled zero modes. The results motivated a conjecture that
local Hamiltonians with exponentially degenerate nullspaces admit
zero-energy eigenstates with area-law entanglement, providing a route
to nonthermal states in the bulk of the spectrum.

Recent exact eigenstate constructions offer a particularly convenient
setting for studying the effects of superposition. In
Ref.~\cite{ZeroEig25}, zero-energy eigenstates formed from entangled
pairs on antipodal sites of periodic spin chains were constructed.
These states can achieve maximal possible entanglement between the
two halves of the chain. A variant of antipodal constructions with
tunable entanglement was also considered in Ref.~\cite{Mesty_2025}.
A complementary approach was developed in Ref.~\cite{Mukherjee2026},
where symmetric tensor scars were constructed using equal-amplitude
superpositions over permutations of antipodal triplet configurations
with fixed occupation numbers. This construction yields families with
volume-law, logarithmic, and area-law entanglement, illustrating how
superposition can alter the entanglement structure of exact eigenstates.

These developments raise a further question: what is the least
entanglement attainable when arbitrary superposition coefficients are
allowed within a specified family of exact eigenstates?
Equal-amplitude superpositions provide natural examples, but need not
minimize entanglement. Moreover, an exponentially large number of
linearly independent states does not by itself guarantee that their
span contains a weakly entangled state. Resolving this question
requires both explicit constructions of candidate minimizers and
lower bounds that hold for every normalized superposition in the
subspace.

Stabilizer theory provides useful algebraic tools for addressing
entanglement and eigenstate structure in many-body systems.
For example, stabilizer methods were used to calculate ground-state
entanglement entropies and their topological contributions in fracton
models, including the X-cube model and Haah's code~\cite{Ma2018}.
Algorithms for minimizing the energy of Pauli Hamiltonians over
stabilizer states were developed in Ref.~\cite{Sun2025}.
Beyond exact stabilizer states, Ref.~\cite{Gu2024} showed that
stabilizer Hamiltonians perturbed by a limited number of Pauli terms
admit an eigenbasis of doped stabilizer states, enabling efficient calculations when the number of perturbing
terms is small.
More recently, a systematic construction of local parent Hamiltonians
hosting stabilizer states as zero-energy quantum many-body scars was
proposed in Ref.~\cite{Dooley2026}.
These applications motivate using stabilizer methods to characterize
the entanglement attainable within families of exact eigenstates.

In this work, we address this problem for the subspace generated by
antipodal states in periodic spin-$1/2$ chains.
Combining stabilizer theory with a polynomial representation of
translation, we characterize the dimension of this subspace and
determine its minimum half-chain entanglement entropy analytically.
We establish how this minimum scales with system size and show that
equal-amplitude superpositions can have substantially larger
entanglement than optimal ones. The optimization concerns the
specified antipodal-state span, which need not exhaust the full
zero-energy eigenspace.

For certain spin Hamiltonians with bounded interaction range, we prove
that every normalized state in the antipodal span has volume-law
half-chain entanglement, even though the span has exponentially large
dimension. No choice of superposition coefficients within this family
can therefore produce an area-law state across this partition.
Nevertheless, by examining the structure of least-entangled
superpositions, we obtain further exact eigenstate constructions.
For suitable models and system sizes, these include zero-energy
states that factorize between the two halves of the ring and hence
have vanishing half-chain entanglement. Whenever the minimum within
the antipodal span is positive, these states necessarily lie outside
it. Our findings thus complement the perspective of
Ref.~\cite{Karle2021}: extensive entanglement can be unavoidable within
an exponentially large subspace while being entirely absent across
the same partition for other states in the full nullspace.
The constructions also extend to exact eigenstates at nonzero energies.

The paper is structured as follows.
Section~\ref{sec:problem_statement} introduces the antipodal
eigenstate construction, defines valid Pauli strings, and formulates
the entanglement optimization problem.
Section~\ref{sec:methods} reviews the stabilizer and symplectic tools
used in our analysis and specifies the entanglement measures.
Section~\ref{sec:results} presents our main results, beginning with
the $XZ$ model and then developing the polynomial framework for
general geometrically $k$-local Hamiltonians. It also presents
further exact eigenstate constructions motivated by the
least-entangled superpositions.
Section~\ref{sec:conclusion} summarizes the findings.
The appendices provide mathematical background, proofs, detailed
calculations, and additional examples.

\section{Problem statement}
\label{sec:problem_statement}

Consider a class of translation invariant spin-$1/2$ Hamiltonians $H(L)$ on a 1D chain of length $L$ with periodic boundary conditions~(PBC):
\begin{equation}\label{eq:half_length_Ham}
    H(L) = \sum_{i=1}^L\,h_i,
\end{equation}
where 
\begin{equation}
    h_i = \sum_{\alpha=1}^t\,J_{\alpha}h^{(\alpha)}_i
\end{equation}
is a linear combination of Pauli strings
$h^{(\alpha)}_i$ (tensor products of Pauli operators),
each acting non-trivially on site $i$ and, possibly,
on subsequent sites within a finite interaction range,
with site indices understood modulo $L$ due to
periodic boundary conditions. Each string acts
as the identity on all sites outside its support. The coefficients $J_{\alpha}$ in these linear combinations are supposed to be real numbers~(coupling strength constants). The Hamiltonian itself is defined on the global Hilbert  space of dimension $2^L$, and each site $i$ can be described with the use of the local computational basis $\{\ket{s_i}\}$, with $s\in\{1,\,2\}$.

We assume the  Hamiltonian $H$ to be \emph{geometrically} $k$-local. This means that each its term $h^{(\alpha)}_i$ contains at most $k$ non-identity Pauli operators and all non-identity operators within the term  act on a contiguous block of at most $k$ sites, i. e.
\begin{equation}
    \text{supp}(h^{(\alpha)}_i) \subseteq \{i,\, i+1,\,\ldots,\,i+k-1\},
\end{equation}
where $\text{supp}(O)$ denotes the support of the operator $O$ (i.e., the set of lattice sites where $O$ does not act as the identity).

We assume $k\le L$ and construct $H(L)$ and $H(2L)$
from the same local Pauli strings and coupling constants,
with periodic site indices taken modulo $L$ and $2L$,
respectively.

Our subsequent considerations are based on the following recently discovered result, which we slightly reformulate for our setting:
\begin{theorem}[Ref.~\cite{ZeroEig25}]
\label{thm:exact_zero}
    Let $\mathcal C=\bigotimes_{i=1}^L\,\mathcal C_i$, with $\mathcal C_i$ being on-site invertible operators. If, for each $h^{(\alpha)}_i$ in the Hamiltonian $H(L)$, it holds that 
    \begin{equation}
\label{eq:zero_en_condition}    h^{(\alpha)}_i\mathcal C + \mathcal C \left(h^{(\alpha)}_i\right)^T = 0,
    \end{equation}
then the state    \begin{align}\label{eq:dimer_main}
      \ket{\Psi}_{\mathcal C}= \mathcal C\left[\bigotimes_{i=1}^L\,\sum_{s=1}^{2}\, \ket{s_i}\otimes\ket{s_{\Bar i}}\right]&\nonumber\\
      = \bigotimes_{i=1}^L\,\left[(\mathcal C_i\otimes I_{\Bar i})\sum_{s=1}^{2}\, \ket{s_i}\otimes\ket{s_{\Bar i}}\right]&, 
    \end{align}
    where $\Bar i\equiv i+L$, is an exact zero-energy eigenstate of the Hamiltonian $H(2L)$ defined on a  chain of size $2L$ with PBC.
\end{theorem}
In Eq.~\eqref{eq:zero_en_condition}, the operation $A^T$ represents the transposition of operator $A$ with respect to computational basis.

A similar result, but for a  different choice of the on-site operators, was  obtained in Ref.~\cite{Mesty_2025}.

We analyze the structure of a family of zero-energy
eigenstates of the physical Hamiltonian $H(2L)$, defined
on a chain of \emph{even} length $2L$. This Hamiltonian
should be distinguished from the original auxiliary
Hamiltonian $H(L)$ in Eq.~\eqref{eq:half_length_Ham},
defined on a chain of length $L$.

\begin{figure}[t]\label{fig:thm_dim}
    \centering
    \includegraphics[width=0.75\linewidth]{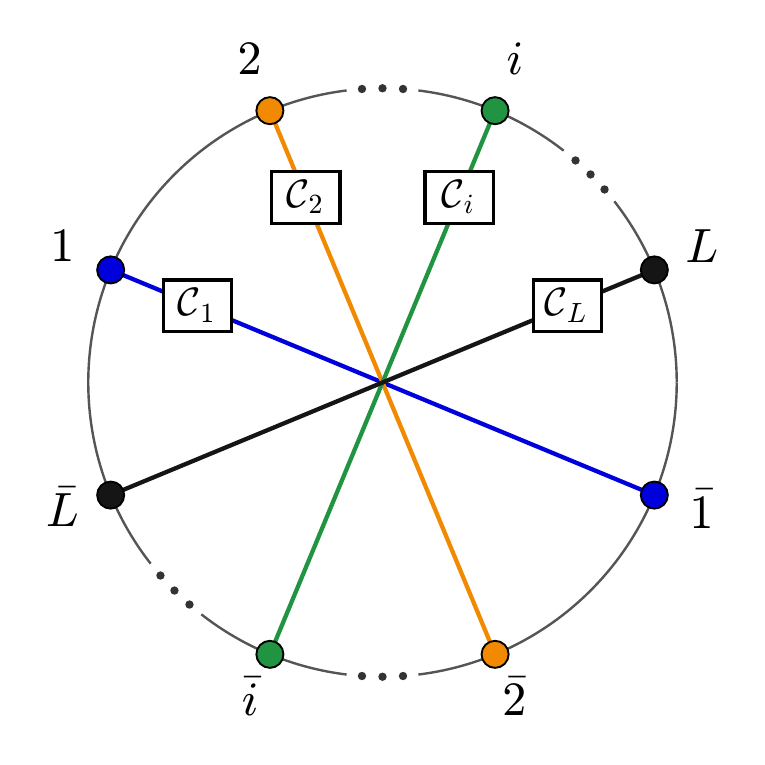}
    \caption{Antipodal dimer construction of Theorem~\ref{thm:exact_zero}
    on a periodic chain of $2L$ sites. Each colored diameter
    joins the two sites supporting one dimer
    $(\mathcal C_i\otimes I_{\bar i})\sum_{s=1}^{2}
    \ket{s}_i\otimes\ket{s}_{\bar i}$.
    The  operators $\mathcal C_i$ act on the unbarred sites.
    }
    \label{fig:antipodal_dimers}
\end{figure}

The exact zero-energy eigenstate $\ket{\Psi}_{\mathcal C}$ in Eq.~\eqref{eq:dimer_main} is   a tensor product of $L$ separate entangled dimers, where the $i$th dimer, represented as
\begin{equation}
    (\mathcal C_i\otimes I_{\Bar i})\sum_{s=1}^{2}\, \ket{s_i}\otimes\ket{s_{\Bar i}},
\end{equation}
has support on antipodal sites $i$ and $\Bar i$. This structure is shown in Fig.~\ref{fig:antipodal_dimers}.

For Pauli strings $h^{(\alpha)}_i$, the zero-energy eigenstate condition~\eqref{eq:zero_en_condition} can be turned into an (anti-)commutation relation. Indeed, for each such term,
\begin{equation}
\left(h^{(\alpha)}_i\right)^T = (-1)^{N_Y\left(\scriptscriptstyle{h^{(\alpha)}_i}\right)}h^{(\alpha)}_i,
\end{equation}
where $N_Y(P)$ is the number of Pauli $Y$ operators in a Pauli string $P$. Condition~\eqref{eq:zero_en_condition} hence can be written as
\begin{equation}
\label{eq:Ham_term_rel}    h^{(\alpha)}_i\mathcal C = \lambda_{\alpha,i}\,\mathcal C\,  h^{(\alpha)}_i,\quad\forall\, i,\,\alpha,
\end{equation}
where 
\begin{equation}
    \lambda_{\alpha,i} = (-1)^{N_Y\left(\scriptscriptstyle{h^{(\alpha)}_i}\right)+1}
\end{equation}
is the factor determining either commutation or anti-commutation character of relation~\eqref{eq:Ham_term_rel}.

For a fixed term~(Pauli string) $h$, let us consider a general operator $O$ satisfying 
\begin{equation}\label{eq:gen_C_rel}
h\,O = \lambda\,O\, h,
\end{equation}
with $\lambda=\pm 1$. Under the Hilbert-Schmidt inner product, the $4^L$ Pauli strings $P^{(l)}$ form an orthogonal basis for the linear space of operators acting on $(\mathbb{C}^2)^{\otimes L}$. The operator $O$ can therefore be uniquely expanded as $O = \sum_{l} a_l P^{(l)}$. Substituting this expansion into~\eqref{eq:gen_C_rel} yields:
\begin{equation}
    \sum_{l} a_l \left(h P^{(l)}\right) = \lambda \sum_{l} a_l \eta_l \left(h P^{(l)}\right),
\end{equation}
where $P^{(l)} h = \eta_l h P^{(l)}$. Since the set $\{P^{(l)} h\}$ remains an orthogonal basis, we equate coefficients, which leads  to condition $a_l (1 - \lambda \eta_l) = 0$. Therefore, $a_l$ can be nonzero only if $\eta_l = \lambda$. The operator $O$ is thus entirely spanned by the subset of Pauli strings that exhibit the same  commutation phase $\lambda$ with $h$. Introducing multiple terms $h^{(\alpha)}_i$ and requiring $h^{(\alpha)}_i\,O = \lambda_{\alpha,i}\,O\, h^{(\alpha)}_i\,\,\forall\, i,\,\alpha$  only decreases the number of possible Pauli strings in the decomposition of $O$, while each of them  retaining completely the same commutation phase $\lambda_{\alpha,i}$ with $h^{(\alpha)}_i$ as the operator $O$. 

In the context of Theorem~\ref{thm:exact_zero}, this means that, without loss of generality, the operator $\mathcal C$  can be assumed to be a linear combination of Pauli strings of length $L$ each satisfying condition~\eqref{eq:zero_en_condition}. More formally,
\begin{equation}\label{eq:C_dec}
\mathcal C = \sum_{l} a_l P^{(l)},    
\end{equation}
where each Pauli string $P^{(l)}$ satisfies
\begin{equation}   h^{(\alpha)}_i P^{(l)} + P^{(l)} \left(h^{(\alpha)}_i\right)^T = 0,\quad\forall\, i,\,\alpha.
\end{equation}
For a given Hamiltonian $H(L)$, we will refer to the set $\{P_l\}$ of all such Pauli strings as the set of \emph{valid} Pauli strings.

Given a decomposition~\eqref{eq:C_dec} of the operator $\mathcal C$, the corresponding zero-energy eigenstate
\begin{equation}\label{eq:psi_c_rep}   \ket{\Psi}_{\mathcal C}= \mathcal C\left[\bigotimes_{i=1}^L\,\sum_{s=1}^{2}\, \ket{s_i}\otimes\ket{s_{\Bar i}}\right]
\end{equation}
is a superposition of states corresponding to individual Pauli strings:
\begin{equation}\label{eq:lin_sup}    \ket{\Psi}_{\mathcal C}= \sum_{l} b_l\ket{\Psi}_{P^{(l)}},
\end{equation}
with
\begin{align}
\label{eq:sing_Pauli}    \ket{\Psi}_{P^{(l)}}= P^{(l)}\left[\bigotimes_{i=1}^L\,\sum_{s=1}^{2}\, \ket{s_i}\otimes\ket{s_{\Bar i}}\right]&\nonumber\\
    = \bigotimes_{i=1}^L\,\left[(P^{(l)}_i\otimes I_{\Bar i})\sum_{s=1}^{2}\, \ket{s_i}\otimes\ket{s_{\Bar i}}\right]&. 
\end{align}
Since each on-site component $P^{(l)}_i$ is unitary~(Pauli operator), for any state $\ket{\Psi}_{P^{(l)}}$,  the reduced density matrix of a contiguous subsystem of length $\leqslant L$ is proportional to identity matrix. In particular, a contiguous subsystem of length $L$, one half of the chain, exhibits maximal possible entanglement with the other half.  

Linear superpositions~\eqref{eq:lin_sup}   constitute a subspace of the zero-energy space of the Hamiltonian $H(2L)$. Certain superpositions may possess  entanglement structure different from  that of the state in Eq.~\eqref{eq:sing_Pauli} - this will be one of the main focuses of the present work. More specifically, for a given Hamiltonian $H(2L)$, we pursue several goals:
\begin{enumerate}
    \item Characterize the set of all valid Pauli strings $\{P^{(l)}\}$, in particular, find the dimension of the subspace generated by the states $\ket{\Psi}_{P^{(l)}}$.
    \item Find superpositions  $\ket{\Psi}_{\mathcal C}$, which correspond to minimal possible entanglement between the two halves of the spin chain; characterize the scaling of this entanglement with $L$. 
    \item Investigate the structure of the least entangled superpositions for small $L$'s. Generalize these structures for arbitrary $L$'s. 
\end{enumerate}

\section{Methods}
\label{sec:methods}

The fact that the states $\ket{\Psi}_{P^{(l)}}$ are defined by Pauli strings motivates invoking the tools of stabilizer theory for our research goals. Stabilizer theory deals with the notion of the Pauli group and its abelian subgroups.
 A  brief review of the main ingredients of the theory is given in Appendix~\ref{sec:appendix_stabilizer_background}.

The Pauli group $\mathcal{P}_L$ consists of all $L$-fold tensor products of the single-qubit Pauli matrices (Pauli strings), together with the global phase factors $\{\pm 1, \pm i\}$. When dealing with Pauli strings, we will denote Pauli matrices together with identity as $\{I, X, Y, Z\}$. 

Every element in the  Pauli group $\mathcal P_L$  admits a useful representation in terms of a binary vector $\mf r=(\mf x\, |\, \mf z)$ of length $2L$ consisting of two parts,  $\mf x = (x_0, \ldots,\,x_{L-1})\in \mathbb F_2^L$ and $\mf z = (z_0, \ldots,\, z_{L-1})\in\mathbb F_2^L$. The vector $\mf r$ is usually referred to as the \emph{check vector}. A Pauli operator at site $k$ of the Pauli string is then expressed as $i^{x_kz_k}X^{x_k}Z^{z_k}$. For any two elements  $g$ and $h$ of the Pauli group~$\mathcal P_L$ the symplectic product of their check vectors
\begin{equation}\label{eq:symp_prod}
    \langle\mf r(g), \mf r(h)\rangle\equiv\sum_{k=0}^{L-1}\,x_k(g)z_k(h)+x_k(h)z_k(g)\:\,\, (\mathrm{mod}\,2)
\end{equation}
can be defined. The product can also be written in the form
\begin{equation}
    \langle\mf r(g), \mf r(h)\rangle = \mf r(g)^T\Omega\,\mf r(h)\quad(\mathrm{mod}\:2),
\end{equation}
where 
\begin{equation}
    \Omega= \begin{pmatrix} 0 & I \\ I & 0 \end{pmatrix}
\end{equation}
is the matrix of the symplectic bilinear form on $\fll$.

This product determines whether the two elements $g$ and $h$ commute or anti-commute:
\begin{eqnarray}\label{eq:comm_cond}
    &[g,\,h]& = 0\quad\mathrm{iff}\quad  \langle\mf r(g), \mf r(h)\rangle = 0,\nonumber\\
&\{g,\,h\}& = 0\quad\mathrm{iff}\quad  \langle\mf r(g), \mf r(h)\rangle =1.
\end{eqnarray}
Consequently, the check vector formalism enables the mapping between the  Pauli group members and the elements of the symplectic vector space. This property will be used extensively in the subsequent sections.

Another useful representation of the Pauli group elements using polynomial framework will be introduced in Section~\ref{sec:gen_theorems}, where the main general results are formulated.  

 For quantification of the entanglement between subsystems of the spin chain, we use two measures. The primary  measure  is the Von Neumann entanglement entropy defined for a pure state $\ket{\Psi}$ with respect to bipartition $A|\Bar{A}$ as follows:
 \begin{equation}
     S(\Psi) = -\Tr{\rho_A \log_2\rho_A},
 \end{equation}
 where $\rho_A = \Tr_{\Bar{A}}{\dyad{\Psi}}$ is the reduced state of subsystem $A$ with respect to bipartition $A|\Bar{A}$. Another measure, which in many settings can be calculated much easier, is the second Rényi entaglement entropy:
 \begin{equation}
     S_2(\Psi)=-\log_2\Tr{\rho_A^2}.
 \end{equation}
 It is known that the  inequality between the two entropies
 \begin{equation}
    S(\Psi)\geqslant S_2(\Psi) 
 \end{equation}
 holds universally.

 Throughout the text, we use the following convention:
physical sites are numbered starting from $1$, as in
Eqs.~\eqref{eq:half_length_Ham}--\eqref{eq:dimer_main},
whereas the components of check vectors corresponding to
Pauli strings of length $L$ are indexed from $0$ to $L-1$.

\section{Results}
\label{sec:results}

We start this section with an explicit example of a Hamiltonian generated by a single term. This example captures our main tools in action  and allows for further extension to general geometrically $k$-local Hamiltonians generated by arbitrary number of independent terms.

%\subsection{Representation in terms of quantum channels}

%Consider three Hilbert spaces $\hil_A,\,\hil_B,\,\hil_C$ with $\dim\hil_A=\dim\hil_B = \dim\hil_C=4$, and  an isometry $V_{3/2}\colon\mathcal H_A\rightarrow\hil_B\otimes\hil_C$, which acts on the computational basis states $\{\ket{k}_A,\:k=0,\,\ldots,\,3\}$ of $\hil_A$ in the following way:
%\begin{equation}
%    V_{3/2}\ket{k}_A = \frac12\sum_{i,\,j=0}^3\,(\gamma_k)_{ij}\ket{i}_B\otimes\ket{j}_C,
%\end{equation}
%i.~e., $V$ maps the $k$-th basis state to the vectorization of $\gamma_k$. 

%Let us introduce the quantum channel $\Phi_{3/2}\colon\mathcal L(\hil_A)\rightarrow\mathcal L(\hil_B)$, which is related to the isometry $V_{3/2}$ via Stinespring's dilation:
%\begin{equation}
%    \Phi_{3/2}(\rho_A) = \Tr_C\left[{V\rho_AV^{\dagger}}\right]
%\end{equation}

\subsection{Motivating example: Hamiltonian with XZ term}
\label{sec:opt_purity}

Consider a Hamiltonian $H(L)$ of the form~\eqref{eq:half_length_Ham}, which is generated by a single term $h = \sigma_x\otimes\sigma_z$~(corresponding length $L$ Pauli string is $X\otimes Z\otimes I^{\otimes(L-2)}$). We analyze superpositions of zero-energy eigenstates of the corresponding physical Hamiltonian on a chain of size $2L$:
\begin{equation}
    H(2L) = \sum_{i=1}^{2L}\,(\sigma_x)_i\otimes (\sigma_z)_{i+1},
\end{equation}
in accordance with Theorem~\ref{thm:exact_zero}.

Any superposition $\ket{\Psi}_{\mathcal C}$ of zero-energy
states in the class under consideration is specified by an
operator $\mathcal C$ of the form given in~\eqref{eq:C_dec},
which is a linear combination of valid Pauli strings.
To characterize the entanglement between the two halves of
the chain in this pure state, we consider the reduced density
operator on sites $1,\ldots,L$, which constitute one half of
the chain. Using the representation in~\eqref{eq:psi_c_rep},
we obtain
\begin{equation}
    \rho_{\mathcal C}\equiv\rho_{1\ldots L}
    =\mathcal C\mathcal C^\dagger,
    \qquad
    \operatorname{Tr}\!\left(
        \mathcal C\mathcal C^\dagger
    \right)=1.
\end{equation}

Since $h$ contains an even number of Pauli $Y$ operators, it
satisfies $h^T=h$. The valid Pauli strings are therefore those
that anticommute with $X\otimes Z\otimes I^{\otimes(L-2)}$ and
each of its lattice translates under periodic boundary
conditions (PBC). This condition is equivalent to the following
nearest-neighbor constraint within each valid string:
\begin{equation}
\begin{aligned}
    I,\;X &\;\longrightarrow\; X \text{ or } Y,\\
    Y,\;Z &\;\longrightarrow\; I \text{ or } Z.
\end{aligned}
\end{equation}
Here, the arrow specifies the allowed operator at the next
site, including across the periodic boundary.

The nearest-neighbor constraint can  be recast in terms of
transitions between check vector components,
$(x_k,z_k)\to(x_{k+1},z_{k+1})$, where $k=0,\ldots,L-1$
and site indices are taken modulo $L$:
\begin{equation}
\begin{aligned}
    I\to X \text{ or } Y
    &\quad\Longleftrightarrow\quad
    (0,0)\to(1,0) \text{ or } (1,1),\\
    X\to X \text{ or } Y
    &\quad\Longleftrightarrow\quad
    (1,0)\to(1,0) \text{ or } (1,1),\\
    Y\to I \text{ or } Z
    &\quad\Longleftrightarrow\quad
    (1,1)\to(0,0) \text{ or } (0,1),\\
    Z\to I \text{ or } Z
    &\quad\Longleftrightarrow\quad
    (0,1)\to(0,0) \text{ or } (0,1).
\end{aligned}
\end{equation}
The transitions above share the common pattern
\begin{equation}
    \label{eq:cv_valid}
    x_{k+1}=z_k\oplus1,
\end{equation}
where $\oplus$ denotes addition modulo two and site indices
are taken modulo $L$. This relation also follows directly
from the anticommutation condition~\eqref{eq:comm_cond}:
applying that condition to a valid check vector
$(\mathbf{x}\mid\mathbf{z})$ and the check vector of
$X_kZ_{k+1}$ gives $z_k\oplus x_{k+1}=1$, which is
equivalent to~\eqref{eq:cv_valid}.

Thus, the valid check vectors are in one-to-one correspondence
with the $2^L$ binary vectors
$\mathbf{b}=(b_0,\ldots,b_{L-1})\in\mathbb{F}_2^L$.
Setting $z_k=b_k$ uniquely determines
$x_k=1\oplus b_{k-1}$. The corresponding Pauli strings are,
up to an overall phase in $\{\pm1,\pm i\}$,
\begin{equation}
    P_{\mathbf{b}}\sim
    \bigotimes_{k=0}^{L-1}
    \left(X^{\,1\oplus b_{k-1}}Z^{\,b_k}\right),
    \qquad \mathbf{b}\in\mathbb{F}_2^L,
\end{equation}
with $b_{-1}=b_{L-1}$.

Factoring out $X^{\otimes L}$, we can write each valid Pauli
string as
\begin{equation}
    \label{eq:Pauli_U_exp}
    P_{\mathbf{b}}\sim X^{\otimes L}U_{\mathbf{b}},
\end{equation}
where
\begin{equation}
    \label{eq:U_prop}
    U_{\mathbf{b}}
    =\bigotimes_{k=0}^{L-1}
    \left(X^{b_{k-1}}Z^{b_k}\right),
    \qquad \mathbf{b}\in\mathbb{F}_2^L.
\end{equation}
All site indices are understood modulo $L$. These operators
satisfy
\begin{equation}
    U_{\mathbf{b}}U_{\mathbf{c}}
    =(-1)^{\sum_{k=0}^{L-1}b_kc_{k-1}}
    U_{\mathbf{b}\oplus\mathbf{c}},
    \qquad \mathbf{b},\mathbf{c}\in\mathbb{F}_2^L,
\end{equation}
where $\oplus$ acts componentwise. Since
$U_{\mathbf{0}}=I^{\otimes L}$ and every $U_{\mathbf{b}}$
has an inverse equal to itself up to a sign, the set
\begin{equation}
    \label{eq:group_ex}
    G=\left\{
        \omega U_{\mathbf{b}}
        \;\middle|\;
        \omega\in\{\pm1,\pm i\},\
        \mathbf{b}\in\mathbb{F}_2^L
    \right\}
\end{equation}
is a subgroup of the Pauli group $\mathcal{P}_L$.
Consequently, by~\eqref{eq:Pauli_U_exp}, the valid Pauli
strings together with all their overall phase multiples
form the left coset $X^{\otimes L}G$.

We expand $\mathcal C$ in the valid Pauli strings:
\begin{equation}
    \mathcal C
    =\sum_{\mathbf b\in\mathbb F_2^L}
    a_{\mathbf b}P_{\mathbf b},
    \qquad
    \sum_{\mathbf b\in\mathbb F_2^L}
    |a_{\mathbf b}|^2=2^{-L}.
\end{equation}
The normalization follows from Pauli orthogonality and
$\operatorname{Tr}(\mathcal C\mathcal C^\dagger)=1$.

Write $F=X^{\otimes L}$. Equation~\eqref{eq:U_prop} gives
\begin{equation}
    FU_{\mathbf b}F
    =(-1)^{\sum_{k=0}^{L-1}b_k}U_{\mathbf b}.
\end{equation}
The phase factors in~\eqref{eq:Pauli_U_exp}, together with
these conjugation signs, can therefore be absorbed into
coefficients $\widetilde a_{\mathbf b}$, yielding
\begin{equation}
    \mathcal C
    =\left(
        \sum_{\mathbf b\in\mathbb F_2^L}
        \widetilde a_{\mathbf b}U_{\mathbf b}
    \right)F.
\end{equation}
Here, $\widetilde a_{\mathbf b}$ differs from $a_{\mathbf b}$
only by a factor in $\{\pm1,\pm i\}$. Since $FF^\dagger=I$,
the reduced density operator is
\begin{equation}
    \rho_{\mathcal C}
    =\mathcal C\mathcal C^\dagger
    =\sum_{\mathbf b,\mathbf b'\in\mathbb F_2^L}
    \widetilde a_{\mathbf b}\widetilde a_{\mathbf b'}^*
    U_{\mathbf b}U_{\mathbf b'}^\dagger.
\end{equation}
Because $G$ is a group of unitary operators,
$\rho_{\mathcal C}$ belongs to the group algebra
$\mathbb C[G]=\operatorname{span}_{\mathbb C}G$.

To analyze the entanglement entropy, we consider the spectral
decomposition
\begin{equation}
    \rho_{\mathcal C}
    =\sum_k\lambda_k\Pi_k,
    \qquad
    \lambda_k\geq0,
    \qquad
    \sum_k\lambda_k\operatorname{rank}\Pi_k=1,
\end{equation}
where the $\Pi_k$ are mutually orthogonal spectral projectors.

Let $\Pi_{\min}\in\mathbb C[G]$ be a nonzero orthogonal
projector of minimum rank $r_{\min}$. As shown below,
the minimum entanglement entropy is attained by
\begin{equation}
    \label{eq:opt_rho_half}
    \rho_{\mathrm{opt}}
    =\frac{1}{r_{\min}}\Pi_{\min}.
\end{equation}
The corresponding entropy is
\begin{equation}
    \label{eq:opt_pur}
    S_{\min}^{\mathrm{ent}}=\log_2 r_{\min}.
\end{equation}
A corresponding normalized operator is
\begin{equation}
    \label{eq:opt_A}
    \mathcal C_{\mathrm{opt}}
    =\frac{1}{\sqrt{r_{\min}}}
    \Pi_{\min}X^{\otimes L}.
\end{equation}
Since $\Pi_{\min}\in\mathbb C[G]$ and
$U_{\mathbf b}X^{\otimes L}\sim P_{\mathbf b}$,
this operator is a linear combination of valid Pauli strings.
It satisfies
$\mathcal C_{\mathrm{opt}}\mathcal C_{\mathrm{opt}}^\dagger
=\rho_{\mathrm{opt}}$.

The problem then reduces to finding a nonzero projector of
minimum rank in the group algebra $\mathbb C[G]$.
For this purpose, let us consider the center of the group
$G$---the set of its elements that commute with every element
of the group. The commutation relation between two elements
$U_{\mf b}$ and $U_{\mf c}$ is determined by the commutation
relations between the operators $X^{b_{k-1}}Z^{b_k}$ and
$X^{c_{k-1}}Z^{c_k}$ at each site $k$. Hence, the two elements
commute if and only if
\begin{equation}
    \label{eq:U_com_cond}
    \sum_{k=0}^{L-1}
    \left(b_k c_{k-1}+b_{k-1}c_k\right)
    \equiv 0 \pmod{2}.
\end{equation}
The left-hand side of~\eqref{eq:U_com_cond} can be regarded
as a bilinear form on the binary vector space $\mathbb F_2^L$:
\begin{equation}
    \label{eq:inn_prod}
    \langle \mf b,\mf c\rangle_K
    =\sum_{i,j=0}^{L-1}b_iK_{ij}c_j,
\end{equation}
with the matrix $K$ defined by
\begin{equation}
    K_{ij}=\delta_{i,j+1}+\delta_{i,j-1}.
\end{equation}
Here, arithmetic is modulo two, and site indices are taken
modulo $L$. A vector $\mf v\in\mathbb F_2^L$ corresponds to
a central element of $G$ if and only if
\begin{equation}
    \langle \mf b,\mf v\rangle_K=0
    \qquad\text{for all }\mf b\in\mathbb F_2^L.
\end{equation}
This is equivalent to
\begin{equation}
    K\mf v=0,
\end{equation}
which, in turn, gives the condition
\begin{equation}
    \label{eq:kernel_center}
    v_{k-1}\oplus v_{k+1}=0.
\end{equation}

Taking periodic boundary conditions into account, the
analysis separates into two cases: even and odd $L$.
When $L$ is even, Eq.~\eqref{eq:kernel_center} imposes
the following structure on the vectors $\mf v$:
\begin{equation}
    \mf v=(x,y,x,y,\ldots).
\end{equation}
In other words, $\mf v$ consists of two interleaved constant
substrings and is thus parametrized by two bits $x$ and $y$.
Consequently, $\dim\ker K=2$ and $\rank K=L-2$.
For odd $L$, Eq.~\eqref{eq:kernel_center} and periodicity
require vectors $\mf v$ of the form
\begin{equation}
    \mf v=(x,x,x,\ldots),
\end{equation}
so $\dim\ker K=1$ and, correspondingly, $\rank K=L-1$.

Both cases can be summarized as
\begin{equation}
    \label{eq:G_gen}
    \begin{gathered}
        \dim\ker K=L-2m,
        \qquad \rank K=2m,\\
        m=\left\lfloor\frac{L-1}{2}\right\rfloor,
    \end{gathered}
\end{equation}
where $\lfloor\cdot\rfloor$ denotes the floor function.

It follows from Eq.~\eqref{eq:G_gen} that the original
$L$-dimensional space of check vectors corresponding to
the elements of $G$ can be decomposed into a direct sum
of two subspaces orthogonal with respect to the bilinear
form~\eqref{eq:inn_prod}.
The first is spanned by $L-2m$ linearly independent check
vectors corresponding to central generators, i.e.,
$L-2m$ group elements that generate the center of $G$
up to scalar phases.
The second is spanned by $2m$ linearly independent check
vectors corresponding to the remaining generators of $G$.

The standard procedure of stabilizer theory
(Appendix~\ref{app:sec:iter_symp}) then allows us to choose
these $2m$ generators in the \emph{canonical generator set}
form. More specifically, they fall into two sets,
$g_1,\ldots,g_m$ and $h_1,\ldots,h_m$.
Choosing their phases so that the generators are Hermitian,
each set consists of $m$ independent generators of a
stabilizer group. Between the two sets, $[g_i,h_j]=0$
for $i\ne j$, whereas $\{g_i,h_i\}=0$.
In other words, there are $m$ anticommuting pairs of
operators $(g_1,h_1),\ldots,(g_m,h_m)$.

We can now choose a maximal commuting set of generators
by taking the $L-2m$ central generators and $m$ additional
operators, one from each of the $m$ pairs.
The resulting set contains $L-2m+m=L-m$ independent
commuting generators.
Choosing Hermitian phases for the central generators
as well, these generators form a stabilizer group $S$
of order $\abs{S}=2^{L-m}$.
The associated stabilizer projector
\begin{equation}\label{eq:max_stabb_proj}
    \Pi_S=\frac{1}{\abs{S}}\sum_{g\in S}g
\end{equation}
has rank
\begin{equation}
    \rank\Pi_S
    =\operatorname{Tr}\Pi_S
    =\frac{\operatorname{Tr}I}{2^{L-m}}
    =\frac{2^L}{2^{L-m}}
    =2^m.
\end{equation}
Since we have chosen a maximal commuting set of generators
(and hence the largest possible $\abs{S}$), this is a
minimum-rank \emph{stabilizer} projector in $\mathbb C[G]$.
We still need to show that no nonzero projector in the
group algebra can have a smaller rank, so that
\begin{equation}
    \label{eq:opt_rank}
    r_{\min}=2^m.
\end{equation}

Another result from stabilizer theory
(Appendix~\ref{app:sec:clifford_background}) tells us that
there is a single Clifford unitary transformation $U_C$
under which the generators take a particular form.
The $2m$ generators are mapped to the operators $X_j$
and $Z_j$, $j=1,\ldots,m$, while the $L-2m$ central
generators are mapped to
$Z_{m+1},\ldots,Z_{L-m}$.
Each transformed generator acts nontrivially on just
one qubit (the \emph{fiducial canonical generator set}
form).

As a result, under $U_C$, all operators in the group
algebra $\mathbb C[G]$ are transformed into operators
acting nontrivially only on the first $L-m$ qubits.
This means that any nonzero projector $\Pi$ in
$\mathbb C[G]$ has the form
\begin{equation}
    U_C\Pi U_C^\dagger
    =\widetilde{\Pi}_{L-m}\otimes I^{\otimes m},
\end{equation}
where $\widetilde{\Pi}_{L-m}$ is a nonzero projector
acting on the first $L-m$ qubits.
Since unitary transformations preserve rank,
\begin{equation}
    \rank\Pi
    =2^m\rank\widetilde{\Pi}_{L-m}
    \geq 2^m,
\end{equation}
which proves the claim.

To complete the entropy-minimization argument, consider an
arbitrary density operator $\rho\in\mathbb C[G]$. It is mapped by the 
Clifford unitary $U_C$ to an operator of the form
\begin{equation}
    U_C\rho U_C^\dagger
    = \rho' \otimes \frac{I^{\otimes m}}{2^m},
\end{equation}
where $\rho'$ is a normalized density operator acting on the
remaining degrees of freedom.

By the unitary invariance of the von Neumann entropy,
$S(\rho)=S(U_C\rho U_C^\dagger)$.
Furthermore, the entropy is additive under tensor products,
so
\begin{align}
    S(\rho)
    &= S\left(\rho'\otimes\frac{I^{\otimes m}}{2^m}\right)
    \nonumber\\
    &= S(\rho')+
       S\left(\frac{I^{\otimes m}}{2^m}\right)
    \nonumber\\
    &= S(\rho')+m
    \geq m,
\end{align}
where the last inequality follows from the nonnegativity
of the von Neumann entropy, $S(\rho')\geq 0$. 

The state proportional to the maximal stabilizer projector $\Pi_S$ from Eq.~\eqref{eq:max_stabb_proj} attains this bound:
\begin{equation}
    \label{eq:opt_rho_half_stab}
    \rho_{\mathrm{opt}}
    =\frac{1}{r_{\min}}\Pi_S.
\end{equation}

For the optimal superposition,
Eqs.~\eqref{eq:opt_pur},~\eqref{eq:G_gen}, 
and~\eqref{eq:opt_rank} give the entropy of entanglement
between the two halves of the chain, measured in bits:
\begin{equation}
    \label{eq:entropy_xz}
    \ent=\left\lfloor\frac{L-1}{2}\right\rfloor.
\end{equation}

To illustrate this result, we provide explicit examples
of the optimal constructions for $L=2$ and $L=3$
in Appendix~\ref{app:XZ}.

It is also interesting to compare this result with the
entanglement scaling of the uniform superposition of the
same states. In this superposition, the states corresponding
to all valid Pauli strings are taken with equal real
coefficients, chosen to ensure normalization.
In Appendix~\ref{sec:appendix_mpo_purity}, we use the transfer
matrix method to calculate how the purity scales with the
half-chain length $L$. 
Our derivation shows that, in the thermodynamic limit
$L\to\infty$, the purity of the half-chain reduced state
$\rho_{\mathrm{unif}}$ scales as
\begin{equation}  
\label{eq:asymp_pur}    \Tr{\rho_{\text{unif}}^2}\sim2^{-\alpha L},
\end{equation}
where 
\begin{equation}
    \alpha = - \log_2\left( \frac{1 + \sqrt{2}}{4} \right) \approx 0.7284.
\end{equation}
The asymptotic purity~\eqref{eq:asymp_pur}
implies
\begin{equation}
    S_2(\rho_{\mathrm{unif}})
    =\alpha L+o(1),
\end{equation}
Since the von Neumann entropy satisfies
$S(\rho)\ge S_2(\rho)$, it follows that
\begin{equation}
    \liminf_{L\to\infty}
    \frac{S(\rho_{\mathrm{unif}})}{L}\ge\alpha.
\end{equation}
%Thus, the uniform superposition has an asymptotic
%entanglement entropy per site strictly greater than
%the optimal value $1/2$.
This indicates that the entanglement entropy of the uniform superposition scales at a substantially higher rate compared to the minimal possible scaling $S_{\min}^{\,\text{ent}}\approx 0.5L$, which is achieved by states with specially structured amplitudes.

\subsection{General case: Hamiltonians with $k$-local terms}
\label{sec:gen_theorems}

In the particular example above, the minimal possible entanglement entropy between the two halves of the chain manifested the  volume law scaling $L/2 - \mathcal O(1)$, while the factorizable state of the two halves existed only for $L\leqslant2$. A question now arises whether other Hamiltonians, possibly with higher locality and larger number of terms, can exhibit different behavior of entanglement. 

We are in a position to formulate the main result of the present paper, which states that the previous example cannot be improved: the lowest possible entropy result scales as $\ent\geqslant L/2 - \mathcal{O}(1)$   for Hamiltonians with non-commuting terms of  locality of at most $k$.

To prove the result, we first consider a Hamiltonian generated by a \emph{single} $k$-local Pauli string $h$ translated along the chain:
\begin{equation}\label{eq:ham2l}
    H(L) = \sum_{i=1}^{L}\,h_i.
\end{equation}
 An example of a $3$-local term could be $XZZ$, $YXZ$, etc.

Building on the key steps of the previous subsection, we develop a general approach based on linear algebra of check vectors in a symplectic space with the inner product~(\ref{eq:symp_prod}). Let the Pauli string $h$ be represented by a check vector $\mathbf h=(\mathbf h_x\,|\, \mathbf h_z)$. The iteratively shifted versions with PBC are given by $T^i\mf h$ for $i = 0,\,\ldots,\,L-1$, where $T$ is the operator of forward shifting~(translation) by $1$ position. 

Recall that our valid string condition was anti-commutation if $h$ contains even number of $Y$'s, and commutation if it has an odd number. Using~(\ref{eq:comm_cond}), both cases can be encapsulated  as follows: for a Pauli string with a check vector $\mf v\in\mathbb{F}_2^{2L}$ to be valid, it must satisfy 
\begin{equation}\label{eq:wind_cond}
    \langle\mf v,\, T^i\mf h\rangle = c, \quad\,i=0,\,\ldots,\,L-1,
\end{equation}
where $c=0$ if $h$ has an odd number of $Y$'s and $c=1$ if  $h$ has an even number of $Y$'s.

The shifted vectors $T^i\mf h$ can be stacked to form an $L\times 2L$ matrix $M$, where the $i$-th row is $\left(T^i\mf h\right)^T\Omega$. The condition~(\ref{eq:wind_cond}) becomes a system of linear equations over $\mathbb F_2$:
\begin{equation}\label{eq:nonh_sys}
    M\mf v = \mf c,
\end{equation}
where $\mf c$ is a length $L$ vector with all the components equal to $c$.

The solutions to this inhomogeneous equation form an affine subspace
\begin{equation}
    \mathcal V = \mathbf{v}_0 + \ker(M),
\end{equation}
where $\mathbf{v}_0$ is a vector corresponding to any single valid Pauli string.

As long as the solution $\mf v_0$ of the inhomogeneous system~\eqref{eq:nonh_sys} exists, the kernel of $M$ becomes the main object of our analysis. The kernel of $M$, which we will call $V$, represents the set of check vectors of all Pauli strings that commute with every shifted operator $T^i\mf h$. These check vectors  generate a subgroup of the Pauli group, which we denote as $G$. Let
\begin{equation}\label{eq:gen_sub}
    \mathcal{W} = \text{span}\{T^0 \mathbf{h}, T^1 \mathbf{h}, \dots, T^{L-1} \mathbf{h}\}
\end{equation}
be the subspace generated by the shifted operators. By definition, $V$ is the symplectic orthogonal complement of $\mathcal{W}$:
\begin{equation}\label{eq:main_group}
    \scv = \mathcal{W}^\perp.
\end{equation}

The dimension of the subspace spanned by
valid Pauli-string states is
\begin{equation}\label{eq:valid_span_dimension}
\dim_{\mathbb C}\mathcal H_{\mathrm{valid}}
=2^{\dim\mathcal W^\perp}.
\end{equation}

As was discussed in the explicit example of the previous subsection, the problem reduces to finding the largest stabilizer subgroup $S_{\mathrm{max}}$ of the group $G$. For that purpose, we need to know the generators of the center of $G$, which translates into finding the  elements of $\scv$, which are orthogonal to any element of $\scv$. All such vectors lie in the intersection
\begin{equation}\label{eq:center_cond}
    \scv\cap\scv^\perp = \mathcal{W}^\perp\cap\mathcal W,
\end{equation}
since $\left(\mathcal W^\perp\right)^\perp = \mathcal W$.

In order to form the maximal commuting subgroup, to the generators of the center~(\ref{eq:center_cond}) we add half of the canonical generators from its complement in $\scv$.  The number of these generators is $\left(\dim\mathcal W^\perp - \dim\mathcal{W}^\perp\cap\mathcal W\right)/2$, and hence the total number of the generators of the maximal commuting subgroup is
\begin{align}\label{eq:stab_order}
\dim(\mathcal S_{\max}) = &\dim\mathcal{W}^\perp\cap\mathcal W  \nonumber\\
&+\frac12\left[\dim\mathcal W^\perp - \dim\mathcal{W}^\perp\cap\mathcal W\right]\nonumber\\
&= \frac12\left[\dim\mathcal W^\perp + \dim\mathcal{W}^\perp\cap\mathcal W\right].
\end{align}
Accordingly, the order of the maximal commuting subgroup is
\begin{equation}\label{eq:gen_opt_ord}
    |S_{\max}| = 2^{\dim(\mathcal S_{\max})},
\end{equation}
and the optimal purity and entanglement entropy  are given by
\begin{align}\label{eq:gen_opt_pur}
    \Tr{\rho_{\mathrm{opt}}^2} &=  2^{-(L-\dim(\mathcal S_{\max}))},\nonumber\\
    \ent &= L-\dim(\mathcal S_{\max}).
\end{align}

The optimality of this construction over arbitrary
superpositions of valid Pauli strings follows by
the same argument as in the $XZ$ example discussed
above. Indeed, the proof of the entropy lower bound
given there relies only on the algebraic structure
of the Pauli group generated by the homogeneous
constraint space $V=\mathcal W^\perp$, and not on
the particular choice of local generators.
Consequently, the maximal stabilizer construction
attains the minimum entropy over the entire
valid-state subspace, rather than merely within
the family of stabilizer states.

Equivalently, using
$\dim(\mathcal W) +\dim(\mathcal W^\perp)=2L$,
the minimum entropy is rewritten as
\begin{equation}\label{eq:general_entropy_min_new}
\ent=\frac12\left[
\dim\mathcal W-
\dim(\mathcal W\cap\mathcal W^\perp)
\right].
\end{equation}

To formulate the solvability conditions for the initial inhomogeneous system~\eqref{eq:nonh_sys} and to obtain closed-form expressions for the dimensions of the subspaces in question, we invoke the polynomial framework over the finite field $\mathbb{F}_2$. Polynomial representations  have a long history in stabilizer
theory and quantum error correction. They have been used to describe
quantum convolutional codes~\cite{OllivierTillich2004},
translation-invariant Clifford dynamics and stabilizer
states~\cite{Schlingemann2008}, and commuting Pauli Hamiltonians
and topological stabilizer codes~\cite{Haah2013,Haah2021}. 

While the comprehensive details are provided in Appendices~\ref{app:pol_intro}-\ref{app:sec:sec:smith_form}, we briefly introduce here the basic objects in terms of which our main results are formulated.
 Let $\mathcal R = \mathbb{F}_2[x] / (x^L - 1)$ be the quotient ring of polynomials of degree less than $L$. The factor $(x^L - 1)$ is introduced to take into account  the  periodic boundary condition, which, in terms of polynomials, reads
\begin{equation}\label{eq:pbc_pol}
    x^L \equiv 1.
\end{equation}
As previously, all arithmetic is performed modulo $2$.

A check vector $\mf v = (\mf v_x\, |\, \mf v_z)$ for a Pauli string of length $L$ can be uniquely represented as a pair of polynomials: $\mf v\rightarrow (a(x), b(x))^T \in\mathcal R^2$, where
\begin{equation}
    a(x) = \sum_{i=0}^{L-1} a_i x^i,\quad b(x) = \sum_{i=0}^{L-1} b_i x^i,
\end{equation}
with the coefficients $a_i$ and $b_i$ being the same as those of the X-part $\mf v_x$ and the Z-part $\mf v_z$ of $\mf v$, respectively. The action of the shift operator $T$ on $\mf v$ is represented simply as the multiplication of both constituents by $x$: $T\mf v\rightarrow (xa(x), xb(x))^T$.

Let our 
 Hamiltonian term $h$ be represented by the fixed polynomial pair:
 \begin{equation}
     \mathbf{h}(x) = \begin{pmatrix} p(x) \\ q(x) \end{pmatrix}.
 \end{equation}
 Since $h$ is geometrically $k$-local,  the degrees of $p(x)$ and $q(x)$ are strictly bounded: $\deg p, \deg q \le k - 1$. For later convenience, let us also introduce the reciprocal polynomials:
 \begin{equation}\label{eq:rec_pol}
     p^*(x) = x^{k-1} p(x^{-1}), \quad q^*(x) = x^{k-1} q(x^{-1}).
 \end{equation}
 Importantly, $\deg p^*, \deg q^* \le k-1$ as well.

To demonstrate the power of the chosen framework, we first formulate the conditions under which valid Pauli strings exist for the anticommutation case, meaning that they anticommute with the Hamiltonian term and all its shifts. This implies that the inhomogeneous system~\eqref{eq:nonh_sys} has at least one solution. For the commutation case, the vector $\mathbf{c}$ on the right-hand side of~\eqref{eq:nonh_sys} is zero, and one immediately proceeds to the analysis of $\ker(M)$. The result is summarized in the following theorem.

\begin{theorem}
\label{thm:anti_commutation_t1}
Let $h$ be a  Hamiltonian term on a periodic 1D lattice of size $L$, characterized by its $X$-type and $Z$-type polynomials  $p(x), q(x) \in \mathbb{F}_2[x]$. Let $L = 2^\nu \cdot n$, where $n$ is an odd integer and $\nu \ge 0$. 

A  Pauli string $P$ that strictly anti-commutes with all $L$ cyclic shifts of $h$ exists if and only if $q(x)$ and $p(x)$ are not simultaneously divisible by $(x+1)^{2^\nu}$. 

This  condition  reduces to the following parity rules:
\begin{enumerate}
    \item \textbf{$L$ is odd $(\nu=0):$} $P$ exists if and only if $h$ contains an odd number of Pauli-$X$ and Pauli-$Y$ operators, or an odd number of Pauli-$Z$ and Pauli-$Y$ operators \textup{(}i.e., $q(1)=1$ or $p(1)=1$\textup{)}.
    \item \textbf{$L$ is even, $(\nu \ge 1):$} $P$ exists if and only if the polynomials  $q(x)$ and $p(x)$ do not simultaneously contain $(x+1)^{2^\nu}$ as a factor over $\mathbb{F}_2$.
\end{enumerate}
\end{theorem}
The proof is provided in Appendix~\ref{app:thm_solv_proof}.

Once the existence of valid Pauli strings is established, the dimension of the solution space can be analyzed using the following result.

\begin{theorem}\label{thm_pol}
The dimensions of the subspaces\eqref{eq:main_group} and~\eqref{eq:center_cond} can be expressed via polynomials in the quotient ring $\mathcal R = \mathbb{F}_2[x] / (x^L - 1)$ as
\begin{equation}\label{eq:thm1_dim1}
    \dim\mathcal W^\perp = L  + \deg K(x) ,
\end{equation}
and
\begin{multline}\label{eq:thm1_dim2}
    \dim(\mathcal{W} \cap \mathcal{W}^\perp) = \deg \gcd(C(x), x^L-1) \\
    - \deg K(x),
\end{multline}
where
\begin{align}
    C(x) &\equiv p(x)q^*(x)+q(x)p^*(x)
    \notag\\[-2pt]
    &\hspace{1em}\pmod{x^L-1},
    \label{eq:c_pol}\\
    K(x) &= \gcd\bigl(p(x),q(x),x^L-1\bigr),
    \label{eq:k_pol}
\end{align}
with $\gcd$ denoting the greatest common divisor of polynomials over $\mathbb{F}_2$, and $\deg$ -- the degree of a polynomial.  
\end{theorem}
The proof is given in Appendix~\ref{app:thm_pol}.

As an important corollary, using~(\ref{eq:stab_order}) and~(\ref{eq:gen_opt_pur}), we obtain the expression for the minimal entanglement entropy~(with $K(x)$ completely eliminated):
\begin{equation}\label{eq:main_ent}
    \ent = \frac{1}{2} \left[ L - \deg \gcd(C(x), x^L-1) \right].
\end{equation}

As a consequence, we can obtain the scaling behavior by exploiting the fact that $k$ is a constant independent of $L$.  We consider only the non-trivial case in which there are non-commuting terms among all  $h_i$ in (\ref{eq:ham2l}). As explained in  Appendix~\ref{app:pol_intro}, this leads to $C(x)\ne 0$. As a consequence, the degree of $\gcd(C(x), x^L-1)$ is bounded by $\deg C(x)\leqslant 2(k-1)=\mathcal{O}(1)$. This further results, according to~(\ref{eq:main_ent}), in the scaling of entanglement entropy as:
\begin{equation}\label{eq:main_ent_scale}
\ent=L/2 - \mathcal{O}(1).
\end{equation}

Now let us consider what happens when the Hamiltonian term is a linear combination of $t$ different Pauli strings: 
\begin{equation}\label{eq:loc_deq}
    h = \sum_{\alpha=1}^t\,J_{\alpha} h^{(\alpha)},
\end{equation}
each having geometric locality at most  $k$. Importantly, we assume \emph{arbitrary} coefficients $J_{\alpha}$  and  enforce commutation/anti-commutation conditions with \emph{every} term $h^{(\alpha)}$. This condition is highly restrictive, and we can demonstrate that introducing multiple terms does not improve the $(L/2)$-scaling of entanglement entropy.

The analogue of the subspace~(\ref{eq:gen_sub}) is no longer generated by shifts of a single operator, but rather by shifts of all $t$ operators in~(\ref{eq:loc_deq}); that is, one should consider the sum of  individual subspaces of each term: 
\begin{equation}
\mathcal W_{\mathrm{tot}} = \mathcal{W}_1 + \mathcal{W}_2 + \dots + \mathcal{W}_t,
\end{equation}
where
\begin{equation}
    \mathcal{W}_{\alpha} = \text{span}\{T^0 \mathbf{h}^{(\alpha)}, T^1 \mathbf{h}^{(\alpha)}, \dots, T^{L-1} \mathbf{h}^{(\alpha)}\},
\end{equation}
with $\mathbf{h}^{(\alpha)}$ being the check vector corresponding to $h^{(\alpha)}$.

%By the previous theorem, a generic $k$-local operator generates a subspace of dimension $\dim(\mathcal{W}_1) = L - \mathcal{O}(1)$. Adding a second, linearly independent term (e.g., $XY$ alongside $XZ$) imposes approximately $L$ additional constraints. Since the total space $\mathbb{F}_2^{2L}$ is only $2L$-dimensional, the union of these subspaces quickly spans the entire space. For generic $k$-local terms, the intersection $\mathcal{W}_i \cap \mathcal{W}_j$ is small, yielding $\dim(\mathcal{W}) = \min(tL, 2L) - \mathcal{O}(1)$. Thus, for $t \ge 2$, $\dim(\mathcal{W}) \approx 2L$. The group $G$ of valid strings is defined by the orthogonal complement $\scv = \mathcal{W}^\perp$, so $\dim(\scv) = 2L - \dim(\mathcal{W})$. For $t \ge 2$, this forces $\dim(\scv) = \mathcal{O}(1)$. The stabilizer subgroup must reside entirely within $G$, which means that its order is bounded by $|G| = 2^{\mathcal{O}(1)}$. Consequently, the maximal stabilizer group order drops to $| S_{max}| = 2^{\mathcal{O}(1)}$, resulting, according to~\eqref{eq:gen_opt_pur}, in the entropy scaling as $\ent = L - \mathcal O(1)$.

%A more rigorous treatment of the case of several terms is presented in Appendix~\ref{app:sev_terms}.

Eq.~\eqref{eq:main_ent} can be extended  to an arbitrary number of generators $t$. Let $p_{\alpha}(x)$ and $q_{\alpha}(x)$ be the $X$ and $Z$ polynomial representations of the $\alpha$th Hamiltonian generator $h^{(\alpha)}$, respectively, and assume that valid states exist (necessary and sufficient conditions for their existence are derived in Appendix~\ref{app:sec:solv_inhom}). Introducing the \emph{cross-commutation} polynomial matrix $C_{\alpha\beta}(x)$ with entries
\begin{equation}
    C_{\alpha\beta}(x) = p_{\beta}(x)q_{\alpha}^*(x) + q_{\beta}(x)p_{\alpha}^*(x),
\end{equation}
the optimal entanglement entropy can be expressed as
\begin{equation}\label{eq:ent_multy}
    \ent = L - \frac{1}{2} \sum_{\alpha=1}^2 \deg \gcd(c_{\alpha}(x), x^L-1),
\end{equation}
where $c_1(x)$ and $c_2(x)$ are the \emph{invariant factors} of $C_{\alpha\beta}(x)$ (at most two, owing to $\operatorname{rank} C_{\alpha\beta}(x) \le 2$). These factors are determined by the  minors of $C_{\alpha\beta}(x)$:
\begin{itemize}
    \item $c_1(x)$ is the greatest common divisor of all $1 \times 1$ minors (i.e., the GCD of all individual entries of $C$).
    \item $c_1(x)c_2(x)$ is the greatest common divisor of all $2 \times 2$ minors.
\end{itemize}
The  nonzero invariant factors have bounded degree, which is $\mathcal O(1)$. The detailed derivation of Eq.~\eqref{eq:ent_multy}, along with an explicit application to  specific Hamiltonians, is provided in Appendix~\ref{app:sec:sec:smith_form}.

If $\operatorname{rank} C_{\alpha\beta}(x) \le 1$, which includes the single-generator case $t=1$, the second invariant factor vanishes, $c_2(x) = 0$. Under the standard polynomial convention $\gcd(0, x^L-1) = x^L-1$, the second term yields $\deg \gcd(0, x^L-1) = \deg(x^L-1) = L$. Equation~\eqref{eq:ent_multy} then evaluates to
\begin{equation}
\begin{aligned}
    \ent &= L - \frac{1}{2} \left[ \deg \gcd(c_1(x), x^L-1) + L \right] \\
    &= \frac{1}{2} \left[ L - \deg \gcd(c_1(x), x^L-1) \right],
\end{aligned}
\end{equation}
For $t=1$, $c_1(x)=C(x)$, recovering Eq.~\eqref{eq:main_ent}.

An interesting example is provided by the Hamiltonian
\begin{align}
    H(L) = &\sum_{i=1}^L\,[J_{xz}(\sigma_x)_i\otimes (\sigma_z)_{i+1}\nonumber\\ 
    &+ J_{xyz} (\sigma_x)_i\otimes (\sigma_y)_{i+1}\otimes (\sigma_z)_{i+2}]
\end{align}
generated by the two distinct Pauli strings
$h^{(1)}=XZI$ and $h^{(2)}=XYZ$. Indeed, writing
$A_i=X_iZ_{i+1}$ and $B_i=X_iY_{i+1}Z_{i+2}$, we have
$A_iA_{i+1}=iB_i$. Every valid Pauli string for the $XZ$ model
anticommutes with both $A_i$ and $A_{i+1}$, and therefore commutes
with $B_i$, exactly as required because $XYZ$ contains one $Y$
operator. Thus, adding the $XYZ$ terms leaves the valid-string set,
the associated group $G$, and the span of valid antipodal states
unchanged. Although the two local generators are distinct and are
not translations of one another, their polynomial vectors satisfy
$\mathbf h^{(2)}(x)=(1+x)\mathbf h^{(1)}(x)$. The cross-commutation matrix
has rank one over $\mathbb F_2(x)$ and hence only one nonzero
invariant factor. The previously derived $XZ$ results therefore
apply unchanged:
$\dim\mathcal W^\perp=L$ and
$\ent=\lfloor(L-1)/2\rfloor$.
This illustrates that the number of distinct local Pauli terms
alone does not determine the entanglement scaling.

When $c_1(x)$ and $c_2(x)$ are \emph{both} non-zero, their
contributions to Eq.~\eqref{eq:ent_multy} are
bounded independently of $L$. Consequently, 
\begin{equation}
    \ent = L - \mathcal{O}(1).
\end{equation}
This is a direct consequence of a dimensional collapse of  the space of valid Pauli strings down to $\mathcal O(1)$ because of highly restrictive character of the commutation/anti-commutation conditions with many Hamiltonian terms  $h^{(\alpha)}$.

\begin{remark}
    For specific coefficients $J_\alpha$ in~(\ref{eq:loc_deq}), there can be an infinite number of operators $A$ such that $A^{\otimes L}$ commutes or anti-commutes with $h$ as a whole, rather than with its individual terms. Such cases, extending beyond single valid Pauli strings, are not investigated here but are explored further in Refs. \cite{Mesty_2025, Mukherjee2026}.    
\end{remark}

\subsection{Illustrative Example: $
3$-Local Hamiltonian}

In this subsection, we apply the results of Theorem~\ref{thm_pol} to the case where the Hamiltonian is generated by the single $3$-local term
\begin{equation}
    h = XZZ.
\end{equation}
The check vector, which corresponds to $h$, has the form $\mf h = (1,\,0,\,0\,|\,0,\,1,\,1)^T$. Translating it to polynomials yields $\mathbf{h}(x) = (p(x), q(x))^T$ with
\begin{equation}
    p(x) = 1,\quad q(x) = x + x^2.
\end{equation}

Since $p(x)=1$ doesn't contain $(x+1)$ as its factor, the conditions of Theorem~\ref{thm:anti_commutation_t1} are met, and hence valid Pauli strings exist for any $L$.

For the polynomial $K(x)$ defined in~(\ref{eq:k_pol}) we have
\begin{equation}
   K(x) =  \gcd(1,\, x+x^2,\, x^L-1) = 1,
\end{equation}
since the only divisor of the first argument is $1$.  Consequently, $\deg K(x) =~0$, and, by~(\ref{eq:thm1_dim1}), 
\begin{equation}
    \dim\mathcal W^\perp = L.
\end{equation}
This result tells us that the number of different Pauli strings in $G$ is $2^{\dim\mathcal W^\perp} = 2^L$, and hence there is an exponential number of valid Pauli strings.

Calculation of the polynomial from~(\ref{eq:c_pol}) yields:
\begin{equation}
    C(x) = 1 + x + x^3 + x^4.
\end{equation}
By Eq.~(\ref{eq:thm1_dim2}),
\begin{equation}\label{eq:ex2_dim}
    \dim(\mathcal{W} \cap \mathcal{W}^\perp) = \deg \gcd(C(x), x^L-1),
\end{equation}
since the second term in the right part of~(\ref{eq:thm1_dim2}) is zero, again, due to $K(x) = 1$.

For arbitrary $L$, the calculation of the greatest common divisor in~(\ref{eq:ex2_dim}) is more involved. The details are provided in Appendix~\ref{app:zxx_center_dimension}, here we present just the final result:
\begin{equation}
    \dim(\mathcal{W} \cap \mathcal{W}^\perp) = \begin{cases} 
1, & L = 1,\,5 \pmod 6 \\
2, & L = 2,\,4 \pmod 6 \\
3, & L = 3\pmod 6 \\
4, & L = 0\pmod 6
\end{cases}
\end{equation}

The entanglement entropy, calculated using~(\ref{eq:main_ent}), exhibits a peculiar staggering behavior:
\begin{equation}
    \ent = \begin{cases} 
\frac{L-1}2, & L = 1,\,5 \pmod 6 \\
\frac{L-2}2, & L = 2,\,4 \pmod 6 \\
\frac{L-3}2, & L = 3\pmod 6 \\
\frac{L-4}2, & L = 0\pmod 6
\end{cases}
\end{equation}
It can be seen that its large $L$ scaling reproduces the general result $L/2 - \mathcal{O}(1)$, while for small values, such as $L=3$, a separable state~($\ent=0$) of the two halves of the spin chain is possible.

\subsection{Exact eigenstate constructions beyond crosscap states and zero energy}
\label{sec:ex_eig}

So far we have considered states at both ends of the entanglement spectrum for superpositions: single states $\ket{\Psi}_{P^{(l)}}$ producing maximal entanglement $S^{\mathrm{\,ent}}=L$ and their optimal linear combinations, in which $\ent\sim L/2$. Inspecting the latter for small $L$ may provide some insight into how novel eigenstate structures emerge for local Hamiltonians within the zero-energy subspace and beyond. In this section, we describe dimer-based configurations that depart from the antipodal construction.  

Let us return to the Hamiltonian with the $XZ$ term from Section~\ref{sec:opt_purity}. As follows from~(\ref{eq:entropy_xz}), a separable state of two halves of the chain exists for $L=2$. It was shown in Appendix~\ref{app:XZ} that this state is a tensor product 
\begin{equation}
\label{eq:prod_st}    \ket{\Phi}_{12\Bar1\Bar2} = \ket{\psi}_{12}\otimes\ket{\phi}_{\Bar1\Bar2}
\end{equation}
of the two maximally entangled states of two sites in each half of the chain:
\begin{equation}\label{eq:xzpsi}   \ket{\psi}_{12}=\frac1{\sqrt2}\sum_{i,\,j=0}^1\,(U_H \sigma_y)_{ij}\ket{i}_1\ket{j}_2,
\end{equation}
and 
\begin{equation}\label{eq:xzphi}    \ket{\phi}_{\Bar1\Bar2}=\frac1{\sqrt2}\sum_{i,\,j=0}^1\,(U_H)_{ij}\ket{i}_{\Bar1}\ket{j}_{\Bar2}, 
 \end{equation}
where 
\begin{equation}\label{eq:had}
    U_H = \frac1{\sqrt2}\begin{pmatrix}
1 & 1 \\
1 & -1
\end{pmatrix}
\end{equation} 
is the Hadamard matrix.

\begin{figure}[t]
            \centering
            \includegraphics[width=0.75\linewidth]{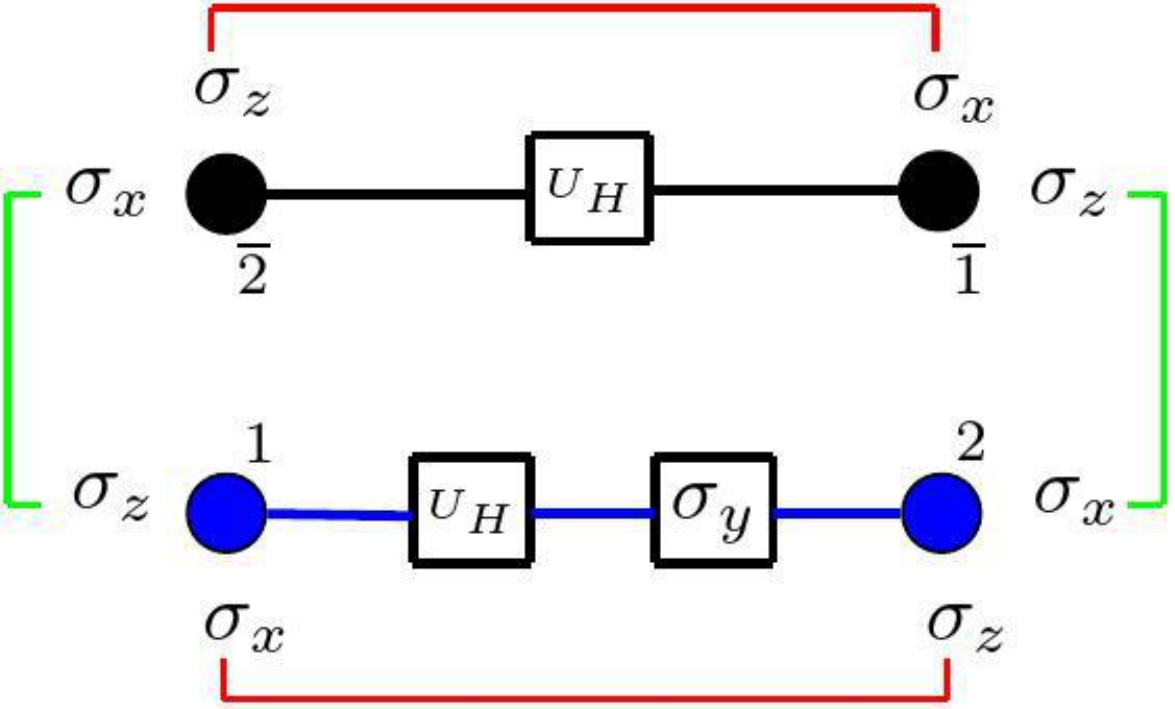}
            \caption{The product zero-energy eigenstate of the XZ Hamiltonian on four sites~($L=2$). The states~(\ref{eq:xzpsi}) (blue) and~(\ref{eq:xzphi}) (black) are shown alongside the four terms of the Hamiltonian and their respective sites. The terms canceling each other out are marked with brackets of the same color~(red and green).
            }
            \label{fig:xzl2}
        \end{figure}

We can examine  how the four terms of the Hamiltonian interact with each other while acting on the product eigenstate~(\ref{eq:prod_st})(Fig.~\ref{fig:xzl2}). 

        The states~(\ref{eq:xzpsi}) and (\ref{eq:xzphi}) exhibit a special behavior when acted upon by Pauli operators on their individual subsystems; namely, for the former, it can be shown that
        \begin{equation}\label{eq:xz_tel1}I_1\otimes(\sigma_x)_2\ket{\psi}_{12} = -(\sigma_z)_1\otimes I_2\ket{\psi}_{12},
        \end{equation}
        while for the latter
        \begin{equation}\label{eq:xz_tel2}
            I_{\Bar2}\otimes (\sigma_z)_{\Bar1}\ket{\phi}_{\Bar2\Bar1} = (\sigma_x)_{\Bar2}\otimes I_{\Bar1}\ket{\phi}_{\Bar2\Bar1},
        \end{equation}
        where the site indices $\Bar1$ and $\Bar2$ are written in reverse order intentionally to match the antipodal configuration in Fig.~\ref{fig:xzl2}~(also note that the state $\ket{\phi}_{\Bar2\Bar1}$ is symmetric under the exchange of its subsystems). These properties are best understood through the lens of quantum information theory, specifically via the operator-state isomorphism, where they correspond to an operator teleportation (or sliding) from one subsystem to another, as shown in Appendix~\ref{app:sec:vec}. 
        
        By applying Eqs.~(\ref{eq:xz_tel1}) and~(\ref{eq:xz_tel2}), it becomes evident that the Hamiltonian term $(\sigma_x)_2\otimes(\sigma_z)_{\Bar1}$—depicted on the right side of Fig.~\ref{fig:xzl2}—is likewise teleported across the composite state to the left side:
        \begin{align}
            I_{1\Bar2}&\otimes(\sigma_x)_2\otimes(\sigma_z)_{\Bar1}\ket{\Phi}_{12\Bar1\Bar2}\nonumber\\
            &=-(\sigma_z)_1\otimes(\sigma_x)_{\Bar2}\otimes I_{2\Bar1}\ket{\Phi}_{12\Bar1\Bar2}.
        \end{align}
        As a result, the two Hamiltonian terms on the left and right sides of Fig.~\ref{fig:xzl2} cancel each other out when acting on the composite state:
        \begin{equation}
           \left[ (\sigma_z)_1\otimes(\sigma_x)_{\Bar2} +(\sigma_x)_2\otimes(\sigma_z)_{\Bar1} \right]\ket{\Phi}_{12\Bar1\Bar2} = 0.
        \end{equation}

        The remaining two Hamiltonian terms are supported on the sets $12$ and $\Bar1\Bar2$, coinciding with the subsystems where the bipartite states $\ket{\psi}_{12}$ and $\ket{\phi}_{\Bar1\Bar2}$ reside. Furthermore, these states are eigenvectors of their respective terms:  
\begin{align}  (\sigma_x)_1\otimes(\sigma_z)_2\ket{\psi}_{12} &= -\ket{\psi}_{12},\nonumber\\
(\sigma_z)_{\Bar2}\otimes(\sigma_x)_{\Bar1}\ket{\phi}_{\Bar2\Bar1} &= \ket{\phi}_{\Bar2\Bar1},
\end{align}
which also results in their mutual cancellation:
\begin{equation}
    \left[(\sigma_x)_1\otimes(\sigma_z)_2 + (\sigma_z)_{\Bar2}\otimes(\sigma_x)_{\Bar1}\right]\ket{\Phi}_{12\Bar1\Bar2} = 0.
\end{equation}
Thus, the pairwise cancellation of the Hamiltonian terms guarantees that $\ket{\Phi}_{12\Bar1\Bar2}$ is a zero-energy eigenstate.

The particular structure of the eigenstate for the four-site chain can be extended to other values of half-length $L$. One possible way to do this is to consider multiple entangled pairs (dimers), given by Eqs.~(\ref{eq:xzpsi}) and~(\ref{eq:xzphi}), arranged together in a ladder configuration. The guiding principle is similar: the dimers, forming the rungs of the ladder, should alternate in such a way that the Hamiltonian terms on the left and right sides of the ladder offset each other. To maintain a zero-energy eigenvalue for the resulting state, the top and bottom dimers must be eigenvectors of their corresponding Hamiltonian terms with eigenvalues of opposite sign. If this latter constraint is lifted, these $\pm 1$ eigenvalues can sum to yield alternative energies of $\pm 2$.

\begin{figure}[t]
            \centering
            \includegraphics[width=0.85\linewidth]{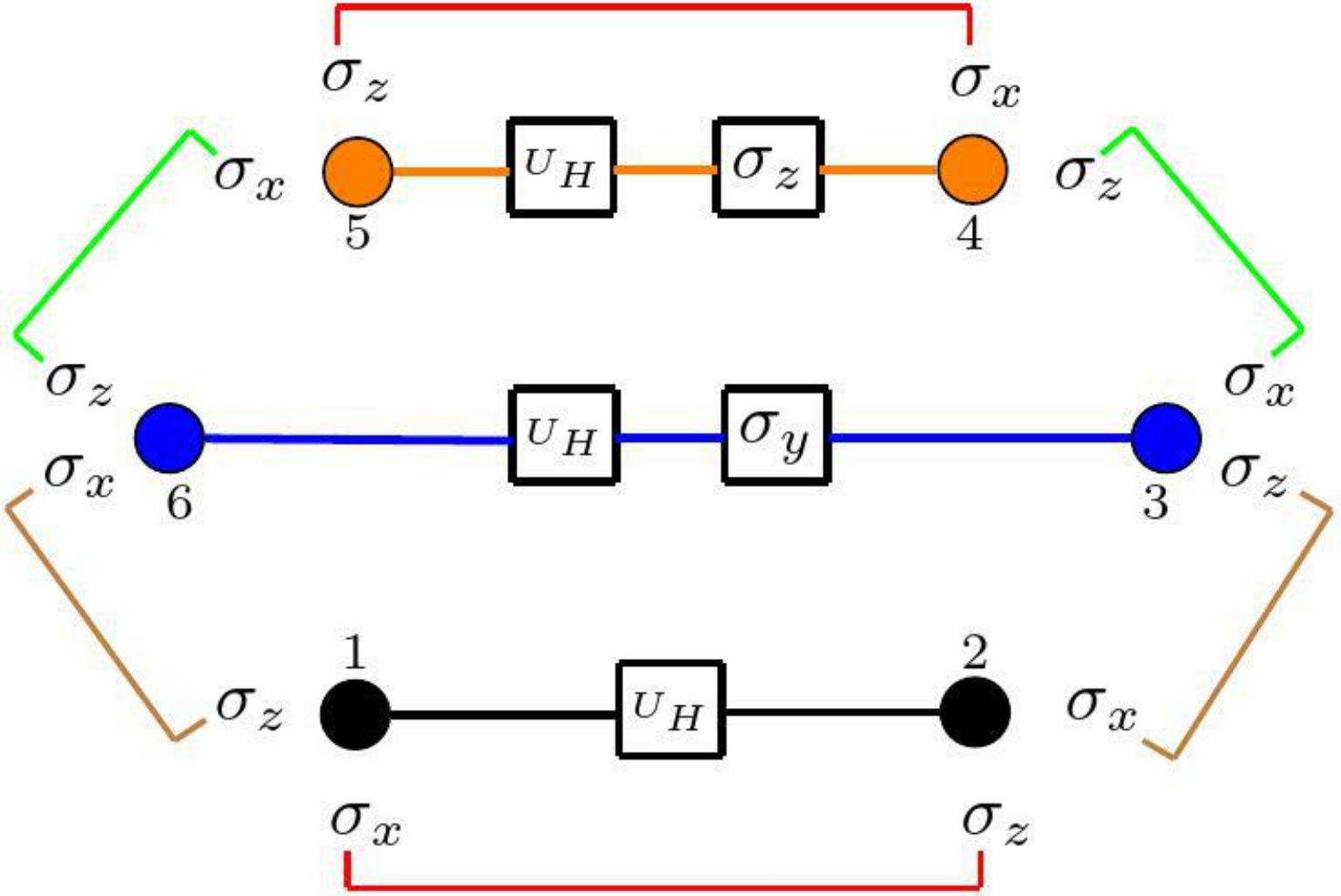}
            \caption{Ladder construction for the eigenstates of the XZ Hamiltonian on a $6$-site spin chain. Mutually counterbalancing Hamiltonian terms are marked with brackets of the same color.
            }
            \label{fig:xzlarge}
        \end{figure}

The construction for $6$ sites~($L=3$) is depicted in Fig.~\ref{fig:xzlarge}. By default, site enumeration for this layout starts at the bottom-left site and continues counterclockwise. Here, another type of dimer is introduced at the top of the ladder:
\begin{equation}\label{eq:xzchi}   \ket{\chi}_{54}=\frac1{\sqrt2}\sum_{i,\,j=0}^1\,(U_H \sigma_z)_{ij}\ket{i}_5\ket{j}_4,
\end{equation}
which has a sliding property identical to that of $\ket{\phi}$ in~(\ref{eq:xz_tel2}):
\begin{equation}\label{eq:xz_tel3}
    I_5\otimes (\sigma_z)_4\ket{\chi}_{54} = (\sigma_x)_5\otimes I_4\ket{\chi}_{54},
\end{equation}
but, unlike $\ket{\phi}$, it is an eigenvector of its corresponding term, $\sigma_z\otimes\sigma_x$, with the opposite eigenvalue, $-1$.
Without introducing the third dimer, the strict alternation of just two dimer types for odd half-length $L$ would prevent the top and bottom Hamiltonian terms from offsetting each other, resulting instead in total energies of $\pm 2$. In contrast, when $L$ is even, alternating between only two types of dimers is sufficient to ensure zero energy.

For even $L$, the ladder configuration also ensures
separability across the half-chain bipartition.
Indeed, one can choose a contiguous block of $L$ sites
containing exactly $L/2$ complete dimers, so that no
dimer crosses the boundary of this block.
The resulting product of dimer states therefore has
zero entanglement entropy across this cut.
By translation invariance of the Hamiltonian,
the eigenstate can be translated to align this
bipartition with the designated halves
$\{1,\ldots,L\}$ and $\{L+1,\ldots,2L\}$.
Consequently, the translated zero-energy eigenstate
satisfies $S^{\mathrm{ent}}=0$.

The results presented above establish that, at least for any even $L$, there exist zero-energy eigenstates that are factorizable across the two halves of the spin chain. In view of Eq.~(\ref{eq:entropy_xz}), these eigenstates do not belong to the span of antipodal dimers for $L>2$, which means that they belong to a  separate structural class.

\begin{figure}[t]
            \centering
            \includegraphics[width=0.82\linewidth]{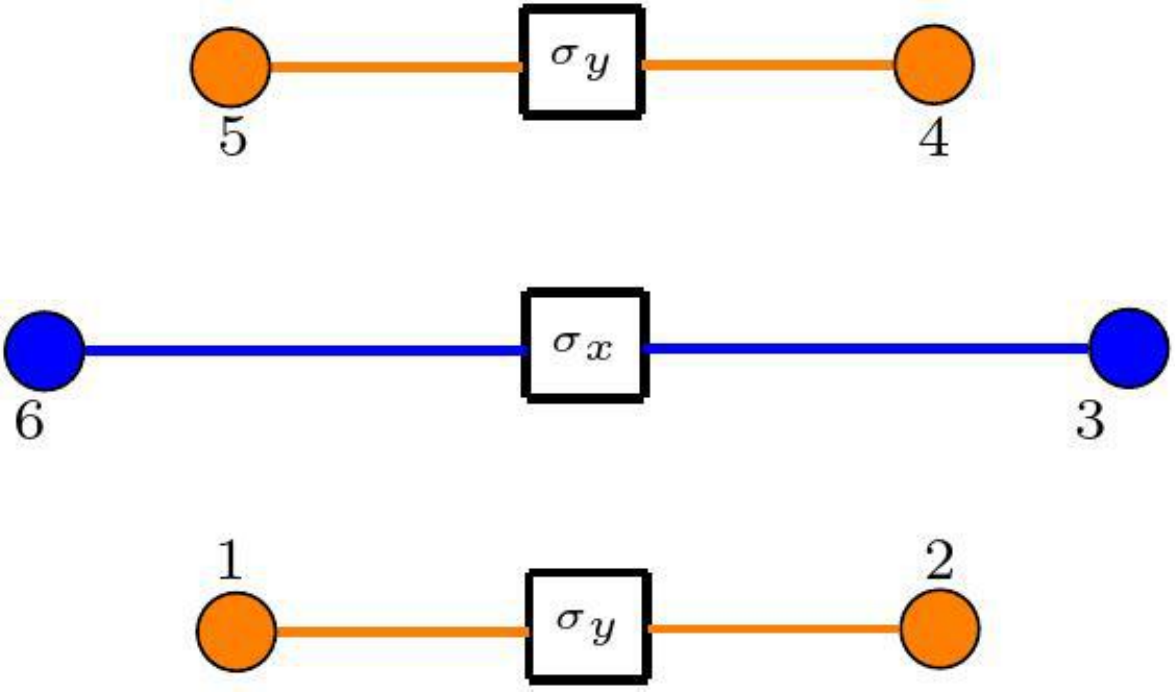}
            \caption{Ladder construction for the eigenstates of the $XY$ model on a $6$-site spin chain, featuring dimer states $\ket{\phi}_X$ (blue) and $\ket{\phi}_Y$ (orange).}
            \label{fig:xydim}
        \end{figure}

The ladder construction can be readily extended to other models, such as the well-known $XY$ spin-$1/2$ chain in a transverse external magnetic field (also with PBC):
\begin{align}\label{eq:xy_model}
    &H = H_{XY} + H_Z,\\
    H_{XY} = \sum_{i=1}^{2L}\,&\left(J_xS^x_i  S^x_{i+1} +  J_yS^y_i S^y_{i+1}\right),\: H_z =\hfex\sum_{i=1}^{2L}\, S^z_i,\nonumber
\end{align}
where $S^\alpha_i = \frac12\sigma^\alpha_i$  $\alpha=x,\,y,\,z$ are the standard spin-$1/2$ operators. This extension is achieved by replacing the dimer states~(\ref{eq:xzpsi}) and~(\ref{eq:xzphi}) with
\begin{equation}\label{eq:xy_xdim}
    \ket{\phi}_X = \frac1{\sqrt2}\sum_{i,\,j=0}^1\,( \sigma_x)_{ij}\ket{i}\ket{j}
\end{equation}
and 
\begin{equation}\label{eq:xy_ydim}
    \ket{\phi}_Y = \frac1{\sqrt2}\sum_{i,\,j=0}^1\,( \sigma_y)_{ij}\ket{i}\ket{j},
\end{equation}
respectively. The state composed of alternating $X$ and $Y$ dimers is shown in Fig~\ref{fig:xydim}. That the construction yields an eigenstate is guaranteed by the properties~(Appendix~\ref{app:sec:dimex})
\begin{align}
    I\otimes\sigma_x\ket{\phi}_{X(Y)} = \pm\,\sigma_x\otimes I\ket{\phi}_{X(Y)},\nonumber\\
    I\otimes\sigma_y\ket{\phi}_{X(Y)} = \pm\,\sigma_y\otimes I\ket{\phi}_{X(Y)},
\end{align}
which result in the simultaneous offsetting of the $XX$ and $YY$ terms on the left and right sides of the ladder. Regarding the top and bottom boundaries, the states~(\ref{eq:xy_xdim}) and~(\ref{eq:xy_ydim}) are joint eigenvectors of both the $XX$ and $YY$ terms, with eigenvalues $\pm J_x/4$ and $\pm J_y/4$. In addition, the $H_Z$ term in Eq.~(\ref{eq:xy_model}) preserves the eigenstate due to the identity
\begin{equation}
    I\otimes\sigma_z\ket{\phi}_{X(Y)} = -\,\sigma_z\otimes I\ket{\phi}_{X(Y)},
\end{equation}
meaning that the magnetic field terms also counterbalance one another. As a result, the construction yields possible total energy eigenvalues of $0$ and $\pm(J_x + J_y)/2$.

It should be noted that the model~(\ref{eq:xy_model}) admits a straightforward extension to the inhomogeneous case where the coefficients $J_{x(y)}$ and $h$ become site-dependent and respect the reflection (mirror) symmetry of the ladder:
\begin{align}
    &J_{x(y),i} = J_{x(y),2L-i+2},\quad i=2,\,\ldots,\,L,\label{eq:J_mirr}\\
    &\hfex_i = \hfex_{2L - i +3},\quad i=2,\,\ldots,\,L+1.\label{eq:h_mirr}
\end{align}

The total energy is determined by the two boundary dimers.
An $X$ ($Y$) dimer contributes
$+(J_x+J_y)/4$ ($-(J_x+J_y)/4$).
Since the dimer types alternate, the boundary signs are
opposite for even $L$ and equal for odd $L$. Consequently,
the two choices of the bottom dimer yield
\begin{equation}
    E_\pm=\pm\frac{
    J_{x,1}+J_{y,1}
    +(-1)^{L-1}\bigl(J_{x,L+1}+J_{y,L+1}\bigr)
    }{4}.
\end{equation}
For homogeneous couplings, this reduces to zero for even $L$
and $\pm(J_x+J_y)/2$ for odd $L$. For even $L$ with
inhomogeneous couplings, zero energy requires
$J_{x,1}+J_{y,1}=J_{x,L+1}+J_{y,L+1}$.

A very similar structure, referred to as mirror dimers, was recently discovered in Ref.~\cite{SashMoudXY26} for the spin-$1$ $XY$ model with $J_x=J_y=J$. The main difference is that  the spin-$1$  generalizations of the proposed maximally entangled  dimers (Appendix~\ref{app:sec:dimex_spinone}) are no longer the joint eigenvectors of the two boundary interactions $XX$ and $YY$; while the dimer states in the bulk can be straightforwardly generalized to this case, the elements on the top and bottom sides of the construction must be modified. Using algebraic manipulations of spin operators, the authors derived special single-site configurations at these boundaries. In essence, however, the entire mechanism can still be described via the slide trick and the mutual offsetting of terms, extending to arbitrary spins (including $S=1/2$) as well as arbitrary coupling constants $J_x$ and $J_y$. Further details of this construction are given in Appendix~\ref{app:sec:mirr}.

The isotropic  $XY$ model without an external field,
\begin{equation}
  H_{XY}^{\text{iso}} = \sum_{i=1}^{2L}\,J_i\left(S^x_i  S^x_{i+1} +  S^y_i S^y_{i+1}\right), \end{equation}
  with $J_i$ satisfying~(\ref{eq:J_mirr}), 
admits a ladder construction involving two additional dimer states. Replacing the operator $U_H$ in Eq.~(\ref{eq:had}) (a $\pi$-rotation around $\mathbf{n}_{xz}=\frac1{\sqrt2}(\mf e_x + \mf e_z)$, a vector in the $XZ$ plane) with
\begin{equation}
    R_{xy} =  \frac{\sigma_x + \sigma_y}{\sqrt{2}} \sim\exp\left(-i \frac{\pi}{2} \mathbf n_{xy} \cdot \bm\sigma\right),
\end{equation}
which generates a $\pi$-rotation around $\mathbf{n}_{xy}=\frac1{\sqrt2}(\mathbf e_x + \mathbf e_y)$, we define the rotated dimer states as
\begin{equation}\label{eq:xy_xrdim}
    \ket{\phi}_{RX} = \frac1{\sqrt2}\sum_{i,\,j=0}^1\,(R_{xy}\, \sigma_x)_{ij}\ket{i}\ket{j}
\end{equation}
and 
\begin{equation}\label{eq:xy_yrdim}
    \ket{\phi}_{RY} = \frac1{\sqrt2}\sum_{i,\,j=0}^1\,(R_{xy}\, \sigma_y)_{ij}\ket{i}\ket{j}.
\end{equation}
The sliding properties~(Appendix~\ref{app:sec:dimex})
\begin{align}
    (\sigma_x)_1\otimes&(\sigma_x)_3\ket{\phi_{12}}_{RX}\otimes\ket{\phi_{34}}_{RY} \nonumber\\&= -(\sigma_y)_2\otimes(\sigma_y)_4\ket{\phi_{12}}_{RX}\otimes\ket{\phi_{34}}_{RY},\nonumber\\   
    (\sigma_y)_1\otimes&(\sigma_y)_3\ket{\phi_{12}}_{RX}\otimes\ket{\phi_{34}}_{RY} \nonumber\\&= -(\sigma_x)_2\otimes(\sigma_x)_4\ket{\phi_{12}}_{RX}\otimes\ket{\phi_{34}}_{RY}
\end{align}
result in the left-side $XX$ terms offsetting the $YY$ terms on the right and vice versa, while the boundary relations
\begin{equation}
    \left(\sigma_x\otimes\sigma_x + \sigma_y\otimes\sigma_y\right)\ket{\phi}_{RX(Y)} = 0
\end{equation}
determine an overall zero eigenvalue for the ladder construction. Remarkably, this state remains an eigenvector of the extended Hamiltonian
\begin{equation}
H = H_{XY}^{\text{iso}} + H_{\text{DM}}^{(z)},
\end{equation}
which incorporates the $z$-component Dzyaloshinskii-Moriya interaction:
\begin{equation}
H_{\text{DM}}^{(z)} = \sum_{i=1}^{2L}\,D_i \left( S_i^x S_{i+1}^y - S_i^y S_{i+1}^x \right),
\end{equation}
where the couplings $D_i$ again satisfy the mirror-symmetry condition~(\ref{eq:J_mirr}). That the state remains an eigenstate follows directly from the properties of the rotated dimer states.

Because of the sliding properties of the $S_z$ operator across the rotated dimers~(Appendix~\ref{app:sec:dimex}), the addition of the external field term $H_z$ preserves the eigenstate if the field parameters adhere to the reflection \emph{antisymmetry} relation
\begin{equation}
\hfex_i = -\hfex_{2L - i +3},\quad i=2,\,\ldots,\,L+1.
\end{equation}
Under this condition, field strengths on opposite legs of the ladder carry opposite signs. Consequently, when traversing the ladder in a (counter)clockwise direction, the field changes sign upon crossing the bottom and top boundaries—a behavior that aligns  with the chiral character of the Dzyaloshinskii-Moriya interaction.

%Dimer-based configurations, and the ladder construction in particular, can be extended to yield eigenstates for a variety of spin layouts beyond 1D, such as 2D and 3D lattices. For the $XY$ model, this direction is further investigated in Appendix~?. 

\section{Conclusion}
\label{sec:conclusion}

In this work, we studied how superpositions change the
entanglement of exact eigenstates constructed from antipodal
dimers. Although each valid Pauli-string state is maximally
entangled between the two halves of the chain, their
superpositions can have substantially smaller entanglement.
The stabilizer description allows us to determine the minimum
half-chain entropy over the entire valid-state span.
The polynomial formulation then relates this minimum to the
local terms of the Hamiltonian and makes its dependence on
system size explicit. In particular, for single-generator
models with a nonzero commutation polynomial and fixed
interaction range, the minimum entropy scales as
$L/2-O(1)$. Thus, even an arbitrary superposition cannot remove
the volume-law entanglement in these subspaces.
The $XZ$ model also illustrates that an equal-amplitude
superposition need not be close to optimal.

These results provide a way to study entanglement optimization
without searching numerically over an exponentially large
set of coefficients. They can also serve as benchmarks for
numerical methods that seek weakly entangled states in
degenerate eigenspaces. Beyond the minimum entropy itself,
the structure of the minimizing states suggests further
exact eigenstate constructions. In the examples considered
here, this leads to states at zero and nonzero energies,
including zero-energy states that are separable between
the two halves of the ring.

Our results are related to the conjecture of
Ref.~\cite{Karle2021}, which proposes the existence of
area-law-entangled zero-energy states for generic local
Hamiltonians with exponentially degenerate nullspaces.
The volume-law bounds obtained here apply to the subspace
spanned by the valid antipodal states, rather than to the
full zero-energy subspace. They therefore do not contradict
the conjecture, but show that an exponentially large family
of exact zero modes may still be insufficient to reach
an area-law state. At the same time, our additional
constructions give examples of zero modes with vanishing
entropy across the half-chain cut. Establishing area-law
behavior for other cuts requires examining their geometry
separately. Subsequent work has also constructed weakly
entangled zero modes in constrained
models~\cite{Nicolau2026}, although these model-specific
results do not settle the conjecture in general.

It is also useful to compare our results with the symmetric
superpositions studied in Ref.~\cite{Mukherjee2026}, where
different choices of states and occupations produce
entanglement ranging from volume-law to area-law behavior.
In their area-law construction using the conventional
triplet basis, only a fixed number of dimers are entangled,
while the remaining dimers are product states. This differs
from our starting point, where every valid Pauli-string
state is maximally entangled across the half-chain cut.
However, this distinction alone does not explain our
volume-law bounds: Ref.~\cite{Mukherjee2026} also finds
logarithmic second R\'enyi entropy for certain symmetric
superpositions built from maximally entangled Bell dimers.
The essential restriction in our problem is the particular
subspace selected by the Hamiltonian's validity constraints.
Within this subspace, we optimize over arbitrary
coefficients, rather than a prescribed family of uniform
superpositions.

There is a further connection with the stabilizer approach
to parent Hamiltonians developed in
Ref.~\cite{Dooley2026}. There, one starts from a stabilizer
state and constructs Hamiltonians for which it is an exact
eigenstate. Here, we start from a Hamiltonian and use its
constraints to find states and optimize their entanglement.
For our qubit stabilizer constructions, these two viewpoints
are closely related. It would be interesting to determine
when the parent-Hamiltonian approach reproduces the local
interactions considered here and whether it suggests
additional exact eigenstates of the same models.

Several extensions remain open. One is to develop an
analogous entanglement-optimization framework for higher
spins, going beyond the explicit constructions considered
here. Another is to study two- and three-dimensional
geometries, where both the arrangement of the elementary
entangled blocks and the choice of bipartition become
important. Polynomial descriptions remain available in
higher dimensions, but the one-variable methods used here
would need to be adapted.

Finally, it would be interesting to go beyond the pairwise
cancellation used in our constructions. Related mechanisms
appear in the stabilizer construction of
Ref.~\cite{Dooley2026} and in some of the
interference-based examples of Ref.~\cite{Nicolau2026}.
A natural question is whether one can find simple local
constructions in which three or more Hamiltonian terms
compensate collectively, without their actions splitting
into separately cancelling pairs. Here the terms are
understood with respect to a fixed local decomposition
of the Hamiltonian. Replacing dimers by trimers or larger
entangled blocks may provide a useful starting point,
although larger blocks alone do not guarantee a different
cancellation mechanism.

Diagrammatic methods may help identify and verify such
relations. Graphical descriptions of quantum processes
and tensor networks
\cite{CoeckeKissinger2017,Biamonte2011,BiamonteBergholm2017},
including the ZX-calculus~\cite{VanDeWetering2020},
provide tools for manipulating multipartite states and
operator identities. They could be particularly useful
for extending the constructions to non-stabilizer states,
where Pauli stabilizer relations no longer give a complete
description. An interesting next step is to use these tools
to search systematically for exact eigenstates while
keeping the Hamiltonian interactions local.

\section*{Acknowledgments}

K. Antipin acknowledges the use of TikZiT, developed and maintained
by Aleks Kissinger and contributors, for the preparation of several
diagrams in this manuscript.

The study of K. Antipin was conducted under
the state assignment of Lomonosov Moscow State University.

\emph{AI statement}: ChatGPT-6 Astra helped with calculations and proofs in Appendices~\ref{sec:appendix_mpo_purity}, \ref{app:thm_pol}, \ref{app:sec:sec:smith_form}. The authors take full
responsibility for the scientific content of the work.

\onecolumngrid

\newpage

\appendix

% --- Dynamic Appendix Figure Numbering Setup ---
\renewcommand{\thefigure}{\thesection\arabic{figure}}
\makeatletter
\@addtoreset{figure}{section}
\makeatother
% ------------------------------------------------

\section{Stabilizer formalism and symplectic geometry}
\label{sec:appendix_stabilizer_background}

This appendix provides the necessary background on the stabilizer formalism, the mapping of Pauli operators to symplectic vector spaces, and the iterative construction of maximal commuting subgroups.

\subsection{The Pauli group and the stabilizer subspace}

The central object of the stabilizer formalism is the $N$-qubit Pauli group, denoted $\mathcal{P}_N$. This group consists of all $N$-fold tensor products of the single-qubit Pauli matrices $\{I, X, Y, Z\}$~(Pauli \emph{strings}), together with the global phase factors $\{\pm 1, \pm i\}$. 

A \emph{stabilizer group} $S \subset \mathcal{P}_N$ is defined as an Abelian (mutually commuting) subgroup of the Pauli group that does not contain the negative identity element $-I$. The main physical concept of this formalism is the \emph{stabilizer subspace}, $\mathcal{V}_S$, defined as the simultaneous $+1$ eigenspace of all operators in the stabilizer group. That is, a state $|\psi\rangle$ lies in the stabilizer subspace if and only if it is stabilized by every element of the group: $P |\psi\rangle = |\psi\rangle$ for all $P \in S$.

The projector $\Pi_S$ onto this stabilizer subspace can be constructed explicitly by summing over all elements of the subgroup:
\begin{equation}
    \Pi_S = \frac{1}{|S|} \sum_{P \in S} P.
\end{equation}
The dimension of the stabilizer subspace is exactly the rank of this projector, which is given by its trace. Because all non-identity Pauli operators are strictly traceless ($\operatorname{Tr}(P) = 0$ for $P \neq I$), only the identity operator $I$ contributes to the sum. Evaluating the trace yields:
\begin{equation}
    \dim(\mathcal{V}_{S}) = \operatorname{Tr}(\Pi_S) = \frac{1}{|S|} \sum_{P \in S} \operatorname{Tr}(P) = \frac{\operatorname{Tr}(I)}{|S|} = \frac{2^N}{|S|}.
\end{equation}
If the stabilizer group $S$ is generated by $k$ independent, mutually commuting Pauli operators, the size of the group is $|S| = 2^k$. Consequently, the stabilized subspace has dimension $2^{N-k}$. When $k=N$, the subspace is one-dimensional, uniquely specifying a single quantum state~(\emph{stabilizer state}).

\subsection{Check vectors and symplectic geometry}

The stabilizer formalism facilitates working with Pauli operators by mapping them to vectors over the binary field $\mathbb{F}_2$. Up to a global phase, any $N$-qubit Pauli operator $P \in \mathcal{P}_N$ can be uniquely factorized into its $X$ and $Z$ components: $P \propto X^{\mathbf{u}} Z^{\mathbf{v}}$, where $\mathbf{u}, \mathbf{v} \in \mathbb{F}_2^N$. This establishes a homomorphism from $\mathcal{P}_N$ to the $2N$-dimensional vector space $\mathbb{F}_2^{2N}$. The operator $P$ is represented by a $2N$-component \emph{check vector}:
\begin{equation}
    \mathbf{r} = \begin{pmatrix} \mathbf{u} \\ \mathbf{v} \end{pmatrix} \in \mathbb{F}_2^{2N}.
\end{equation}
Under this mapping, the product of two Pauli operators corresponds (up to a global phase) to the element-wise addition of their check vectors modulo 2: $P_1 P_2 \mapsto \mathbf{r}_1 + \mathbf{r}_2 \pmod 2$. 

Remarkably, the commutation relations of the Pauli operators are connected to the geometry of this binary vector space. Two Pauli operators $P_1$ and $P_2$ commute if and only if their corresponding check vectors $\mathbf{r}_1$ and $\mathbf{r}_2$ are orthogonal under the \emph{symplectic inner product}:
\begin{equation}
    \langle \mathbf{r}_1, \mathbf{r}_2 \rangle = \mathbf{r}_1^T \Omega \, \mathbf{r}_2 = \mathbf{u}_1 \cdot \mathbf{v}_2 + \mathbf{v}_1 \cdot \mathbf{u}_2 \pmod 2,
\end{equation}
where 
\begin{equation}
    \Omega = \begin{pmatrix} 0 & I \\ I & 0 \end{pmatrix}
\end{equation}
is the standard symplectic form matrix.

In the geometric language, finding a set of mutually commuting Pauli operators is equivalent to finding a specific type of subspace in $\mathbb{F}_2^{2N}$. We require the following formal definitions from symplectic geometry:

\begin{itemize}
    \item \textbf{Symplectic Orthogonal Complement ($V^\perp$):} For any subspace $V \subseteq \mathbb{F}_2^{2N}$, its orthogonal complement is the set of all vectors that symplectically commute with every vector in $V$:
    \begin{equation}
        V^\perp = \{ \mathbf{u} \in \mathbb{F}_2^{2N} \mid \langle \mathbf{u}, \mathbf{v} \rangle = 0 \quad \forall \mathbf{v} \in V \}.
    \end{equation}
    
    \item \textbf{Isotropic Subspace ($\mathcal S$):} A subspace $\mathcal S$ is \emph{isotropic} if $\mathcal S \subseteq \mathcal S^\perp$. Physically, this means every operator in the corresponding Pauli subgroup $S$~(note the difference between  $\mathcal S\subset\mathbb{F}_2^{2N}$ and  $S\subset\mathcal{P}_N$) commutes with every other operator in $S$. An isotropic subspace thus serves as the exact geometric representation of the stabilizer group introduced in the previous section.
    
    \item \textbf{The Center / Radical ($C$):} The center of a subspace $V$ consists of the elements inside $V$ that commute with everything else in $V$. It is defined as $C = V \cap V^\perp$. If $C = \{0\}$, the space $V$ is called \emph{non-degenerate} or \emph{symplectic}.
\end{itemize}

\subsection{Iterative construction of maximal commuting subgroups}\label{app:sec:iter_symp}

In applications, it is often needed to construct the largest possible set of commuting observables drawn from a restricted, physically motivated set of operators. Let $V \subseteq \mathbb{F}_2^{2N}$ be an arbitrary vector space representing such a set (in the context of our paper,
$V = \mathcal{W}^\perp$, the space of Pauli check vectors
commuting with every translated Hamiltonian generator).

We wish to construct a maximal isotropic subspace $\mathcal S_{\max} \subseteq V$. This is achieved through a constructive procedure which is a symplectic Gram-Schmidt orthogonalization:
\begin{enumerate}
    \item First, identify the center $C = V \cap V^\perp$. Because the elements of $C$ commute with \emph{all} operators in $V$, they strictly commute with any mutually commuting subset we could possibly build. Therefore, any maximal commuting set \emph{must} contain the entire center. We initialize our stabilizer group with the center: $\mathcal S_0 = C$.

    \item To add more generators, we look at the remaining degrees of freedom by moving to the quotient space $V / C$. Because we have factored out the center, this remaining space is strictly non-degenerate: no non-trivial operator here commutes with everything else.

    We pick an arbitrary basis vector $\mathbf{e}_1 \in V \setminus C$. Because the space is non-degenerate, there must exist at least one other vector $\mathbf{f}_1 \in V$ such that $\langle \mathbf{e}_1, \mathbf{f}_1 \rangle = 1$. Physically, this means $\mathbf{e}_1$ and $\mathbf{f}_1$ are a pair of anti-commuting operators (analogous to a canonical logical $X_1$ and $Z_1$ pair).

    \item We add $\mathbf{e}_1$ to the initial subspace: $\mathcal S_1 = \mathcal S_0 \oplus \operatorname{span}\{\mathbf{e}_1\}$. Because $\mathbf{f}_1$ anti-commutes with $\mathbf{e}_1$, it cannot be added to our subspace. We have thus extracted a two-dimensional symplectic plane spanned by $\{\mathbf{e}_1, \mathbf{f}_1\}$. We now restrict our search to the subspace of $V$ that is symplectically orthogonal to both $\mathbf{e}_1$ and $\mathbf{f}_1$.

    \item We repeat the pairing process in the restricted subspace. We find another anti-commuting pair $\{\mathbf{e}_2, \mathbf{f}_2\}$ that commutes with the first pair. We add $\mathbf{e}_2$ to our stabilizer $\mathcal S_2 = \mathcal S_1 \oplus \operatorname{span}\{\mathbf{e}_2\}$ and discard $\mathbf{f}_2$. 
\end{enumerate}

We can continue this iterative process until the non-degenerate degrees of freedom are entirely exhausted, decomposing $V$ into the center and a direct sum of mutually orthogonal symplectic planes.

\subsection{The subspace dimension formula}

The iterative construction directly proves the dimensional formulas used in the main text. 

Let the dimension of the initial space of operators be $\dim V$, and the dimension of its center be $\dim C$. The dimension of the non-degenerate part of the space (the operators participating in the pairing process) is exactly $\dim V - \dim C$.

In every step of the iteration, we consume exactly two vectors from the non-degenerate space (one $\mathbf{e}_i$ and one $\mathbf{f}_i$), but we only add one commuting generator (the $\mathbf{e}_i$) to our maximal stabilizer subgroup. Therefore, the number of independent generators we can extract from the non-degenerate space is exactly half of its dimension:
\begin{equation}
    \#[\mathrm{added\,generators}] = \frac{1}{2} (\dim V - \dim C).
\end{equation}
The total dimension of the maximal isotropic subspace $\mathcal S_{\max}$ is the sum of the initial center generators and the iteratively added generators:
\begin{equation}
    \dim(\mathcal S_{\max}) = \dim C + \frac{1}{2} (\dim V - \dim C) = \frac{1}{2} (\dim V + \dim C).
\end{equation}

In the context of the present paper, by substituting the valid constraint space $V = \mathcal{W}^\perp$ and the generated center $C = \mathcal{W} \cap \mathcal{W}^\perp$, we obtain the exact dimensional formula used in the main text:
\begin{equation}
    \dim(\mathcal S_{\max}) = \frac{\dim(\mathcal{W}^\perp) + \dim(\mathcal{W} \cap \mathcal{W}^\perp)}{2}.
\end{equation}
The subspace $\mathcal S_{\max}$ corresponds to the maximal stabilizer subgroup $S_{\max}$ of order $|S_{\max}| = 2^{\dim(\mathcal S_{\max})}$.

\subsection{Clifford unitaries and the fiducial canonical form}
\label{app:sec:clifford_background}

In the main text, we utilized a unitary $U_C$ to map a set of $L$ independent Pauli generators into a standard \emph{fiducial canonical form} acting on single qubits. This transformation belongs to the Clifford group, which is the natural set of operations for manipulating stabilizer groups.

The $N$-qubit Clifford group, denoted $\mathcal{C}_N$, is defined as the normalizer of the Pauli group $\mathcal{P}_N$ within the unitary group $U(2^N)$. That is, a unitary operator $U$ is a Clifford operator if and only if $U P U^\dagger \in \mathcal{P}_N$ for all $P \in \mathcal{P}_N$. 

The Clifford group is generated by three elementary gates: the Hadamard gate ($H$), the Phase gate ($S$), and the controlled-NOT gate ($\text{CNOT}$). The single-qubit $H$ and $S$ gates are essential for changing bases. Because $H = H^\dagger$, its action by conjugation swaps the $X$ and $Z$ bases:
\begin{equation}
    H X H = Z, \quad H Z H = X.
\end{equation}
The $S$ gate leaves $Z$ unchanged but rotates the $X$ and $Y$ bases:
\begin{equation}
    S X S^\dagger = Y, \quad S Z S^\dagger = Z.
\end{equation}

Because $U_C \in \mathcal{C}_N$ maps Pauli operators to Pauli operators, its action on the Pauli group can be tracked entirely via the $2N$-dimensional check vectors introduced in Appendix~\ref{sec:appendix_stabilizer_background}. Under a Clifford unitary, the mapping of check vectors takes the form
\begin{equation}
    \mathbf{r} \mapsto M \mathbf{r} \pmod 2,
\end{equation}
where $M \in \operatorname{Sp}(2N, \mathbb{F}_2)$ is a symplectic matrix (i.e., it preserves the symplectic inner product $M^T \Omega M = \Omega$).  

The canonical generator set described in the main text consists of $m$ anti-commuting pairs $(g_j, h_j)$ and $L-2m$ central generators $c_k$. The target fiducial set consists of $(X_j, Z_j)$ for $j=1,\ldots,m$, and $Z_{m+k}$ for $k=1,\ldots,L-2m$. Because the source generators and the target generators share the same commutation relations, the existence of the Clifford unitary $U_C$  is mathematically guaranteed by the symplectic geometry.

The  guarantee  rests on a fundamental result  known as \emph{Witt's Extension Theorem}\cite{artin1988geometric,grove2002classical}~ (specifically its adaptation to symplectic spaces). In the check vector formalism, the commutation relation between any two Pauli operators is defined by the symplectic inner product of their binary vectors. Therefore, stating that the source generators and the target fiducial generators have identically matching mutual commutation graphs is  equivalent to stating that their check vectors have the  same symplectic Gram matrix. 

If we let the source check vectors be $\{\mathbf{v}_1, \dots, \mathbf{v}_L\}$ and the target check vectors be $\{\mathbf{u}_1, \dots, \mathbf{u}_L\}$, the matching commutation relations enforce $\langle \mathbf{v}_i, \mathbf{v}_j \rangle = \langle \mathbf{u}_i, \mathbf{u}_j \rangle$ for all $i, j$. This implies that the linear map defined by $T(\mathbf{v}_i) = \mathbf{u}_i$ is a local  isometry between the $L$-dimensional subspaces spanned by these sets of vectors. 

Witt's Extension Theorem states that in a non-degenerate symplectic space, any such local isometry between subspaces can be extended to a global symplectic transformation $M \in \operatorname{Sp}(2N, \mathbb{F}_2)$ acting on the entire $2N$-dimensional space. Finally, because there is a surjective homomorphism from the Clifford group $\mathcal{C}_N$ to the symplectic group $\operatorname{Sp}(2N, \mathbb{F}_2)$~(where the kernel is the Pauli group up to global phases), the existence of the global symplectic matrix $M$ strictly guarantees the existence of a Clifford unitary $U_C$ that implements this transformation. Any residual sign mismatches~(for example, if a generator is mapped to $-X_j$ rather than $+X_j$) can be trivially corrected by composing $U_C$ with an appropriate Pauli operator, which is itself a Clifford operation.

The unitary $U_C$ can also be explicitly constructed as a quantum circuit using a procedure analogous to Gaussian elimination over the binary field $\mathbb{F}_2$. In this algorithmic construction, single-qubit $H$ and $S$ gates are used to rotate an arbitrary Pauli operator on a given qubit into either $X$ or $Z$. This establishes the necessary pivots for two-qubit $\mathrm{CNOT}$ gates, which perform row additions across different qubits. Rather than detailing the full procedure here, we refer the reader to standard literature on the stabilizer formalism~\cite{Gottesman1997, AaronsonGottesman2004}.

\section{Optimal superpositions for the
$XZ$ Hamiltonian at $L=2$ and $L=3$}
\label{app:XZ}

In this appendix, we construct optimal superpositions explicitly
for $L=2$ and $L=3$. We use
\begin{equation}
    \mathcal C_{\mathrm{opt}}
    =\frac{1}{\sqrt{r_{\min}}}
     X^{\otimes L}\Pi_{\min},
\end{equation}
with the multiplication order reversed relative to
Eq.~\eqref{eq:opt_A}. This also gives a linear combination
of valid Pauli strings, since left multiplication by
$X^{\otimes L}$ maps the homogeneous Pauli group to the
valid-string coset, up to phases. Moreover,
\begin{align}
    \mathcal C_{\mathrm{opt}}
    \mathcal C_{\mathrm{opt}}^\dagger
    &=\frac{1}{r_{\min}}
      X^{\otimes L}\Pi_{\min}X^{\otimes L},\\
    \mathcal C_{\mathrm{opt}}^\dagger
    \mathcal C_{\mathrm{opt}}
    &=\frac{1}{r_{\min}}\Pi_{\min}.
\end{align}
Thus, $\operatorname{Tr}(
\mathcal C_{\mathrm{opt}}\mathcal C_{\mathrm{opt}}^\dagger)=1$,
and the minimum entanglement is unchanged.

\subsection{Case $L=2$}

By Eq.~\eqref{eq:G_gen}, $m=0$. A maximal stabilizer group
has order $2^{L-m}=4$, and its projector has rank
$r_{\min}=2^m=1$. The valid Pauli strings are
\begin{equation}
    XX,\quad YI,\quad IY,\quad ZZ.
\end{equation}
Left multiplication by $X^{\otimes2}=XX$ gives, up to
scalar phases, the representatives
\begin{equation}
    II,\quad ZX,\quad XZ,\quad YY.
\end{equation}
These representatives form a stabilizer group generated
by $ZX$ and $XZ$, with projector
\begin{equation}
    \Pi_{\min}
    =\frac14\left(II+ZX+XZ+YY\right).
\end{equation}
The corresponding normalized optimal operator is
\begin{equation}
    \mathcal C_{\mathrm{opt}}
    =X^{\otimes2}\Pi_{\min}
    =\frac14\left(XX-iYI-iIY-ZZ\right).
\end{equation}

The reduced density operator on sites $1$ and $2$ is
\begin{align}
    \rho_{12}
    &=\mathcal C_{\mathrm{opt}}
      \mathcal C_{\mathrm{opt}}^\dagger \nonumber\\
    &=X^{\otimes2}\Pi_{\min}X^{\otimes2} \nonumber\\
    &=\frac14\left(II-ZX-XZ+YY\right).
\end{align}
This is a rank-one stabilizer projector with generators
$-ZX$ and $-XZ$. It projects onto
\begin{equation}
    \ket{\psi}_{12}
    =\frac{1}{\sqrt2}
      \sum_{i,j=0}^{1}(U_HY)_{ij}
      \ket{i}_1\otimes\ket{j}_2,
\end{equation}
where
\begin{equation}
    \label{eq:app:hadard}
    U_H=\frac{1}{\sqrt2}
    \begin{pmatrix}
        1&1\\
        1&-1
    \end{pmatrix}
\end{equation}
is the Hadamard matrix. The stabilizer relations can be
verified directly using the vectorization identities
in Appendix~\ref{app:sec:vec}.

The reduced density operator on the opposite half is
\begin{equation}
    \rho_{\Bar{1}\Bar{2}}
    =\left(
      \mathcal C_{\mathrm{opt}}^\dagger
      \mathcal C_{\mathrm{opt}}
      \right)^T
    =\Pi_{\min},
\end{equation}
where $\Pi_{\min}^T=\Pi_{\min}$. It projects onto
\begin{equation}
    \ket{\phi}_{\Bar{1}\Bar{2}}
    =\frac{1}{\sqrt2}
      \sum_{i,j=0}^{1}(U_H)_{ij}
      \ket{i}_{\Bar{1}}\otimes\ket{j}_{\Bar{2}},
\end{equation}
which is stabilized by $ZX$ and $XZ$.

Consequently, up to an overall phase, the optimal
superposition is
\begin{equation}
    \ket{\Psi}_{\mathcal C_{\mathrm{opt}}}
    =\ket{\psi}_{12}\otimes
     \ket{\phi}_{\Bar{1}\Bar{2}}.
\end{equation}
The two halves are therefore unentangled, although each
half contains a maximally entangled two-qubit state.
Thus, $S_{\min}^{\mathrm{ent}}=0$.

\subsection{Case $L=3$}

For $L=3$, we have $m=1$. A maximal stabilizer group again
has order $2^{L-m}=4$, while its projector now has rank
$r_{\min}=2^m=2$. The eight valid Pauli strings are
\begin{equation}
    \begin{gathered}
        XXX,\quad XYI,\quad YIX,\quad YZI,\\
        IXY,\quad IYZ,\quad ZIY,\quad ZZZ.
    \end{gathered}
\end{equation}
Left multiplication by $X^{\otimes3}=XXX$ gives the
following representatives of $G$, up to scalar phases:
\begin{equation}
    \begin{gathered}
        III,\quad IZX,\quad ZXI,\quad ZYX,\\
        XIZ,\quad XZY,\quad YXZ,\quad YYY.
    \end{gathered}
\end{equation}
The commuting operators $YXZ$ and $IZX$ generate a
stabilizer subgroup of order four. Since their product
is $YYY$, the associated projector is
\begin{equation}
    \Pi_{\min}
    =\frac14\left(III+YYY+YXZ+IZX\right).
\end{equation}
The normalized optimal operator is therefore
\begin{align}
    \mathcal C_{\mathrm{opt}}
    &=\frac{1}{\sqrt2}X^{\otimes3}\Pi_{\min} \nonumber\\
    &=\frac{1}{4\sqrt2}
      \left(XXX-iZZZ+ZIY-iXYI\right).
\end{align}
Its reduced density operator on the first half is
\begin{equation}
    \rho_{123}
    =\mathcal C_{\mathrm{opt}}
     \mathcal C_{\mathrm{opt}}^\dagger
    =\frac12 X^{\otimes3}\Pi_{\min}X^{\otimes3}.
\end{equation}
Because $\Pi_{\min}$ has rank two, the two nonzero
eigenvalues of $\rho_{123}$ are both $1/2$. Hence,
\begin{equation}
    \operatorname{Tr}\rho_{123}^2=\frac12,
    \qquad
    S_{\min}^{\mathrm{ent}}=1,
\end{equation}
in agreement with Eqs.~\eqref{eq:opt_rank}
and~\eqref{eq:entropy_xz}.

\section{Uniform superposition of valid Pauli strings for the  Hamiltonian generated by the $XZ$ term}
\label{sec:appendix_mpo_purity}

In this appendix, we analyze the purity of the state corresponding to the uniform superposition of all Pauli strings that satisfy the anti-commutation constraints of the $XZ$ Hamiltonian. We encode this sum as a matrix product operator (MPO) and calculate its purity  using the transfer matrix approach.

\subsection{MPO representation and vectorization}

Consider the set of all valid $N$-qubit Pauli strings that strictly anti-commute with the Hamiltonian term $XZ$ and all its cyclic shifts. We define the operator $A$ as the uniform superposition over these strings. As shown in the main text, valid Pauli strings are defined by the following transition rules between Pauli operators on adjacent sites:
\begin{align}\label{app:eq:transit_rules}
    I &\longrightarrow X \text{ or } Y, & X &\longrightarrow X \text{ or } Y, \nonumber\\
    Y &\longrightarrow I \text{ or } Z, & Z &\longrightarrow Z \text{ or } I.
\end{align}

To encode the transition rules, we use the  mapping function $f: \{0,1\}^2 \to \{I, X, Y, Z\}$ defined as:
\begin{equation}
    f(0,0) = X, \quad f(0,1) = Y, \quad f(1,0) = I, \quad f(1,1) = Z.
\end{equation}
This can also be expressed as $f(i,\,j)=M_{ij}$, where 
\begin{equation}\label{app:eq:encod_mat}
    M = \begin{pmatrix}
        X & Y\\
        I & Z
    \end{pmatrix}.
\end{equation}

Introducing binary strings $\mathbf{a}\in\mathbb{F}_2^{L}$  with $a_{L+1}\equiv a_1$~(PBC), we represent the uniform superposition of valid Pauli strings in the form:
\begin{equation}\label{app:eq:unif_direct}
    A = \frac1{2^L}\sum_{\mathbf{a}\in\mathbb{F}_2^{L}}\bigotimes_{i=1}^L\,\sigma_{f(a_i a_{i+1})}.
\end{equation}
One can check that each term in this sum, which corresponds to one of $2^L$ possible binary strings, produces a valid Pauli string, i.~e., the mapping $f$ correctly encodes the transition rules of Eq.~\eqref{app:eq:transit_rules}.

By utilizing the auxiliary $2\times2$ matrices $M^{(p)}$ with entries $M^{(p)}_{ij} = \delta_{p, f(i,j)}$, for $i,\,j\in\{0,\,1\}$, the operator $A$ can be rewritten in the MPO form:
\begin{equation}\label{app:eq:mpo_xz}
    A = \frac{1}{2^L} \sum_{p_1, \dots, p_L \in \{I, X, Y, Z\}} \Tr(M^{(p_1)} M^{(p_2)} \dots M^{(p_L)}) \, \sigma_{p_1} \otimes \sigma_{p_2} \otimes \dots \otimes \sigma_{p_L}.
\end{equation}
By construction, this is an MPO with bond dimension $2$. Although we will use Eq.~\eqref{app:eq:unif_direct} directly, the MPO expression can be useful for further generalizations.

When vectorized, $A$ becomes a quantum state $|A\rangle\!\rangle$. Since uniform superposition $A$ is Hermitian, the purity of this state is given by the trace of the fourth power of the operator, $P = \Tr(A^4)$. To evaluate $A^4$, we take four independent copies of $A$ from Eq.~\eqref{app:eq:unif_direct}. The full trace factors into a product of local traces over the  Pauli matrices at each site:
\begin{equation}\label{app:eq:mpo_pur}
    P = \frac1{2^{4L}}\sum_{\mathbf a,\mathbf b,\mathbf c,\mathbf d\in\mathbb{F}_2^L}\prod_{i=1}^L\Tr(\sigma_{f(a_i a_{i+1})}\sigma_{f(b_i b_{i+1})}\sigma_{f(c_i c_{i+1})}\sigma_{f(d_i d_{i+1})}).
\end{equation}

Introducing the $16 \times 16$ transfer matrix $\mathbb{T}$, whose rows and columns are labeled by the multi-indices $\vec{i} = (i_1, i_2, i_3, i_4)$ and $\vec{j} = (j_1, j_2, j_3, j_4)$, we define its elements as:
\begin{equation}\label{app:eq:transfer_xz}
    \mathbb{T}_{\vec{i}, \vec{j}} = \frac12\Tr\Big( \sigma_{f(i_1, j_1)} \sigma_{f(i_2, j_2)} \sigma_{f(i_3, j_3)} \sigma_{f(i_4, j_4)} \Big),
\end{equation}
where the factor $1/2$ is introduced for normalization of the trace expression.

The expression for the purity \eqref{app:eq:mpo_pur} now can be written in a compact form as:
\begin{equation}
    P = \frac{2^L}{2^{4L}} \Tr(\mathbb{T}^L)=\frac1{2^{3L}} \Tr(\mathbb{T}^L).
\end{equation}

\subsection{Exact diagonalization of the transfer matrix}

\subsubsection{Reduction to the $8 \times 8$ matrix}
The transfer matrix  acts initially on the  space of $4$ binary variables, corresponding to a $16 \times 16$ matrix defined on the computational basis $\{|i_1 i_2 i_3 i_4\rangle\}$. The matrix elements are given by the trace of four constituent operators:
\begin{equation}
    \mathbb{T}_{\vec{i}, \vec{j}} = \frac{1}{2} \operatorname{Tr}( M_{i_1 j_1} M_{i_2 j_2} M_{i_3 j_3} M_{i_4 j_4} ),
\end{equation}
where the local components $M_{ij} \in \{X, Y, I, Z\}$ are defined by \eqref{app:eq:encod_mat}.

To determine which matrix elements are non-vanishing, we can decompose any Pauli operator into the standard $X^a Z^b$ representation: $I \propto X^0 Z^0$, $X \propto X^1 Z^0$, $Z \propto X^0 Z^1$, and $Y \propto X^1 Z^1$. Observing the index mapping, we can compactly express the operator selection as:
\begin{equation}
    M_{i, j} \propto X^{(1-i)} Z^{j}.
\end{equation}
The trace of a product of Pauli matrices is non-zero if and only if the product is proportional to the identity. This requires that the total number of $X$ operations and the total number of $Z$ operations must both be even. Substituting the operator mapping into this trace rule imposes two  parity constraints on the input string $\vec{i}$ and output string $\vec{j}$:
\begin{align}
    \text{X-parity:} \quad \sum_{k=1}^4 (1-i_k) &= 4 - \sum_{k=1}^4 i_k \equiv 0 \pmod 2 \implies |\vec{i}| \text{ is even}, \\
    \text{Z-parity:} \quad \sum_{k=1}^4 j_k &= |\vec{j}| \equiv 0 \pmod 2 \implies |\vec{j}| \text{ is even},
\end{align}
where $|\vec{v}|$ denotes the Hamming weight (number of $1$s) of the bitstring. 

These constraints show that $\mathbb{T}_{\vec{i}, \vec{j}} = 0$ if either $\vec i$ or $\vec j$ has an odd Hamming weight. Consequently, the original $16 \times 16$ transfer matrix  partitions into a null $8 \times 8$ odd-parity block, null off-diagonal blocks, and a single non-trivial $8 \times 8$ block operating  on the even-weight subspace. The   computational basis of this latter subspace consists  of the $8$ states with weight $w \in \{0, 2, 4\}$, which we order as:
\begin{equation}
    \mathcal{H}_{\text{even}} = \{ |0000\rangle, |0101\rangle, |1010\rangle, |0011\rangle, |0110\rangle, |1100\rangle, |1001\rangle, |1111\rangle \}.
\end{equation}
Evaluating the trace expression~\eqref{app:eq:transfer_xz} over this ordered basis  yields the  reduced $8 \times 8$ transfer matrix:
\begin{equation}
    \mathbb{T}_8 = \begin{pmatrix}
    1 & -1 & -1 &  1 &  1 &  1 &  1 &  1 \\
    1 & -1 &  1 &  i & -i &  i & -i & -1 \\
    1 &  1 & -1 & -i &  i & -i &  i & -1 \\
    1 &  i & -i &  1 &  i &  1 & -i &  1 \\
    1 & -i &  i & -i &  1 &  i &  1 &  1 \\
    1 &  i & -i &  1 & -i &  1 &  i &  1 \\
    1 & -i &  i &  i &  1 & -i &  1 &  1 \\
    1 &  1 &  1 &  1 &  1 &  1 &  1 &  1
    \end{pmatrix}.
\end{equation}

\subsubsection{Further reduction via  the cyclic symmetry}
To  diagonalize the $8 \times 8$ transfer matrix, we exploit its commutation with the  cyclic shift operator $P_{\text{cyc}}$ (defined such that it shifts bits to the right). This symmetry originates from the cyclic property of trace in expression~\eqref{app:eq:transfer_xz}.  Because $[ \mathbb{T}_8, P_{\text{cyc}} ] = 0$, the matrix $\mathbb{T}_8$ takes block diagonal form in the eigen-basis of  the cyclic shift operator. 

We construct such eigen-basis using the Discrete Fourier Transform  operator:
\begin{equation}\label{app:eq:proj_dft}
    \Pi_k = \frac{1}{d} \sum_{m=0}^{d-1} e^{-i \frac{2\pi}{d} k m} P_{\text{cyc}}^m,
\end{equation}
where $d=4$ is the order of the cyclic shift operator and $k \in \{0, 1, 2, 3\}$. By applying $\Pi_k$ to a representative computational basis state $|s_{a,0}\rangle$ belonging to an orbit $O_a$ of length $N_a$~(under the action of $P_{\text{cyc}}$), we generate the unnormalized momentum eigenstate. The properly normalized basis state $|v_a^{(k)}\rangle$ is thus defined as:
\begin{equation}\label{app:eq:mom_base}
    |v_a^{(k)}\rangle = \frac{1}{\sqrt{N_a}} \sum_{m=0}^{N_a-1} e^{-i \frac{2\pi}{d} k m} |s_{a,m}\rangle,
\end{equation}
where $|s_{a,m}\rangle \equiv P_{\text{cyc}}^m |s_{a,0}\rangle$. The normalization constant $1/\sqrt{N_a}$ ensures that $\langle v_a^{(k)} | v_a^{(k)} \rangle = 1$, which naturally follows from the orthogonality of the distinct computational states within the orbit ($\langle s_{a,m} | s_{a,m'} \rangle = \delta_{m,m'}$). By construction, the eigenvalues of $P_{\text{cyc}}$ are defined by the momentum number:
\begin{equation}
    P_{\text{cyc}} |v_a^{(k)}\rangle = e^{i \frac{2\pi k}{d}} |v_a^{(k)}\rangle.
\end{equation}

It is crucial to note that the exponent's denominator in~\eqref{app:eq:proj_dft} and~\eqref{app:eq:mom_base} is exactly the global parameter $d=4$, not the local orbit length $N_a$. The momentum  number $k$ is a representation of the global cyclic group $\mathbb{Z}_d$, which means that the phase acquired by shifting one site~(bit) is strictly $2\pi/d$, regardless of the period of the specific orbit.

For an orbit of length $N$, the transformation $P_{\text{cyc}}^N$ acts as the identity. If a non-vanishing eigenstate $|v^{(k)}\rangle$ is constructed from this orbit, applying $N$ shifts must leave the state invariant:
\begin{equation}
    P_{\text{cyc}}^N |v^{(k)}\rangle = \left(e^{i \frac{2\pi k}{d}}\right)^N |v^{(k)}\rangle = |v^{(k)}\rangle,
\end{equation}
which imposes a condition on the momentum number:
\begin{equation}
    e^{i 2\pi \frac{k N}{d}} = 1 \implies \frac{k N}{d} \in \mathbb{Z}.
\end{equation}

Applying this condition  to all possible orbits yields the eigenvectors within the following momentum sectors:  
\begin{itemize}
    \item \textbf{Length $N=1$ orbits} ($O_1=\{|0000\rangle\}$, $O_2=\{|1111\rangle\}$): the condition $k/4 \in \mathbb{Z}$ implies $k=0$.
    \item \textbf{Length $N=2$ orbit} ($O_3 = \{|0101\rangle, |1010\rangle\}$): the condition $2k/4 = k/2 \in \mathbb{Z}$ restricts this orbit to the $k=0$ and $k=2$ sectors.
    \item \textbf{Length $N=4$ orbit} ($O_4 = \{|0011\rangle, |1001\rangle, |1100\rangle, |0110\rangle\}$): the condition $4k/4 = k \in \mathbb{Z}$ is satisfied for all $k\in\{0,\,1,\,2,\,3\}$. This orbit contributes one vector to each of the four $k$-sectors.
\end{itemize}
As a result, the $8$-dimensional subspace $\mathcal{H}_{\text{even}}$ is partitioned into smaller subspaces~(momentum sectors): a $4$-dimensional sector $k=0$, a $2$-dimensional sector  $k=2$, and two $1$-dimensional sectors for $k=1$ and $k=3$.

\subsubsection{Derivation of the expression for matrix element}

We only need to evaluate matrix elements of the transfer matrix $\mathbb{T}_8$ between basis states within the same momentum sector. This is a direct consequence of the commutation relation $[\mathbb{T}_8, P_{\text{cyc}}] = 0$, which implies $\mathbb{T}_8 = P_{\text{cyc}}^\dagger \mathbb{T}_8 P_{\text{cyc}}$. Evaluating a general transition amplitude between states of momenta $k$ and $k'$ yields:
\begin{equation}\label{app:eq:aux_block_mom}
    \langle v_a^{(k')} | \mathbb{T}_8 | v_b^{(k)} \rangle = \langle v_a^{(k')} | P_{\text{cyc}}^\dagger \mathbb{T}_8 P_{\text{cyc}} | v_b^{(k)} \rangle = e^{i \frac{2\pi}{d}(k - k')} \langle v_a^{(k')} | \mathbb{T}_8 | v_b^{(k)} \rangle.
\end{equation}
If $k\ne k'$, the phase factor $e^{i \frac{2\pi}{d}(k - k')} \neq 1$: because $d=4$ and $k$ and $k'$ are restricted to values $\{0,\,1,\,2,\,3\}$, the maximum possible difference $\abs{k-k'}$ is $3$, and so the phase factor is only $1$ when $k-k'$ is a multiple of $4$, which means $k=k'$. Consequently, for the equality~\eqref{app:eq:aux_block_mom} to hold, the matrix element $\langle v_a^{(k')} | \mathbb{T}_8 | v_b^{(k)} \rangle$ must be identically zero. Therefore, $\mathbb{T}_8$ is exactly block-diagonalized by the momentum sectors  of the basis~\eqref{app:eq:mom_base}, and we only need to construct the sub-matrices where $k = k'$.

To compute the transfer matrix elements between any two normalized symmetry-adapted basis states, $|v_a^{(k)}\rangle$ (from orbit $O_a$) and $|v_b^{(k)}\rangle$ (from orbit $O_b$), we evaluate the double sum generated by their definitions:
\begin{equation}
    \langle v_a^{(k)} | \mathbb{T}_8 | v_b^{(k)} \rangle = \frac{1}{\sqrt{N_a N_b}} \sum_{m=0}^{N_a-1} \sum_{n=0}^{N_b-1} e^{i \frac{2\pi}{d} k (m-n)} \langle s_{a,m} | \mathbb{T}_8 | s_{b,n} \rangle.
\end{equation}
We now exploit the translational invariance of the transfer matrix, $[ \mathbb{T}_8, P_{\text{cyc}} ] = 0$, which implies $\mathbb{T}_8 = (P_{\text{cyc}}^\dagger)^m \mathbb{T}_8 P_{\text{cyc}}^m$. Applying this to the bare matrix element allows us to shift both states backwards by $m$:
\begin{equation}
    \langle s_{a,m} | \mathbb{T}_8 | s_{b,n} \rangle = \langle P_{\text{cyc}}^m s_{a,0} | \mathbb{T}_8 | P_{\text{cyc}}^n s_{b,0} \rangle = \langle s_{a,0} | \mathbb{T}_8 | P_{\text{cyc}}^{n-m} s_{b,0} \rangle.
\end{equation}

When evaluating the inner sum over $n$, we introduce the relative shift variable $r' = n - m$. For a fixed  position $m$, as $n$ runs from $0$ to $N_b - 1$, the new index $r'$ originally runs from $-m$ to $N_b - 1 - m$. The matrix element thus becomes:
\begin{equation}
    \langle v_a^{(k)} | \mathbb{T}_8 | v_b^{(k)} \rangle = \frac{1}{\sqrt{N_a N_b}} \sum_{m=0}^{N_a-1} \left( \sum_{r'=-m}^{N_b-1-m} e^{-i \frac{2\pi}{d} k r'} \langle s_{a,0} | \mathbb{T}_8 | s_{b,r'} \rangle \right).
\end{equation}
To rewrite this inner sum from $0$ to $N_b - 1$, we must prove that the summand is strictly periodic with period $N_b$. Let $f(r') = e^{-i \frac{2\pi}{d} k r'} \langle s_{a,0} | \mathbb{T}_8 | s_{b,r'} \rangle$ and evaluate $f(r' + N_b)$:
\begin{equation}
    f(r' + N_b) = e^{-i \frac{2\pi}{d} k (r' + N_b)} \langle s_{a,0} | \mathbb{T}_8 | P_{\text{cyc}}^{r' + N_b} s_{b,0} \rangle.
\end{equation}
We examine the two components of this expression separately.  First, by the definition of an orbit of length $N_b$, shifting the state by $N_b$ acts as the identity. Therefore, $P_{\text{cyc}}^{r' + N_b} |s_{b,0}\rangle = P_{\text{cyc}}^{r'} |s_{b,0}\rangle = |s_{b,r'}\rangle$. Second, the phase factor separates into $e^{-i \frac{2\pi}{d} k r'} \cdot e^{-i 2\pi \frac{k N_b}{d}}$. Recall the momentum number condition, which requires that $\frac{k N_b}{d} \in \mathbb{Z}$. Therefore, the factor $e^{-i 2\pi \frac{k N_b}{d}}$ evaluates exactly to $1$.

Consequently, $f(r' + N_b) = f(r')$. Because the summand is periodic with period $N_b$, summing over \textit{any} contiguous sequence of $N_b$ integers yields the  same value. We can therefore  shift the summation window from $[-m, N_b - 1 - m]$ to the standard $[0, N_b - 1]$, replacing $r'$ with $r \equiv (n - m) \pmod{N_b}$:
\begin{equation}
    \sum_{r'=-m}^{N_b-1-m} f(r') = \sum_{r=0}^{N_b-1} f(r).
\end{equation}
Because this inner sum no longer depends on $m$, the outer sum over $m$  contributes a factor of $N_a$, leading  to the simplified expression for the matrix element:
\begin{equation}
    \langle v_a^{(k)} | \mathbb{T}_8 | v_b^{(k)} \rangle = \frac{N_a}{\sqrt{N_a N_b}} \sum_{r=0}^{N_b-1} e^{-i \frac{2\pi}{d} k r} \langle s_{a,0} | \mathbb{T}_8 | s_{b,r} \rangle.
\end{equation}
Simplifying the prefactor and substituting $d=4$  yields the final expression:
\begin{equation} \label{eq:matrix_element}
    \langle v_a^{(k)} | \mathbb{T}_8 | v_b^{(k)} \rangle = \sqrt{\frac{N_a}{N_b}} \sum_{r=0}^{N_b-1} e^{-i \frac{\pi}{2} k r} \langle s_{a,0} | \mathbb{T}_8 | s_{b,r} \rangle.
\end{equation}
This formula shows that we only need to evaluate the transitions from the fixed representative state $|s_{a,0}\rangle$ to the shifted states of the target orbit $|s_{b,r}\rangle$.

\subsubsection{Evaluation of the symmetric sector \textup{(}$k=0$\textup{)}}

Calculating Eq.~\eqref{eq:matrix_element} for the $k=0$ basis $\{|v_1^{(0)}\rangle, |v_2^{(0)}\rangle, |v_3^{(0)}\rangle, |v_4^{(0)}\rangle\}$  produces the  $4 \times 4$ block:
\begin{equation}
    \mathbb{T}^{(0)} = \begin{pmatrix}
    1 & 1 & -\sqrt{2} & 2 \\
    1 & 1 & \sqrt{2} & 2 \\
    \sqrt{2} & -\sqrt{2} & 0 & 0 \\
    2 & 2 & 0 & 2
    \end{pmatrix}.
\end{equation}
The characteristic equation $\det(\mathbb{T}^{(0)}-\lambda I)=0$ yields the polynomial
\begin{equation}
    (\lambda^2+4)(\lambda^2 - 4\lambda -4)=0.
\end{equation}

As a result, in the symmetric sector the transfer matrix  has eigenvalues $\lambda^{(0)}_{1,2} = 2 \pm 2\sqrt{2}$ and $\lambda^{(0)}_{3,4} = \pm 2i$.

\subsubsection{The 1D sectors: $k=1$ and $k=3$}
The $k=1$ and $k=3$ sectors are one-dimensional, parameterized entirely by the $O_4 \to O_4$ transitions. Using Eq.~\eqref{eq:matrix_element}, we evaluate the four raw transition amplitudes from the inputs $s_{4,r}$ to the fixed output $s_{4,0} = |0011\rangle$. Assuming $P_{\text{cyc}}$ shifts bits to the right (i.e., $s_{4,1} = |1001\rangle$):
\begin{align}
    \langle 0011 | \mathbb{T}_8 | 0011 \rangle &= \frac{1}{2} \operatorname{Tr}(XXZZ) = 1, \\
    \langle 0011 | \mathbb{T}_8 | 1001 \rangle &= \frac{1}{2} \operatorname{Tr}(YXIZ) = -i, \\
    \langle 0011 | \mathbb{T}_8 | 1100 \rangle &= \frac{1}{2} \operatorname{Tr}(YYII) = 1, \\
    \langle 0011 | \mathbb{T}_8 | 0110 \rangle &= \frac{1}{2} \operatorname{Tr}(XYZI) = i.
\end{align}
For the $k=1$ sector, the phases $e^{-i \frac{\pi}{2} (1) r} = (-i)^r$ produce   eigenvalue:
\begin{equation}
    \lambda^{(1)} = (1)(1) + (-i)(-i) + (-1)(1) + (i)(i) = -2.
\end{equation}
For the $k=3$ sector, the phase factor $e^{-i \frac{\pi}{2} (3) r} = (i)^r$ yields:
\begin{equation}
    \lambda^{(3)} = (1)(1) + (i)(-i) + (-1)(1) + (-i)(i) = 2.
\end{equation}

\subsubsection{The 2D sector: $k=2$}
The $k=2$ sector is spanned by basis vectors from $O_3$ and $O_4$, acquiring alternating signs $e^{-i \pi r} = (-1)^r$. Evaluating the diagonal element for $O_3 \to O_3$ yields:
\begin{equation}
    \mathbb{T}^{(2)}_{3,3} = \sum_{r=0}^1 (-1)^r \langle 0101 | \mathbb{T}_8 | s_{3,r} \rangle = (1)(-1) + (-1)(1) = -2.
\end{equation}
Reusing the $O_4 \to O_4$ amplitudes from above, the $O_4$ diagonal element is:
\begin{equation}
    \mathbb{T}^{(2)}_{4,4} = \sum_{r=0}^3 (-1)^r \langle 0011 | \mathbb{T}_8 | s_{4,r} \rangle = (1)(1) + (-1)(-i) + (1)(1) + (-1)(i) = 2.
\end{equation}
The normalized $O_4 \to O_3$ and $O_3 \to O_4$ elements both evaluate  to $2i\sqrt{2}$. Assembling the $2 \times 2$ block:
\begin{equation}
    \mathbb{T}^{(2)} = \begin{pmatrix} -2 & 2i\sqrt{2} \\ 2i\sqrt{2} & 2 \end{pmatrix}.
\end{equation}
The characteristic equation $\det(\mathbb{T}^{(2)} - \lambda I) = \lambda^2 + 4 = 0$ provides the complex conjugate pair:
\begin{equation}
    \lambda^{(2)}_{1,2} = \pm 2i.
\end{equation}

\subsubsection{Summary of the exact transfer matrix spectrum}
We have obtained the complete spectrum of $\mathbb{T}_8$ analytically. Grouping the eigenvalues by momentum sectors reads:
\begin{itemize}
    \item $k=0$ (Symmetric): $\lambda = \{ 2+2\sqrt{2}, \ 2-2\sqrt{2}, \ 2i, \ -2i \}$.
    \item $k=1$: $\lambda = \{ -2 \}$.
    \item $k=2$ : $\lambda = \{ 2i, \ -2i \}$.
    \item $k=3$: $\lambda = \{ 2 \}$.
\end{itemize}

\subsection{Dominant eigenvalue and asymptotic purity}

By analyzing the spectrum, we  identify the dominant eigenvalue, which determines the asymptotic behavior of the purity in the thermodynamic limit ($L \to \infty$):
\begin{equation}
    \lambda_{\max} = 2 + 2\sqrt{2}.
\end{equation}

Substituting this dominant eigenvalue back into the original expression for the purity, we obtain the asymptotic limit:
\begin{equation}
    P \simeq \frac{2^L}{16^L} \lambda_{\max}^L = \left( \frac{2(2 + 2\sqrt{2})}{16} \right)^L = \left( \frac{1 + \sqrt{2}}{4} \right)^L.
\end{equation}
To express this scaling in the standard form $2^{-\alpha L}$, we evaluate the logarithm of the base:
\begin{equation}
    \alpha = - \log_2\left( \frac{1 + \sqrt{2}}{4} \right) \approx 0.7284.
\end{equation}
Thus, the purity of the uniform superposition state scales as:
\begin{equation}
    P \sim 2^{-0.728L}.
\end{equation}
This demonstrates that a uniform superposition over all valid Pauli strings does not achieve the optimal purity scaling.
 In contrast, the procedure used in the main text isolates a maximal commuting stabilizer subgroup, which results in the family of states with the  purity scaling of $2^{-0.5L}$.

\section{Polynomial framework}
\label{app:pol_intro}

As mentioned in the main text, we employ the polynomial framework over the finite field $\mathbb{F}_2$. The central object of this framework is the quotient ring of polynomials of degree less than $L$, defined as $\mathcal{R} = \mathbb{F}_2[x] / (x^L - 1)$. The identification $x^L \equiv 1$ represents the periodic boundary conditions for the Hamiltonian terms.

The check vectors $\mathbf{v} = (\mathbf{v}_x \,|\, \mathbf{v}_z)$ for a Pauli string of length $L$ are uniquely represented as a pair of polynomials:
\begin{align}
    a(x) &= \sum_{i=0}^{L-1} a_i x^i, & b(x) &= \sum_{i=0}^{L-1} b_i x^i,
\end{align}
with coefficients $a_i = v_{x,i}$ and $b_i = v_{z,i}$.

Under this mapping, $\mathbf{v} \rightarrow (a(x), b(x))^T \in \mathcal{R}^2$, the version of $\mathbf{v}$ shifted by one site is conveniently represented as $T\mathbf{v} \rightarrow (xa(x), xb(x))^T$.

In this framework, it is also convenient to express dot products between vectors and their shifted versions. Let $\mathbf{a}$ and $\mathbf{b}$ be two binary vectors of length $L$. Their standard dot product over $\mathbb{F}_2$ is given by
\begin{equation}
    \mathbf{a} \cdot \mathbf{b} = \sum_{i=0}^{L-1} a_i b_i \pmod 2.
\end{equation}
Although these vectors can be represented by the polynomials $a(x) = \sum_{i=0}^{L-1} a_i x^i$ and $b(x) = \sum_{i=0}^{L-1} b_i x^i$, the standard product $a(x)b(x)$ does not yield this dot product.

To resolve this, we define the reciprocal (or dual) polynomial of $b(x)$ as
\begin{equation}
    b^*(x) = b(x^{-1}) = \sum_{i=0}^{L-1} b_i x^{-i},
\end{equation}
noting that over the quotient ring $\mathcal{R} = \mathbb{F}_2[x] / (x^L - 1)$, negative powers of $x$ are not independent elements: since $x^L \equiv 1$, multiplication by $x^{-i}$ is algebraically equivalent to multiplication by $x^{L-i}$.

Expanding the product
\begin{equation}
    a(x)b^*(x) = \left( \sum_{i=0}^{L-1} a_i x^i \right) \left( \sum_{j=0}^{L-1} b_j x^{-j} \right) = \sum_{i,j=0}^{L-1} a_i b_j x^{i-j} \pmod{x^L - 1},
\end{equation}
we observe that the constant coefficient ($x^0$) is precisely the standard dot product $\mathbf{a} \cdot \mathbf{b}$. Furthermore, the coefficient of $x^l$ is given by $\sum_{j=0}^{L-1} a_{j+l} b_j \pmod 2$ (with indices evaluated modulo $L$), which corresponds to the dot product of $\mathbf{a}$ with $\mathbf{b}$ shifted by $l$ sites. In other words, the $x^l$-coefficient of $a(x)b^*(x) \pmod{x^L - 1}$ represents the dot product between $\mathbf{a}$ and $T^l \mathbf{b}$, where $T$ denotes the single-site translation operator.

In the context of the present paper, let $\mathbf{v} \rightarrow (a(x), b(x))^T$ and $\mathbf{h} \rightarrow (p(x), q(x))^T$ represent a  Pauli string from the group $G$ and a fixed Hamiltonian term, respectively. The symplectic inner product between $\mathbf{v}$ and $\mathbf{h}$ is a sum of two dot products, and the string $\mathbf{v}$ must commute with every possible shift of $\mathbf{h}$:
\begin{align}\label{app:eq:symp_and_com}
    \langle \mathbf{v}, \mathbf{h} \rangle &= \sum_{k=0}^{L-1} \left( v_{x,k}h_{z,k} + v_{z,k}h_{x,k} \right) \pmod 2, & \langle \mathbf{v}, T^l\mathbf{h} \rangle &= 0, \quad l = 0, \ldots, L-1.
\end{align}
Using the considerations from the previous paragraph, one can easily see that these inner products for all $l$ are given by the coefficients of the expression:
\begin{equation}\label{app:eq:self_comm}
    a(x) q^*(x) + b(x) p^*(x) \pmod{x^L-1}.
\end{equation}
Consequently, the commutation condition translates into:
\begin{equation}\label{app:eq:wor_cond}
    a(x) q^*(x) + b(x) p^*(x) \equiv 0 \pmod{x^L-1}.
\end{equation}
This equation defines the symplectic orthogonal complement
$\mathcal{W}^\perp$, consisting of the check vectors of
Pauli strings commuting with all translations of the
Hamiltonian term.

The expression in \eqref{app:eq:self_comm} can also be employed to characterize how the Hamiltonian term $\mathbf{h}$ commutes or anticommutes with its own shifted copies. By substituting the components $p(x)$ and $q(x)$ of the vector $\mathbf{h}$ in place of $a(x)$ and $b(x)$, respectively, one obtains the cross-commutation relation:
\begin{equation}\label{app:eq:self_com_poly}
    C(x) \equiv p(x)q^*(x) + q(x)p^*(x) \pmod{x^L-1}.
\end{equation}
The $x^l$ coefficient of $C(x)$ determines precisely whether the operator $\mathbf{h}$ commutes or anticommutes with its copy shifted by $l$ sites. Accordingly, $C(x) = 0$ implies that $\mathbf{h}$ commutes with \emph{all} of its shifted copies. The expression $C(x)$ will be referred to as the \emph{cross-commutation polynomial}. This polynomial plays an important role in the formulation of Theorem~\ref{thm_pol}.

To avoid dealing with negative powers, we redefine the reciprocal polynomial as follows:
\begin{equation}\label{app:eq:rec_pol}
    b^*(x) = b(x^{-1}) \quad \longrightarrow \quad b^*(x) = x^{k-1}b(x^{-1}),
\end{equation}
where $k$ denotes the locality of the term $\mathbf{h}$ (so that $k-1$ is the maximum degree of its constituent polynomials $p(x)$ and $q(x)$). Such a global shift by $k-1$ sites preserves both the physical constraints and all subspace dimensions. More precisely, the monomial $x^{k-1}$
is strictly coprime to $x^L-1$ because $x=0$ is never a root of $x^L-1$. Consequently, for a divisibility condition involving some polynomial expression $A(x)$, it holds:
\begin{equation}
    \gcd(x^{k-1}A(x), x^L-1)=\gcd(A(x), x^L-1).
\end{equation}
 
 Definition~\eqref{app:eq:rec_pol} now matches that in Eq.~\eqref{eq:rec_pol} from the main text.

\section{Proof of Theorem~\ref{thm:anti_commutation_t1}}\label{app:thm_solv_proof}

\begin{proof}
For a Pauli operator represented by  $\mathbf{v}(x) = (a(x), b(x))^T$ to  (anti-)commute with every shift of the Hamiltonian term $\mathbf{h}(x) = (p(x), q(x))^T$, it must satisfy  the equation
\begin{equation}\label{app:eq:inh_eq}
    a(x) q^*(x) + b(x) p^*(x) \equiv s(x) \pmod{x^L-1},
\end{equation}
where 
\begin{equation}\label{app:eq:syn_pol1d}
s(x) = x^{k-1}\sum_{l=0}^{L-1} s_{l}\, x^l \in\mathcal R.
\end{equation}
The additional factor $x^{k-1}$ comes from our redefinition~\eqref{app:eq:rec_pol}.
The coefficient $s_{l} \in \{0, 1\}$ denotes whether $\mathbf{v}(x)$ commutes ($0$) or anti-commutes ($1$) with the $l$-th spatial translation $x^l \mathbf{h}(x)$ of the Hamiltonian term.

Since we are interested only in the anti-commutation case, all  the coefficients $s_l$ equal to $1$, and the polynomial $s(x)$ reduces to 
\begin{equation}
    E(x) = \sum_{j=0}^{L-1} x^j,
\end{equation}
the
geometric series polynomial. Here we have used $x^{k-1}E(x)\equiv E(x)$, since multiplication by $x$ cyclically shifts the polynomial coefficients, and $E(x)$ has every coefficient equal to $1$.

By Lemma \ref{app:lem:scalar_solvability}, the inhomogeneous equation~\eqref{app:eq:inh_eq}  admits a solution if and only if $s(x)$  lies in the ideal generated by the polynomials in the left part. In our particular case, this requires:
\begin{equation}
    g(x) = \gcd(q^*(x), p^*(x), x^L-1) \Big\vert E(x), \label{eq:gcd_div_E}
\end{equation}
where $q^*(x)$ and $p^*(x)$ are the reciprocal polynomials of $q(x)$ and $p(x)$, respectively.

Over the field $\mathbb{F}_2$, we can use the following  identity:
\begin{equation}
    x^L - 1 = (x+1)E(x) \label{eq:modulus_factor}.
\end{equation}
Because $g(x)$ divides $x^L-1$ by definition, every irreducible factor of $g(x)$ is also a factor of $x^L-1$. For Eq.~\eqref{eq:gcd_div_E} to hold, the multiplicity of every irreducible factor in $g(x)$ must be less than or equal to its multiplicity in $E(x)$. 

From~\eqref{eq:modulus_factor} it follows that $E(x)$ shares all roots with $x^L-1$ at identical multiplicities, except for the root $x=1$, whose multiplicity is reduced by exactly one. Therefore, the divisibility condition reduces entirely to evaluating the multiplicity of the factor $(x+1)$.

To find this multiplicity, we decompose the lattice size as $L = 2^\nu \cdot n$ for an odd integer $n$. Over $\mathbb{F}_2$~(or any field of characteristic $2$), for any elements $A$ and $B$ it holds that
\begin{equation}
    (A+B)^{2^\nu} = A^{2^\nu} + B^{2^\nu}.
\end{equation}
Consequently,
\begin{equation}
    x^L - 1 = (x^n - 1)^{2^\nu}.
\end{equation}
Because $n$ is odd, the formal derivative of $x^n-1$ at a point $x=1$ is $nx^{n-1}|_{x=1} \neq 0$, confirming that $x=1$ is a simple root of $x^n-1$. Consequently, the multiplicity of the root $x=1$ in $x^L-1$ is exactly $2^\nu$. 
By~\eqref{eq:modulus_factor}, the multiplicity of $(x+1)$ in $E(x)$ is strictly $2^\nu - 1$.

For $g(x)$ to divide $E(x)$, the multiplicity of $(x+1)$ in $g(x)$ cannot exceed $2^\nu-1$. This is equivalent to stating that $g(x)$ is not divisible by $(x+1)^{2^\nu}$. 
Since $g(x) = \gcd(q^*(x), p^*(x), x^L-1)$, this requires that $q^*(x)$ and $p^*(x)$ are not simultaneously divisible by $(x+1)^{2^\nu}$.

Finally, we map the reciprocal polynomials back to the original ones. A polynomial $f(x)$ is divisible by $(x+1)^m$ if and only if its reciprocal $f^*(x)$ is divisible by $(x+1)^m$, because $(x+1)$ is symmetric under coefficient reversal. Thus, the condition is equivalent to $q(x)$ and $p(x)$ not being simultaneously divisible by $(x+1)^{2^\nu}$.

For the odd lattice case ($\nu=0$), the condition yields that $q(x)$ and $p(x)$ are not both divisible by $(x+1)^1$. Over $\mathbb{F}_2$, the condition $(x+1) \mid f(x)$ is equivalent to $f(1) = 0$. Since evaluating a polynomial at $x=1$ sums its coefficients modulo $2$, the condition $p(1)=0$ or $q(1)=0$ corresponds to an even number of Pauli operators. Thus, a solution exists if and only if $q(1)=1$ or $p(1)=1$, corresponding to odd parity in either the $X$ or $Z$ components of $h$.
\end{proof}

\section{Proof of Theorem~\ref{thm_pol}}
\label{app:thm_pol}

\begin{proof}
We first compute the dimension of the subspace $\mathcal{W}$, defined as the span of all shifts of the vector $\mathbf{h}$. Consider the mapping $\phi \colon \mathcal{R} \to \mathcal{R}^2$ given by
\begin{equation}
    \phi(f(x)) = f(x)\mathbf{h}(x) \pmod{x^L-1} =  \begin{pmatrix} f(x)p(x) \\ f(x)q(x) \end{pmatrix} \pmod{x^L-1}.
\end{equation}
By construction, the subspace $\mathcal{W}$ is the image of $\phi$, i.e., $\mathcal{W} = \operatorname{im}(\phi)$.

The domain $\mathcal{R} = \mathbb{F}_2[x]/(x^L-1)$ has dimension $L$ as an $\mathbb{F}_2$-vector space, with standard basis $\{1, x, \dots, x^{L-1}\}$. Applying the rank-nullity theorem to $\phi$ as a linear map over $\mathbb{F}_2$ yields
\begin{equation}
    L = \dim(\mathcal{W}) + \dim(\ker(\phi)).
\end{equation}

The kernel $\ker(\phi)$ consists of all $f(x) \in \mathcal{R}$ mapping to zero, which requires
\begin{equation}\label{app:eq:sim_null}
    f(x)p(x) \equiv 0 \pmod{x^L-1} \quad \text{and} \quad f(x)q(x) \equiv 0 \pmod{x^L-1}.
\end{equation}
We stress that the kernel of $\phi$, in general, does not coincide with $\mathcal W^{\perp}$ as the latter is defined by the commutation condition~\eqref{app:eq:wor_cond} having a different structure.

To rewrite~\eqref{app:eq:sim_null}, let us define the polynomial greatest common divisor
\begin{equation}
    K(x) = \gcd(p(x), q(x), x^L-1).
\end{equation}
By Lemma~\ref{app:lem:simultaneous_kernel}, $f(x)$ must be a multiple of $\frac{x^L - 1}{K(x)}$ in $\mathcal{R}$:
\begin{equation}
    f(x) \in \left\langle \frac{x^L - 1}{K(x)} \right\rangle.
\end{equation}
The dimension of the space of these multiples over $\mathbb{F}_2$ is equal to the degree of $K(x)$:
\begin{equation}
    \dim(\ker(\phi)) = \deg K(x).
\end{equation}
Applying the rank-nullity result, the dimension of $\mathcal{W}$ is given by
\begin{equation}
    \dim(\mathcal{W}) = L - \deg K(x).
\end{equation}

Since the symplectic inner product on $\mathbb{F}_2^{2L}$ is non-degenerate, the dimension of the symplectic complement $\mathcal{W}^\perp$ is given by
\begin{equation}
    \dim(\mathcal{W}^\perp) = 2L - \dim(\mathcal{W}) = L + \deg(K(x)).
\end{equation}

We proceed to finding the dimension of the intersection $\mathcal W\cap\mathcal W^{\perp}$. Returning to the map $\phi(f(x)) = f(x)\mathbf{h}(x)$, the image of the entire ring $\mathcal R$ gives the whole space $\mathcal W$: $\operatorname{im}(\phi) = \mathcal W$. Now we need only the operators in $\mathcal W$ that commute with everything, i.e., those in $\mathcal W\cap\mathcal W^{\perp}$. To characterize them, we substitute their components $f(x)p(x)$ and $f(x)q(x)$ in place of $a(x)$ and $b(x)$ in the commutation condition~\eqref{app:eq:wor_cond}:
\begin{equation}
    f(x)p(x) q^*(x) + f(x)q(x) p^*(x) \equiv 0 \pmod{x^L-1}.
\end{equation}
Factoring out the common polynomial $f(x)$, we obtain the condition:
\begin{equation}\label{app:eq:self_C}
    f(x)C(x) \equiv 0 \pmod{x^L-1},
\end{equation}
where $C(x)=p(x)q^*(x) + q(x)p^*(x)$ is the cross-commutation polynomial introduced in~\eqref{app:eq:self_com_poly}.

Let us restrict the domain of the map to only the polynomials that satisfy the above cross-commutation condition:
\begin{equation}
    U = \{ f(x) \in \mathcal R \mid f(x)C(x) \equiv 0 \pmod{x^L-1} \}.
\end{equation}
By construction, the image of this restricted domain is exactly the center: $\phi(U) = \mathcal W\cap\mathcal W^{\perp}$.

Applying the rank-nullity theorem to $\phi$ restricted to the domain $U$ yields:
\begin{equation}\label{app:eq:dim_res}
    \dim(\phi(U)) = \dim(U) - \dim(U \cap \ker(\phi)).
\end{equation}
By Lemma~\ref{app:lem:kernel_structure},
\begin{equation}\label{app:eq:dim_U}
    \dim(U) = \deg\left(\gcd(C(x), x^L - 1)\right).
\end{equation}

To calculate $\dim(U \cap \ker(\phi))$, observe that a polynomial $f(x)\in \ker(\phi)$, i.~e. that satisfying~\eqref{app:eq:sim_null}, automatically satisfies~\eqref{app:eq:self_C}, since $C(x)$ is a linear combination of $p(x)$ and $q(x)$. Consequently, $\ker(\phi)\subset U$ and 
\begin{equation}\label{app:eq:dim_U_ker}
    \dim(U \cap \ker(\phi)) =\dim(\ker(\phi))  = \deg\left(K(x)\right).
\end{equation}

Finally, using the results~\eqref{app:eq:dim_res}, \eqref{app:eq:dim_U}, and~\eqref{app:eq:dim_U_ker}, we obtain:
\begin{equation}
   \dim(\mathcal W\cap\mathcal W^{\perp}) = \dim(\phi(U)) = \deg\left(\gcd(C(x), x^L - 1)\right) - \deg\left(K(x)\right).
\end{equation}
\end{proof}

\section{Auxiliary polynomial lemmas}
\label{app:sec:pol_lem}

In this appendix, we list several auxiliary results regarding polynomial divisibility and vector space dimensions in the quotient ring $\mathcal{R}$. These results are used in the derivation of the main results; complete proofs are included here for self-containment.
\begin{lemma}[Divisibility and nullspace dimension in $\mathcal{R}$]
\label{app:lem:kernel_structure}
Let $\mathcal{R} = \mathbb{F}_2[x]/(x^L - 1)$ be the ring of  polynomials over $\mathbb{F}_2$. For a fixed non-zero element $h(x) \in \mathcal{R}$, consider the equation in $f(x) \in \mathcal{R}$:
\begin{equation}
    f(x)h(x) \equiv 0 \pmod{x^L - 1}.
\end{equation}
Let $g(x) = \gcd(h(x), x^L - 1) \in \mathbb{F}_2[x]$. Then, the solution space $S \subseteq \mathcal{R}$ is given by
\begin{equation}
    S = \left\{ f(x) \in \mathcal{R} \;\middle|\; f(x) \equiv k(x) \frac{x^L - 1}{g(x)} \pmod{x^L - 1},\, k(x) \in \mathbb{F}_2[x] \right\} = \left\langle \frac{x^L - 1}{g(x)} \right\rangle,
\end{equation}
where $\left\langle \cdot \right\rangle$ denotes the subspace of polynomial multiples generated in $\mathcal{R}$. Furthermore, as a vector space over $\mathbb{F}_2$,
\begin{equation}
    \dim_{\mathbb{F}_2}(S) = \deg(g(x)) = \deg\left(\gcd(h(x), x^L - 1)\right).
\end{equation}
\end{lemma}

\begin{proof}
Set $N(x) = x^L - 1$. In $\mathbb{F}_2[x]$, the condition $f(x)h(x) \equiv 0 \pmod{N(x)}$ is equivalent to $N(x) \mid f(x)h(x)$.

Factoring out $g(x) = \gcd(h(x), N(x))$, define
\begin{equation}
    \tilde{h}(x) = \frac{h(x)}{g(x)}, \quad \tilde{N}(x) = \frac{N(x)}{g(x)},
\end{equation}
so that $\gcd(\tilde{h}(x), \tilde{N}(x)) = 1$. The divisibility condition becomes
\begin{equation}
    g(x)\tilde{N}(x) \;\Big|\; f(x)g(x)\tilde{h}(x),
\end{equation}
and, after factoring  $g(x)$ out,
\begin{equation}
    \tilde{N}(x) \;\Big|\; f(x)\tilde{h}(x).
\end{equation}
Since $\gcd(\tilde{N}(x), \tilde{h}(x)) = 1$, Euclid's Lemma yields $\tilde{N}(x) \mid f(x)$. Thus $f(x) = k(x)\tilde{N}(x)$ for $k(x) \in \mathbb{F}_2[x]$, which gives $S = \left\langle \tilde{N}(x) \right\rangle = \left\langle \frac{N(x)}{g(x)} \right\rangle$.

To compute $\dim_{\mathbb{F}_2}(S)$, note that unique elements of $\mathcal{R}$ correspond to polynomials in $\mathbb{F}_2[x]$ of degree strictly less than $L$. For $f(x) = k(x)\tilde{N}(x)$ to satisfy $\deg(f) < L$, the multiplier $k(x)$ must satisfy
\begin{equation}
     \deg(k) < L - \deg(\tilde{N}).
\end{equation}
Substituting $\deg(\tilde{N}) = L - \deg(g)$ gives $\deg(k) < \deg(g)$. The space of multipliers $k(x) \in \mathbb{F}_2[x]$ with $\deg(k) < \deg(g)$ is spanned by $\{1, x, \dots, x^{\deg(g)-1}\}$, which has dimension $\deg(g)$. Since each such $k(x)$ maps to a distinct $f(x) \in \mathcal{R}$, $S \cong \mathbb{F}_2^{\deg(g)}$ and $\dim_{\mathbb{F}_2}(S) = \deg(g)$.
\end{proof}

\begin{lemma}[LCM of quotients]
\label{app:lem:lcm_identity}
Let $A(x), B_1(x), B_2(x), \ldots, B_m(x) \in \mathbb{F}_2[x]$ be polynomials such that $B_i(x) \mid A(x)$ for all $i \in \{1, \ldots, m\}$. The least common multiple of their quotients is given by:
\begin{equation}
    \text{lcm}\left( \frac{A(x)}{B_1(x)}, \frac{A(x)}{B_2(x)}, \ldots, \frac{A(x)}{B_m(x)} \right) = \frac{A(x)}{\gcd(B_1(x), B_2(x), \ldots, B_m(x))}.
\end{equation}
\end{lemma}

\begin{proof}
We proceed by induction on $m$. First, consider the base case $m=2$ with divisors $B_1(x), B_2(x)$. Let $D(x) = \gcd(B_1(x), B_2(x))$, and factor $B_1(x) = D(x)\tilde{B}_1(x)$ and $B_2(x) = D(x)\tilde{B}_2(x)$, where $\gcd(\tilde{B}_1(x), \tilde{B}_2(x)) = 1$. Let $U_1(x) = \frac{A(x)}{B_1(x)}$ and $U_2(x) = \frac{A(x)}{B_2(x)}$. 

Observe that $\frac{A(x)}{D(x)} = U_1(x)\tilde{B}_1(x) = U_2(x)\tilde{B}_2(x)$, making it a common multiple of $U_1(x)$ and $U_2(x)$. For any other common multiple $M(x) = k_1(x) U_1(x) = k_2(x) U_2(x)$, substituting the expressions for $U_1(x)$ and $U_2(x)$ yields:
\begin{equation}
    k_1(x) \frac{A(x)}{B_1(x)} = k_2(x) \frac{A(x)}{B_2(x)},
\end{equation}
which results in
\begin{equation}
    k_1(x) B_2(x) = k_2(x) B_1(x).
\end{equation}
Substituting the factored forms of $B_1(x)$ and $B_2(x)$ and dividing by $D(x)$ leaves $k_1(x) \tilde{B}_2(x) = k_2(x) \tilde{B}_1(x)$. Since $\gcd(\tilde{B}_1(x), \tilde{B}_2(x)) = 1$, Euclid's Lemma requires $\tilde{B}_1(x) \mid k_1(x)$. Thus, $k_1(x) = m(x)\tilde{B}_1(x)$ for some polynomial $m(x)$. Substituting this back into $M(x)$ gives:
\begin{equation}
    M(x) = k_1(x) U_1(x) = m(x)\tilde{B}_1(x)\frac{A(x)}{D(x)\tilde{B}_1(x)} = m(x)\frac{A(x)}{D(x)},
\end{equation}
which implies $M(x)$ is a multiple of $\frac{A(x)}{D(x)}$. Thus, the base case holds:
\begin{equation}
    \text{lcm}\left( \frac{A(x)}{B_1(x)}, \frac{A(x)}{B_2(x)} \right) = \frac{A(x)}{\gcd(B_1(x), B_2(x))}.
\end{equation}

Now assume the statement holds for $m-1$ terms. By the associativity of the $\text{lcm}$ operator:
\begin{equation}
    \text{lcm}\left( \frac{A(x)}{B_1(x)}, \ldots, \frac{A(x)}{B_m(x)} \right) = \text{lcm}\left( \text{lcm}\left(\frac{A(x)}{B_1(x)}, \ldots, \frac{A(x)}{B_{m-1}(x)}\right), \frac{A(x)}{B_m(x)} \right).
\end{equation}
Applying the induction hypothesis to the inner $\text{lcm}$ term yields:
\begin{equation}
    \text{lcm}\left( \frac{A(x)}{\gcd(B_1(x), \ldots, B_{m-1}(x))}, \frac{A(x)}{B_m(x)} \right).
\end{equation}
Setting $B'(x) = \gcd(B_1(x), \ldots, B_{m-1}(x))$, both $B'(x)$ and $B_m(x)$ divide $A(x)$. Applying the $m=2$ base case to $B'(x)$ and $B_m(x)$ gives:
\begin{equation}
    \frac{A(x)}{\gcd(B'(x), B_m(x))} = \frac{A(x)}{\gcd\left( \gcd(B_1(x), \ldots, B_{m-1}(x)), B_m(x) \right)} = \frac{A(x)}{\gcd(B_1(x), \ldots, B_m(x))},
\end{equation}
which completes the proof.
\end{proof}

\begin{lemma}[Simultaneous nullspace dimension in $\mathcal{R}$]
\label{app:lem:simultaneous_kernel}
Let $\mathcal{R} = \mathbb{F}_2[x]/(x^L - 1)$ be the ring of polynomials over $\mathbb{F}_2$. For a collection of fixed polynomials $p_1(x), p_2(x), \ldots, p_m(x) \in \mathcal{R}$, consider the system of simultaneous equations in $f(x) \in \mathcal{R}$:
\begin{equation}
    f(x)p_i(x) \equiv 0 \pmod{x^L - 1} \quad \text{for all } i \in \{1, \ldots, m\}.
\end{equation}
Let $g(x) = \gcd(p_1(x), \dots, p_m(x), x^L - 1) \in \mathbb{F}_2[x]$. Then, the solution space $S' \subseteq \mathcal{R}$ is given by
\begin{equation}
    S' = \left\langle \frac{x^L - 1}{g(x)} \right\rangle,
\end{equation}
where $\left\langle \cdot \right\rangle$ denotes the subspace of polynomial multiples generated in $\mathcal{R}$. Furthermore, as a vector space over $\mathbb{F}_2$,
\begin{equation}
    \dim_{\mathbb{F}_2}(S') = \deg(g(x)) = \deg\left(\gcd(p_1(x), \ldots, p_m(x), x^L - 1)\right).
\end{equation}
\end{lemma}

\begin{proof}
Set $N(x) = x^L - 1$. Applying Lemma~\ref{app:lem:kernel_structure} to each constraint independently requires $f(x) \in \left\langle \frac{N(x)}{\gcd(p_i(x), N(x))} \right\rangle$ for every $i \in \{1, \ldots, m\}$.

To satisfy all $m$ constraints simultaneously, $f(x)$ must be a multiple  of each of these $m$ generators, which means it must be a multiple of their joint least common multiple:
\begin{equation}
    f(x) \in \left\langle \text{lcm}\left( \frac{N(x)}{\gcd(p_1(x), N(x))}, \ldots, \frac{N(x)}{\gcd(p_m(x), N(x))} \right) \right\rangle.
\end{equation}
Applying Lemma~\ref{app:lem:lcm_identity} with $A(x) = N(x)$ and $B_i(x) = \gcd(p_i(x), N(x))$, the generator simplifies to:
\begin{equation}
    \frac{N(x)}{\gcd\left( \gcd(p_1(x), N(x)), \ldots, \gcd(p_m(x), N(x)) \right)} = \frac{N(x)}{\gcd(p_1(x), \ldots, p_m(x), N(x))} = \frac{N(x)}{g(x)}.
\end{equation}
This establishes that $S' = \left\langle \frac{N(x)}{g(x)} \right\rangle$.

Defining $\tilde{N}(x) = \frac{N(x)}{g(x)}$, elements of $S'$ are uniquely represented by polynomials $f(x) = k(x)\tilde{N}(x)$ with $\deg(f) < L$, which imposes $\deg(k) < L - \deg(\tilde{N}) = \deg(g)$. The space of multipliers $k(x) \in \mathbb{F}_2[x]$ satisfying $\deg(k) < \deg(g)$ is spanned by $\{1, x, \ldots, x^{\deg(g)-1}\}$, yielding $\dim_{\mathbb{F}_2}(S') = \deg(g)$.
\end{proof}

\begin{lemma}[Multi-variable inhomogeneous solvability in $\mathcal{R}$]
\label{app:lem:scalar_solvability}
For any given sequence of polynomials $h_1(x), \dots, h_k(x)$ and a target polynomial $s(x) \in \mathbb{F}_2[x]$, the linear equation 
\begin{equation}
    h_1(x) y_1(x) + \dots + h_k(x) y_k(x) \equiv s(x) \pmod{x^L-1}
\end{equation}
admits a polynomial solution vector $\big(y_1(x), \dots, y_k(x)\big)$ if and only if the greatest common divisor of all constraint polynomials and the boundary condition divides $s(x)$: 
\begin{equation}
    \gcd\big(h_1(x), \dots, h_k(x), x^L-1\big) \mid s(x).
\end{equation}
\end{lemma}

\begin{proof}
A solution vector $\big(y_1(x), \dots, y_k(x)\big)$ exists if and only if there is some polynomial $w(x) \in \mathbb{F}_2[x]$ such that the equation holds exactly over the polynomial ring:
\begin{equation}
    h_1(x)y_1(x) + \dots + h_k(x)y_k(x) + (x^L-1)w(x) = s(x) \label{eq:generalized_bezout_target}.
\end{equation}
The left-hand side of Equation \eqref{eq:generalized_bezout_target} defines the ideal $\mathcal{I} = \langle h_1(x), \dots, h_k(x), x^L-1 \rangle$, which is the set of all possible polynomial linear combinations of the $k$ constraint polynomials and the periodicity polynomial $x^L-1$. Let $g(x) = \gcd(h_1(x), \dots, h_k(x), x^L-1)$. We prove that $\mathcal{I}$ is exactly the set of all polynomial multiples of $g(x)$.

First, we show that every combination in $\mathcal{I}$ is a multiple of $g(x)$. By the definition of the greatest common divisor, $g(x)$ simultaneously divides all generators of the ideal: $h_j(x) = g(x)\tilde{h}_j(x)$ for each $j \in \{1, \dots, k\}$, and $x^L-1 = g(x)\tilde{Q}(x)$ for some polynomials $\tilde{h}_j(x), \tilde{Q}(x) \in \mathbb{F}_2[x]$. Substituting these into an arbitrary linear combination yields:
\begin{equation}
    \sum_{j=1}^k h_j(x)A_j(x) + (x^L-1)B(x) = g(x) \left[ \sum_{j=1}^k \tilde{h}_j(x)A_j(x) + \tilde{Q}(x)B(x) \right].
\end{equation}
Because the bracketed term is simply another polynomial in $\mathbb{F}_2[x]$ formed by sums and products, any such combination is strictly a multiple of $g(x)$. It is impossible to form a combination smaller than the collective greatest common divisor.

Second, we show that \textit{every} multiple of $g(x)$ can be constructed as such a combination. In $\mathbb{F}_2[x]$, Bézout's identity guarantees that there exist specific polynomial coefficients $u_1(x), \dots, u_k(x)$ and $v(x)$ such that $g(x)$ itself is obtained from the linear combination of the generators:
\begin{equation}
    h_1(x)u_1(x) + \dots + h_k(x)u_k(x) + (x^L-1)v(x) = g(x).
\end{equation}
To construct any arbitrary multiple $g(x)m(x)$ for some $m(x) \in \mathbb{F}_2[x]$, we multiply this generalized Bézout's identity entirely by $m(x)$:
\begin{equation}
    h_1(x)\big[u_1(x)m(x)\big] + \dots + h_k(x)\big[u_k(x)m(x)\big] + (x^L-1)\big[v(x)m(x)\big] = g(x)m(x).
\end{equation}
This shows that any multiple of $g(x)$ can be successfully constructed from some linear combination of the constraint polynomials and $x^L-1$. 

Consequently, the multi-generator ideal $\mathcal{I} = \langle h_1(x), \dots, h_k(x), x^L-1 \rangle$ is identical to the principal ideal $\langle g(x) \rangle$. For Equation \eqref{eq:generalized_bezout_target} to hold, the polynomial $s(x)$ must be a member of this ideal. Thus, a solution exists if and only if $s(x)$ is a multiple of $g(x)$, which is precisely the condition that $\gcd(h_1(x), \dots, h_k(x), x^L-1)$ divides $s(x)$.
\end{proof}

\section{Exact  subspace dimensions for the case of Hamiltonian generated by $t$ terms} 
\label{app:sec:sec:smith_form}

\subsection{Formulation of the inhomogeneous system}

Let the Hamiltonian term be a linear combination of $t$  Pauli strings:
\begin{equation}
    h = \sum_{\alpha=1}^t\,J_{\alpha} h^{(\alpha)}.
\end{equation}
We consider the space of operators $\mathbf{v}(x) = (a(x), b(x))^T$ satisfying  commutation or anti-commutation condition  independently with each term $h^{(\alpha)}$. Let each term $h^{(\alpha)}$ be represented by a fixed polynomial vector $\mathbf{h}_{\alpha}(x) = (p_{\alpha}(x), q_{\alpha}(x))^T$. Because they are geometrically $k$-local, we have strict degree bounds for all $\alpha$:
\begin{equation}\label{app:eq:k_bound}
    \deg(p_\alpha), \deg(q_\alpha) \le k - 1.
\end{equation}

The condition that each vector $\mathbf{v}(x)$ (anti-)commutes with all shifted copies of all $t$ terms translates to a system of $t$ simultaneous polynomial equations, which is a generalization of~\eqref{app:eq:inh_eq} and~\eqref{app:eq:syn_pol1d}:
\begin{equation}\label{app:eq:lin_sys_scal}
    a(x) q_\alpha^*(x) + b(x) p_\alpha^*(x) \equiv s_\alpha(x) \pmod{x^L-1} \quad \text{for } \alpha \in \{1, 2, \dots, t\},
\end{equation}
where 
\begin{equation}\label{app:eq:syn_pol}
s_\alpha(x) = x^{k-1}\sum_{l=0}^{L-1} s_{\alpha,l} x^l \in\mathcal R.
\end{equation}
The coefficient $s_{\alpha,l} \in \{0, 1\}$ denotes whether $\mathbf{v}(x)$ commutes ($0$) or anti-commutes ($1$) with the $l$-th spatial translation $x^l \mathbf{h}_\alpha(x)$ of the $\alpha$-th Hamiltonian term.

By introducing the $2 \times t$ generator matrix:
\begin{equation}
    W(x) = \begin{pmatrix} p_1(x) & p_2(x) & \dots & p_t(x) \\ q_1(x) & q_2(x) & \dots & q_t(x) \end{pmatrix}
\end{equation}
and the vector of polynomials $\mathbf{s}(x) = (s_1(x), \dots, s_t(x))^T$, the inhomogeneous system~\eqref{app:eq:lin_sys_scal} rewrites as
\begin{equation}\label{app:eq:main_inhom}
    H(x) \mathbf{v}(x) \equiv \mathbf{s}(x) \pmod{x^L-1},
\end{equation}
where $H(x) = W^*(x)^T \Omega$ is a $t \times 2$  matrix with 
\begin{equation}
\Omega = \begin{pmatrix} 0 & 1 \\ 1 & 0 \end{pmatrix}.
\end{equation}

To analyze the solutions of~\eqref{app:eq:main_inhom}, we diagonalize the polynomial matrix $H(x)$ using the Smith Normal Form, which we briefly review in the next subsection.

\subsection{Smith Normal Form}

For any $m \times n$ polynomial matrix $M(x)$ of rank $r \le \min(m, n)$, there exist invertible (unimodular) matrices $U(x) \in \text{GL}(m, \mathbb{F}_2[x])$ and $V(x) \in \text{GL}(n, \mathbb{F}_2[x])$ with $\det U(x) = \det V(x) = 1$ that diagonalize $M(x)$ into its \emph{Smith Normal Form} (SNF):
\begin{equation}\label{app:eq:SNF}
    S(x) = U(x) M(x) V(x) = \operatorname{diag}(d_1(x), d_2(x), \dots, d_r(x), 0, \dots, 0)
\end{equation}
The unique  polynomials $d_k(x)$ are the \emph{invariant factors}, satisfying the  divisibility chain $d_1(x) \mid d_2(x) \mid \dots \mid d_r(x)$. 

The invariant factors are determined directly by the minors of $M(x)$:
\begin{itemize}
    \item $d_1(x)$ is the greatest common divisor of all $1 \times 1$ minors (i.e., the GCD of all individual entries in $M(x)$).
    \item $d_1(x)d_2(x) \dots d_k(x)$ is the GCD of all $k \times k$ minors of $M(x)$ (the $k$-th determinantal divisor $D_k(x)$).
\end{itemize}

When solving a coupled nullspace constraint system $M(x)\mathbf{v}(x) = 0 \pmod{x^L-1}$, the SNF allows us to perform a change of basis. Defining transformed  variables $\mathbf{v}'(x) = V^{-1}(x)\mathbf{v}(x)$ and left-multiplying the system by $U(x)$, the  system takes the form:
\begin{equation}
    S(x) \mathbf{v}'(x) = 0 \pmod{x^L-1}.
\end{equation}
The system decouples into $r$ independent scalar constraints
\begin{equation}
    d_k(x)v'_k(x)=0\pmod{x^L-1},
    \quad k=1,\ldots,r,
\end{equation}
while the remaining $n-r$ coordinates
$v'_{r+1}(x),\ldots,v'_n(x)$ are unconstrained.
By Lemma~\ref{app:lem:kernel_structure}, each constrained
coordinate contributes
$\deg\gcd(d_k(x),x^L-1)$ binary degrees of freedom,
whereas each unconstrained coordinate contributes $L$.
Therefore,
\begin{equation}
\dim_{\mathbb F_2}\ker M
=(n-r)L+
\sum_{k=1}^{r}\deg\gcd(d_k(x),x^L-1),
\end{equation}
where the kernel is taken modulo $x^L-1$.

\subsection{Solvability of the inhomogeneous system}
\label{app:sec:solv_inhom}

First we need to establish whether the inhomogeneous system~\eqref{app:eq:main_inhom} has at least one solution.

We decouple the system~\eqref{app:eq:main_inhom} by computing the Smith Normal Form (SNF) of $H(x)$ over $\mathbb{F}_2[x]$. 
Let $r=\operatorname{rank}_{\mathbb F_2(x)}H(x)
\leq\min(t,2)$. There exist unimodular matrices
$U(x)$ and $V(x)$ such that
\[
U(x)H(x)V(x)=
D(x)=\operatorname{diag}
(d_1(x),\ldots,d_r(x),0,\ldots,0),
\]
where $d_1(x)\mid\cdots\mid d_r(x)$ are the
nonzero invariant factors. Setting
$\mathbf s'=U\mathbf s$ and
$\mathbf v'=V^{-1}\mathbf v$, the system becomes
\[
D(x)\mathbf v'(x)\equiv
\mathbf s'(x)\pmod{x^L-1}.
\]
By Lemma~\ref{app:lem:scalar_solvability},
the system is solvable if and only if
\begin{equation}\label{app:eq:solv_m}
\begin{aligned}
\gcd(d_j(x),x^L-1)&\mid s'_j(x),
&&j=1,\ldots,r,\\
s'_j(x)&\equiv0\pmod{x^L-1},
&&j=r+1,\ldots,t.
\end{aligned}
\end{equation}

For the $t=1$ case,  the constraint matrix reduces to a  $1 \times 2$ row vector: $H(x) = \begin{pmatrix} q^*(x) & p^*(x) \end{pmatrix}$. There is only $1$  invariant factor $d_1(x)$  defined as the greatest common divisor of all $1 \times 1$ minors (which are simply the entries of the matrix itself). Consequently, $d_1(x) = \gcd(q^*(x), p^*(x))$. Furthermore, the row-transformation matrix $U(x)$ is just the $1 \times 1$ identity matrix, which means that $\mathbf{s}'(x) = s(x)$. Using Eq.~\eqref{app:eq:solv_m}, the solvability condition  reduces to a single requirement:
\begin{equation}
\gcd(q^*(x), p^*(x), x^L-1) \mid s(x).
\end{equation}

Hence, when this greatest common divisor equals
$1$, the system is solvable for any prescribed
commutation or anticommutation pattern $s(x)$.

\subsection{Derivation of subspace dimensions}

\subsubsection{Space  $\mathcal{W}^\perp$}

The space $\mathcal{W}^\perp$ consists of all Pauli
check vectors
$\mathbf{v}(x) = (a(x), b(x))^T$
that commute with all $t$ translated Hamiltonian terms. This imposes the  commutation constraint
\begin{equation}
H(x)\mathbf{v}(x) = 0 \pmod{x^L-1}.
\end{equation}
This is the homogeneous version of system~\eqref{app:eq:main_inhom}, which we analyze only after establishing that the inhomogeneous system has at least one solution.

Transforming to the decoupled basis $(a'(x), b'(x))^T = V(x)^{-1}(x)\mathbf{v}(x)$ again,  the  system reduces to two independent conditions:
\begin{align}
    d_1(x) a'(x) &= 0 \pmod{x^L-1}, \\
    d_2(x) b'(x) &= 0 \pmod{x^L-1}.
\end{align}

Summing the dimensions of spaces of independent solutions for $a'(x)$ and $b'(x)$, which can be obtained from Lemma~\ref{app:lem:kernel_structure}, gives the exact dimension of the valid subspace:
\begin{equation}\label{app:eq:dimwp}
    \dim \mathcal{W}^\perp = \deg \gcd(d_1(x), x^L-1) + \deg \gcd(d_2(x), x^L-1).
\end{equation}

Let us denote by $w_1(x)$ and $w_2(x)$ the two invariant factors of the matrix $W(x)$, which are determined as follows:
\begin{align}
    w_1(x) &= \gcd_{\alpha=1\dots t} \big(p_\alpha(x), q_\alpha(x)\big), \\
    w_1(x)w_2(x) &= \gcd_{1 \le \alpha < \beta \le t} \big(p_\alpha(x)q_\beta(x) + p_\beta(x)q_\alpha(x)\big).
\end{align}
Since $H(x)=x^{k-1}W(x^{-1})^T\Omega$, its nonzero
invariant factors are obtained from those of $W(x)$
by replacing $x$ with $x^{-1}$, up to powers of $x$.
These powers are coprime to $x^L-1$ and therefore do
not affect the relevant greatest common divisors.
Inversion preserves factor multiplicities and maps
$x^L-1$ to a power of $x$ times itself. Consequently,
\begin{equation}
    \deg\gcd(d_\alpha(x),x^L-1)
    =
    \deg\gcd(w_\alpha(x),x^L-1),
    \qquad \alpha=1,2.
\end{equation}
For a zero invariant factor, both sides equal $L$.

Substituting this identity into~\eqref{app:eq:dimwp},
we obtain the dimension of the homogeneous solution
space $\mathcal{W}^\perp$ in terms of the invariant
factors of $W(x)$:
\begin{equation}\label{app:eq:ker_gen_main}
    \dim \mathcal{W}^\perp = \deg \gcd(w_1(x), x^L-1) + \deg \gcd(w_2(x), x^L-1).
\end{equation}

When $t=1$, the second invariant factor $w_2(x)$ (also $d_2(x)$) is zero, which leads to $\deg \gcd(w_2(x), x^L-1)=L$. As a result, the formula~\eqref{eq:thm1_dim1} from the formulation of Theorem~\ref{thm_pol} is recovered:
\begin{equation}
        \dim \mathcal{W}^\perp = \deg\gcd(p(x), q(x), x^L-1) +L=L+\deg K(x).
\end{equation}

\subsubsection{Center space  $\mathcal{W} \cap \mathcal{W}^\perp$}

The space $\mathcal{W}$ is the image of the mapping $\Phi: \mathcal{R}^t \to \mathcal{R}^2$ given by $\Phi(\mathbf{f}) = W(x)\mathbf{f} \pmod{x^L-1}$, for the coefficient vector $\mathbf{f}(x) = (f_1(x), \dots, f_t(x))^T$.

To find operators in the intersection $\mathcal{W} \cap \mathcal{W}^\perp$, we need vectors $\mathbf{f} \in \mathcal{R}^t$ such that $\Phi(\mathbf{f})$ commutes with all $t$ Hamiltonian operators. The commutations between all pairs of Hamiltonian terms form a $t \times t$ symplectic Gram matrix $C(x)$:
\begin{equation}
    C(x) = W^*(x)^T \begin{pmatrix} 0 & 1 \\ 1 & 0 \end{pmatrix} W(x).
\end{equation}
The entries of $C(x)$ are the cross-commutation polynomials:
\begin{equation}
C_{\alpha\beta}(x) = p_\beta(x)q_\alpha^*(x) + q_\beta(x)p_\alpha^*(x). 
\end{equation}

For $\Phi(\mathbf{f})$ to be in $\mathcal{W}^\perp$, the vector $\mathbf{f}$ must lie in the nullspace of this Gram matrix: $C(x)\mathbf{f} \equiv 0 \pmod{x^L-1}$. In this connection, we consider the  domain  $U = \ker(C(x))$. By the  restricted rank-nullity theorem, the dimension of the image $\Phi(U)$, which is exactly $\mathcal{W} \cap \mathcal{W}^\perp$, is given by:
\begin{equation}
\dim(\mathcal{W} \cap \mathcal{W}^\perp) = \dim(U) - \dim(U \cap \ker(\Phi)).
\end{equation}
 If $\mathbf{f} \in \ker(\Phi)$, then $W(x)\mathbf{f} = 0$. Since $C(x) = W^*(x)^T \Omega W(x)$, it follows that $C(x)\mathbf{f} = 0$, which means $\mathbf{f}$ is also in $\ker(C(x))$.
Therefore, $\ker(\Phi) \subseteq U$, which simplifies the intersection:
\begin{equation}\label{app:eq:dim_cen}
    \dim(\mathcal{W} \cap \mathcal{W}^\perp) = \dim(\ker(C(x))) - \dim(\ker(\Phi)).
\end{equation}

Since $\Phi$ maps from $\mathcal{R}^t$ to $\mathcal{R}^2$, the standard rank-nullity theorem yields $\dim(\ker(\Phi)) = tL - \dim(\mathcal{W})$. Recalling also that $\dim(\mathcal{W}^\perp) = 2L - \dim(\mathcal{W})$, we can rewrite the kernel's dimension as:
\begin{equation}\label{app:eq:phi_ker}
    \dim(\ker(\Phi)) = (t-2)L + \dim(\mathcal{W}^\perp).
\end{equation}

Now we proceed to the $t \times t$ matrix $C(x)$.
Because $C(x)$ is formed by multiplying a $t \times 2$ matrix by a $2 \times t$ matrix, $\operatorname{rank}(C(x)) \le 2$. Thus, $C(x)$ has at most two non-zero invariant factors, $c_1(x)$ and $c_2(x)$.

Let $r=\operatorname{rank}_{\mathbb F_2(x)}C(x)
\leq\min(t,2)$ be the polynomial rank.
After the unimodular Smith transformation, the
kernel condition reduces to $r$ nonzero diagonal
equations
\[
c_j(x)f'_j(x)\equiv0\pmod{x^L-1},
\qquad j=1,\ldots,r,
\]
while the remaining $t-r$ coordinates are free.
Therefore,
\[
\dim\ker C(x)=(t-r)L+
\sum_{j=1}^{r}\deg\gcd(c_j(x),x^L-1).
\]
Equivalently, padding the list of invariant factors
by zeros up to two entries, and using
$\gcd(0,x^L-1)=x^L-1$, we obtain
\begin{equation}\label{app:eq:ker_C}
\dim\ker C(x)=(t-2)L+
\sum_{j=1}^{2}\deg\gcd(c_j(x),x^L-1).
\end{equation}
The padded expression remains valid for $t=1$.

Substituting~\eqref{app:eq:phi_ker} and~\eqref{app:eq:ker_C} into~\eqref{app:eq:dim_cen}, we obtain the dimention of the center, with the $(t-2)L$ terms completely eliminated:
\begin{equation}\label{app:eq:dim_cent}
    \dim(\mathcal{W} \cap \mathcal{W}^\perp) = \deg \gcd(c_1(x), x^L-1) + \deg \gcd(c_2(x), x^L-1) - \dim \mathcal{W}^\perp.
\end{equation}
Similarly, $t=1$ case recovers Eq.~\eqref{eq:thm1_dim2} from the formulation of Theorem~\ref{thm_pol}, since $c_1(x)$ coincides with the cross-commutation polynomial $ p(x)q^*(x) + q(x)p^*(x)$ and $\deg \gcd(c_2(x), x^L-1) = \deg \gcd(0, x^L-1)=L$. 

\subsubsection{ Maximal stabilizer subgroup}
By the relation~\eqref{eq:stab_order}, the maximal stabilizer subgroup $S_{max}$ dimension~(in terms of the check vector subspace):
\begin{equation}\label{app:eq:dim_stabb_max}
    \dim(\mathcal S_{max}) = \frac{1}{2} \left[ \deg \gcd(c_1(x), x^L-1) + \deg \gcd(c_2(x), x^L-1) \right].
\end{equation}

The minimal entanglement entropy is obtained from~\eqref{eq:gen_opt_pur}:
\begin{equation}\label{app:eq:ent_gen}
    \ent = L - \frac{1}{2} \left[ \deg \gcd(c_1(x), x^L-1) + \deg \gcd(c_2(x), x^L-1) \right].
\end{equation}

When $c_1(x)$ and $c_2(x)$ are both non-zero, the system size $L$ completely drops out of the invariant factor bounds, proving an $\mathcal{O}(1)$  collapse of $\dim(\mathcal S_{max})$ and the scaling behavior $\ent = L - \mathcal{O}(1)$ of the minimal entanglement entropy.

\subsection{Example: $XY$ model}

Consider the Hamiltonian generated by two Pauli operators: $XX$ and $YY$ ($t=2$). Over $\mathbb{F}_2[x]$, the  corresponding  generators are $(p_1, q_1) = (1+x, 0)$ for $XX$ and $(p_2, q_2) = (1+x, 1+x)$ for $YY$.

The generator matrix is:
\begin{equation}\label{app:eq:gen_XY}
    W(x) = \begin{pmatrix} 1+x & 1+x \\ 0 & 1+x \end{pmatrix}.
\end{equation}

\subsubsection{Existence of solution}

We first determine whether there exists a solution to~\eqref{app:eq:main_inhom}, with the right part represented by $\mathbf{s}(x) = (E(x), E(x))^T$, where $E(x) = \sum_{j=0}^{L-1} x^j$. This solution corresponds to Pauli strings that  anti-commute with every cyclic shift of both the $XX$ and $YY$ terms. 

The  constraint matrix $H(x) = W^*(x)^T \Omega$ is given by:
\begin{equation}
    H(x) = \begin{pmatrix} 0 & 1+x \\ 1+x & 1+x \end{pmatrix}.
\end{equation}
To apply condition~\eqref{app:eq:solv_m}, we diagonalize $H(x)$ into its Smith Normal Form (SNF) via  matrix $U(x)$. One can easily verify that choosing 
\begin{equation}
    U(x) = \begin{pmatrix} 1 & 1 \\ 1 & 0 \end{pmatrix}
    \end{equation}
gives:
\begin{equation}
    U(x) H(x) = \begin{pmatrix} 1 & 1 \\ 1 & 0 \end{pmatrix} \begin{pmatrix} 0 & 1+x \\ 1+x & 1+x \end{pmatrix} = \begin{pmatrix} 1+x & 0 \\ 0 & 1+x \end{pmatrix}.
\end{equation}
Thus, the invariant factors are $d_1(x) = d_2(x) = 1+x$. The transformed vector $\mathbf s(x)$ reads:
\begin{equation}
    \mathbf{s}'(x) = U(x) \mathbf{s}(x) = \begin{pmatrix} 1 & 1 \\ 1 & 0 \end{pmatrix} \begin{pmatrix} E(x) \\ E(x) \end{pmatrix} = \begin{pmatrix} E(x)+E(x) \\ E(x) \end{pmatrix} = \begin{pmatrix} 0 \\ E(x) \end{pmatrix} \pmod 2.
\end{equation}
Condition~\eqref{app:eq:solv_m} requires $\gcd(d_\alpha(x), x^L-1) \mid s'_\alpha(x)$ for  $\alpha=1,\,2$. 
\begin{itemize}
    \item For $\alpha=1$, $\gcd(1+x, x^L-1) \mid 0$, which is trivially satisfied.
    \item For $\alpha=2$, we require $\gcd(1+x, x^L-1) \mid E(x)$. Since $(1+x)$  divides $x^L-1$ for all $L$, this simplifies to $(1+x) \mid E(x)$. 
\end{itemize}
Over $\mathbb{F}_2$, $(1+x) \mid E(x)$ if and only if $E(1) = 0$. For odd $L$, $E(1) = 1$, the condition fails, and no solution exists. For even $L$, $E(1) = 0$, the divisibility holds, and the existence of  anti-commuting Pauli operator is guaranteed.

\subsubsection{Space $\mathcal{W}^\perp$ and maximal stabilizer subspace}

We only need to consider the case of even $L$.

Calculation of invariant factors of $W(x)$ from Eq.~\eqref{app:eq:gen_XY} yields:

\begin{itemize}
    \item $w_1(x) = \gcd(1+x, 1+x, 0, 1+x) = 1+x$.
    \item $w_1(x)w_2(x) = \det W(x) = (1+x)^2$, which implies $w_2(x) = 1+x$.
\end{itemize}
Since $(1+x)$ divides $(x^L-1)$ for any $L$, we have $\deg \gcd(1+x, x^L-1) = 1$, and, by~\eqref{app:eq:ker_gen_main}:
\begin{equation}
    \dim \mathcal{W}^\perp = \deg(1+x) + \deg(1+x) = 1 + 1 = 2
\end{equation}
The homogeneous solution space $\mathcal{W}^\perp$
has binary dimension $2$ for all even $L$ and is
generated by the check vectors of
$X^{\otimes L}$ and $Y^{\otimes L}$.
The set of valid strings forms an affine coset
$\mathcal{V} = \mathbf{v}_0 + \mathcal{W}^\perp$
and therefore contains four distinct
Pauli strings.

Computing cross-commutation matrix $C_{\alpha\beta}(x) = p_\beta(x)q_\alpha^*(x) + q_\beta(x)p_\alpha^*(x)$  yields:
\begin{equation}
    C(x) = \begin{pmatrix} 0 & x+x^{-1} \\ x+x^{-1} & 0 \end{pmatrix} \cong \begin{pmatrix} 0 & (1+x)^2 \\ (1+x)^2 & 0 \end{pmatrix},
\end{equation}
where we factored out the multiplier $x^{-1}$. The invariant factors are:
\begin{itemize}
    \item $c_1(x)  = (1+x)^2$.
    \item $c_1(x)c_2(x) = \det C(x) = (1+x)^4$, which implies $c_2(x) = (1+x)^2$.
\end{itemize}

The dimensions of the center and stabilizer are then determined  by $\deg \gcd((1+x)^2, x^L-1)$. For even $L$, the polynomial $x^L-1 = (x^{L/2}-1)^2$ contains repeated roots, so $(1+x)^2 \mid (x^L-1)$. Thus, $\gcd((1+x)^2, x^L-1) = (1+x)^2$. Applying~\eqref{app:eq:dim_cent} and~\eqref{app:eq:dim_stabb_max}, we obtain:
\begin{equation}
    \dim(\mathcal{W} \cap \mathcal{W}^\perp) = \dim(\mathcal S_{max}) = 2.
\end{equation}
This reflects the fact that the whole group $G$ is abelian and hence it has  the same complex linear span as the maximal stabilizer subgroup generated by commuting Pauli strings $X^{\otimes L}$ and $Y^{\otimes L}$.

The entanglement entropy, calculated by~\eqref{app:eq:ent_gen}, reads:
\begin{equation}
    S_{\min} = L-2.
\end{equation}
This expression shows that for $L=2$ there exists a factorized state of the  two halves of the spin chain~($S_{\min}=0$). Indeed, the minimal stabilizer projector in this case is  generated by $XX$ and $YY$:
\begin{equation}\label{app:eq:projxy}
    \Pi_{\min}=\frac14\left(II + XX + YY - ZZ\right).
\end{equation}
Using the vectorization framework of Appendix~\ref{app:sec:vec}, it can be easily obtained that this is a projector on a pure state:
\begin{equation}
\frac1{\sqrt2}\ket{\sigma_x}\!\rangle=\frac1{\sqrt2}\sum_{i,\,j=0}^1\,( \sigma_x)_{ij}\ket{i}\ket{j}.
\end{equation}
This is an eigenstate of each of the four operators in the right part of~\eqref{app:eq:projxy}, with eigenvalue $1$.

The optimal superposition operator can be taken in the form: 
\begin{equation}
   A_{\mathrm{opt}} = V \Pi_{\min},
\end{equation}
where $V$ - one of the valid strings, for example, $V=IZ$. The operator $V \Pi_{\min}$ will hence be supported on a coset of the group $G$ generated by $V$. 

The reduced state of one of the two halves is then 
\begin{equation}
    \rho_{1,2} = A_{\mathrm{opt}} A_{\mathrm{opt}}^{\dagger} = V \Pi_{\min}V = \frac14\left(II - XX - YY - ZZ\right),
\end{equation}
which is a projector on another vectorization state~(Appendix~\ref{app:sec:vec}):
\begin{equation}   \frac1{\sqrt2}\ket{\sigma_y}\!\rangle=\frac1{\sqrt2}\sum_{i,\,j=0}^1\,( \sigma_y)_{ij}\ket{i}\ket{j}.
\end{equation}
The other half of the chain is in the state $\ket{\sigma_x}\!\rangle$:
\begin{equation}    \rho_{\Bar{1},\Bar{2}}=\left(A_{\mathrm{opt}}^{\dagger}A_{\mathrm{opt}}\right)^T=\Pi_{\min}=\frac12\ket{\sigma_x}\!\rangle \langle\!\bra{\sigma_x}.
\end{equation}
Consequently, the whole chain is in the factorized state $\ket{\sigma_y}\!\rangle_{1,2}\otimes \ket{\sigma_x}\!\rangle_{\Bar{1},\Bar{2}}$, which is a particular $L=2$ case of the ladder dimer construction for $XY$ model discussed in Section~\ref{sec:ex_eig}.

\section{Dimension of the center for the $XZZ$ generator}
\label{app:zxx_center_dimension}

For a single generator $XZZ$, the dimension of the center  is given by:
\begin{equation}
    \dim(\mathcal{W} \cap \mathcal{W}^\perp) = \deg \gcd(C(x), x^L-1),
\end{equation}
where the cross-commutation polynomial evaluates to:
\begin{equation}
    C(x) = x^4 + x^3 + x + 1.
\end{equation}
To obtain the dimension of the center for an arbitrary lattice size $L$, we first determine the irreducible factorization of $C(x)$ over  $\mathbb{F}_2$, and then analyze which of its factors divide the boundary condition polynomial $x^L-1$.

\subsubsection{Factorization of $C(x)$}
By grouping the terms of the polynomial, we can factor $C(x)$ as:
\begin{equation}
    C(x) = x^3(x + 1) + 1(x + 1) = (x^3 + 1)(x + 1).
\end{equation}
The sum of cubes $x^3+1$  over $\mathbb{F}_2$ factors as $(x+1)(x^2+x+1)$, and so  the complete  factorization of $C(x)$ reads:
\begin{equation}
    C(x) = (x+1)(x+1)(x^2+x+1) = (x+1)^2(x^2+x+1).
\end{equation}
Thus, the  roots of $C(x)$ include a repeated root at $x=1$ (multiplicity $2$) and a pair of roots satisfying $x^2+x+1=0$.

\subsubsection{Divisibility of $x^L-1$}
The greatest common divisor is constructed from the irreducible factors of $C(x)$ that simultaneously divide $x^L-1$. We evaluate the divisibility conditions for each factor and its power independently:

\begin{itemize}
    \item \textbf{Factor $(x+1)$:} 
    Evaluating the periodicity polynomial at the root $x=1$ yields $1^L - 1 = 0 \pmod 2$ for all $L \ge 1$. Thus, $(x+1)$ divides $x^L-1$ universally for all lattice sizes.
    
    \item \textbf{Factor $(x+1)^2$:} 
    A polynomial over $\mathbb{F}_2$ contains a repeated root if and only if its formal derivative vanishes at that root. The formal derivative of $x^L-1$ is $L x^{L-1}$. This derivative evaluates to $0 \pmod 2$ if and only if $L$ is even. Consequently, the multiplicity of the root $x=1$ is strictly greater than $1$, which means that $(x+1)^2 \mid (x^L-1)$ if and only if $L \equiv 0 \pmod 2$.
    
    \item \textbf{Factor $(x^2+x+1)$:}
Since neither $0$ nor $1$ is a root of this quadratic, $x^2+x+1$ is irreducible over $\mathbb{F}_2$. Extending the field to $\mathbb{F}_4 = \mathbb{F}_2[\omega]$, the roots may be expressed in terms of $\omega$, which satisfies $\omega^2 + \omega + 1 = 0$. Every element in $\mathbb{F}_4$ can be written uniquely as $a + b\omega$ with $a, b \in \{0, 1\}$. Since $x^3+1 = (x+1)(x^2+x+1)$ over $\mathbb{F}_2$, the roots of $x^2+x+1$ satisfy $\omega^3 = 1$. 
No smaller positive power equals $1$: $\omega \neq 1$ because $1^2+1+1 = 1 \neq 0$ in $\mathbb{F}_2$, and $\omega^2 = \omega + 1 \neq 1$ (which would otherwise imply $\omega = 0$). Thus, the roots of $x^2+x+1$ are primitive third roots of unity (where a primitive $n$-th root of unity has $n$ as the minimal positive integer for which $\omega^n = 1$). One can  check that if $\omega$ is a root of $x^2+x+1$, then $\omega^2$ is the second root.

To determine when these roots satisfy the condition $\omega^L - 1 = 0$, we write the system size as $L = 3l + r$, where $l$ is the quotient and the remainder $r \in \{0, 1, 2\}$. Using $\omega^3 = 1$, we obtain
\begin{equation}
    \omega^L = \left(\omega^3\right)^l \omega^r = \omega^r.
\end{equation}
For $\omega^L$ to evaluate to $1$, the remainder must be $r = 0$. Consequently, the roots satisfy $\omega^L - 1 = 0$ if and only if $L$ is an integer multiple of $3$. Therefore, $x^2+x+1$ divides $x^L-1$ in $\mathbb{F}_2[x]$ if and only if $L \equiv 0 \pmod{3}$.
\end{itemize}

\subsubsection{Calculation of the center dimension}
Because the divisibility of the factors depends on the parity of $L$ (mod $2$) and its divisibility by $3$ (mod $3$), the degree of the GCD cycles  with modulo $6$. Let $g(x) = \gcd(C(x), x^L-1)$. We have the four possible cases:

\begin{enumerate}
    \item \textbf{$L$ is odd and not divisible by 3} ($L \equiv 1, 5 \pmod 6$): \\
    Only the single root $(x+1)$ divides $x^L-1$. \\
    $g(x) = x+1$ implies $\dim(\mathcal{W} \cap \mathcal{W}^\perp) = 1$.

    \item \textbf{$L$ is even and not divisible by 3} ($L \equiv 2, 4 \pmod 6$): \\
    The root $x=1$ has multiplicity $2$, but the third roots of unity  are not shared with $x^L-1$. \\
    $g(x) = (x+1)^2 = x^2+1$ implies $\dim(\mathcal{W} \cap \mathcal{W}^\perp) = 2$.

    \item \textbf{$L$ is odd and divisible by 3} ($L \equiv 3 \pmod 6$): \\
    The root $x=1$ has multiplicity $1$, but the primitive third roots are supported. \\
    $g(x) = (x+1)(x^2+x+1) = x^3+1$ implies $\dim(\mathcal{W} \cap \mathcal{W}^\perp) = 3$.

    \item \textbf{$L$ is even and divisible by 3} ($L \equiv 0 \pmod 6$): \\
    Both the degenerate root at $x=1$ and the third roots of unity are shared with $x^L-1$. The GCD is equal to $C(x)$. \\
    $g(x) = (x+1)^2(x^2+x+1) = x^4+x^3+x+1$ implies $\dim(\mathcal{W} \cap \mathcal{W}^\perp) = 4$.
\end{enumerate}

\section{Vectorization and Operator Sliding}\label{app:sec:vec}
\subsection{General formulation}

The operator-state isomorphism (or matrix vectorization) allows any bipartite pure state to be uniquely expressed by acting with a local single-system operator $M$ on one subsystem of an unnormalized maximally entangled Bell state.

A bipartite pure state $\ket{\psi} \in \mathcal{H}_A \otimes \mathcal{H}_B$ can be expanded in the computational bases $\{\ket{i}_A\}$ and $\{\ket{j}_B\}$ as
\begin{equation}
\ket{\psi} = \sum_{i=0}^{d_A-1} \sum_{j=0}^{d_B-1} M_{ij} \ket{i}_A \otimes \ket{j}_B,
\end{equation}
where $d_A$ and $d_B$ denote the Hilbert space dimensions of subsystems $A$ and $B$, respectively, and $M \in \mathbb{C}^{d_A \times d_B}$ is the coefficient matrix with elements $M_{ij} = \langle i, j \vert \psi \rangle$. Defining the canonical (unnormalized) maximally entangled state as
\begin{equation}
\ket{\Phi^+} = \sum_{k=0}^{d_B-1} \ket{k} \otimes \ket{k} \in \mathcal{H}_B \otimes \mathcal{H}_B,
\end{equation}
we may express $\ket{\psi}$ by acting with $M$ locally on the first subsystem:
\begin{equation}
\ket{\psi} = (M \otimes I_B) \ket{\Phi^+} = \sum_{k} (M\ket{k}) \otimes \ket{k} = \sum_{i,k} M_{ik} \ket{i} \otimes \ket{k}.
\end{equation}
In connection with this representation for bipartite pure states, a special double-ket notation is introduced:
\begin{equation}
    \ket{M}\!\rangle \equiv (M \otimes I_B) \ket{\Phi^+}.
\end{equation}
The state $\ket{M}\!\rangle$ is referred to as the vectorization of the operator $M$. As a bipartite pure state, any dimer state can be described as the vectorization of its corresponding operator $M$.

For equal subsystem dimensions $d_A=d_B=d$, the sliding
trick allows an operator $M$ acting on the left subsystem
to be shifted onto the right subsystem via transposition:
\begin{equation}
(M \otimes I_B)\ket{\Phi^+} = (I_A \otimes M^T)\ket{\Phi^+}.
\end{equation}
This identity is easily verified by expanding in the computational basis:
\begin{equation}
(I_A \otimes M^T)\ket{\Phi^+} = \sum_k \ket{k} \otimes M^T \ket{k} = \sum_{k,j} (M^T)_{jk} \ket{k} \otimes \ket{j} = \sum_{k,j} M_{kj} \ket{k} \otimes \ket{j} = (M \otimes I_B) \ket{\Phi^+}.
\end{equation}

Similarly, one easily establishes that applying local operators $A$ and $B$ to both subsystems of $\ket{M}\!\rangle$ corresponds to matrix multiplication inside the double-ket:
\begin{equation}\label{eq:app:gen_sliding}
(A \otimes B)\ket{M}\!\rangle = \ket{A M B^T}\!\rangle.
\end{equation}
Identity~(\ref{eq:app:gen_sliding}) plays a central role when evaluating the action of various Hamiltonian terms on dimer-based constructions, as described in the next subsection.

\subsection{Working examples}
Here we list the properties of various dimer states used in the construction of exact eigenstates for the Hamiltonians presented in the main text. 

\subsubsection{Spin-$1/2$}\label{app:sec:dimex}

We consider the operator generating a $\pi$-rotation around the axis $\mathbf{n}_{xz} = \frac{1}{\sqrt{2}}(\mathbf{e}_x + \mathbf{e}_z)$:
\begin{equation}
    R_{xz} = \frac{\sigma_x + \sigma_z}{\sqrt{2}} \sim \exp\left(-i \frac{\pi}{2} \mathbf{n}_{xz} \cdot \bm{\sigma}\right),
\end{equation}
also known as the Hadamard operator $U_H$ introduced in Eqs.~(\ref{eq:had}) and~(\ref{eq:app:hadard}). In addition to being Hermitian and unitary, this operator satisfies key algebraic transformation relations:
\begin{equation}\label{app:eq:xztf_prop}
    R_{xz}\sigma_x = \sigma_z R_{xz},\quad R_{xz}\sigma_z = \sigma_x R_{xz},\quad R_{xz}\sigma_y = -\sigma_y R_{xz}.
\end{equation}

The three dimer states from the ladder construction for the $XZ$ Hamiltonian, given in Eqs.~(\ref{eq:xzpsi}), (\ref{eq:xzphi}), and~(\ref{eq:xzchi}) (and also shown in Fig.~\ref{fig:xzlarge}), can be concisely expressed via vectorization:
\begin{equation}
    \ket{\psi} = \ket{R_{xz}\sigma_y}\!\rangle,\quad\ket{\phi} = \ket{R_{xz}}\!\rangle,\quad\ket{\chi} = \ket{R_{xz}\sigma_z}\!\rangle.
\end{equation}
Here and in what follows, when connecting previously introduced states with vectorization, the identities hold up to an overall normalization factor~($1/\sqrt{2}$ for spin-$1/2$ case).

Combining Eqs.~(\ref{app:eq:xztf_prop}) and~(\ref{eq:app:gen_sliding}) with the algebraic properties of the Pauli matrices yields the following operator sliding relations:
\begin{align}
    (I \otimes \sigma_x)\ket{R_{xz}}\!\rangle &= (\sigma_z \otimes I)\ket{R_{xz}}\!\rangle, & (I \otimes \sigma_x)\ket{R_{xz}\sigma_z}\!\rangle &= -(\sigma_z \otimes I)\ket{R_{xz}\sigma_z}\!\rangle,\nonumber \\
    (I \otimes \sigma_z)\ket{R_{xz}}\!\rangle &= (\sigma_x \otimes I)\ket{R_{xz}}\!\rangle, & (I \otimes \sigma_z)\ket{R_{xz}\sigma_z}\!\rangle &= (\sigma_x \otimes I)\ket{R_{xz}\sigma_z}\!\rangle, \nonumber\\
    (I \otimes \sigma_x)\ket{R_{xz}\sigma_y}\!\rangle &= -(\sigma_z \otimes I)\ket{R_{xz}\sigma_y}\!\rangle, & (I \otimes \sigma_z)\ket{R_{xz}\sigma_y}\!\rangle &= -(\sigma_x \otimes I)\ket{R_{xz}\sigma_y}\!\rangle,
\end{align}

some of which correspond directly to Eqs.~(\ref{eq:xz_tel1}), (\ref{eq:xz_tel2}), and~(\ref{eq:xz_tel3}) from the main text.

The next group of dimer states consists of those introduced in Eqs.~(\ref{eq:xy_xdim}) and~(\ref{eq:xy_ydim}) for the $XY$ model:
\begin{equation}\label{eq:app:xydim}
    \ket{\phi}_X = \ket{\sigma_x}\!\rangle,\quad \ket{\phi}_Y = \ket{\sigma_y}\!\rangle.
\end{equation}
Applying Eq.~(\ref{eq:app:gen_sliding}) alongside the Pauli matrix algebra leads to the identities
\begin{align}
    (I \otimes \sigma_x)\ket{\sigma_x}\!\rangle &= (\sigma_x \otimes I)\ket{\sigma_x}\!\rangle, & (I \otimes \sigma_x)\ket{\sigma_y}\!\rangle &= -(\sigma_x \otimes I)\ket{\sigma_y}\!\rangle,\nonumber \\
    (I \otimes \sigma_y)\ket{\sigma_x}\!\rangle &= (\sigma_y \otimes I)\ket{\sigma_x}\!\rangle, & (I \otimes \sigma_y)\ket{\sigma_y}\!\rangle &= -(\sigma_y \otimes I)\ket{\sigma_y}\!\rangle,\nonumber \\
    (I \otimes \sigma_z)\ket{\sigma_x}\!\rangle &= -(\sigma_z \otimes I)\ket{\sigma_x}\!\rangle, & (I \otimes \sigma_z)\ket{\sigma_y}\!\rangle &= -(\sigma_z \otimes I)\ket{\sigma_y}\!\rangle,
\end{align}

where we have used the fact that $\sigma_y^T = -\sigma_y$.

By using the Hermitian and unitary operator 
\begin{equation}
    R_{xy} = \frac{\sigma_x + \sigma_y}{\sqrt{2}} \sim \exp\left(-i \frac{\pi}{2} \mathbf{n}_{xy} \cdot \bm{\sigma}\right),
\end{equation}
which generates a $\pi$-rotation around $\mathbf{n}_{xy} = \frac{1}{\sqrt{2}}(\mathbf{e}_x + \mathbf{e}_y)$, we obtain the following transformation properties:
\begin{equation}\label{app:eq:xytf_prop}
    R_{xy}\sigma_x = \sigma_y R_{xy}, \quad R_{xy}\sigma_y = \sigma_x R_{xy}, \quad R_{xy}\sigma_z = -\sigma_z R_{xy}.
\end{equation}
Consequently, the rotated $X$- and $Y$-dimers from Eqs.~\eqref{eq:xy_xrdim} and~\eqref{eq:xy_yrdim} can be expressed as
\begin{equation}\label{eq:app:xyrdim}
    \ket{\phi}_{RX} = \ket{R_{xy}\sigma_x}\!\rangle, \quad \ket{\phi}_{RY} = \ket{R_{xy}\sigma_y}\!\rangle.
\end{equation}
Similarly, combining Eqs.~\eqref{eq:app:gen_sliding} and~\eqref{app:eq:xytf_prop} yields the sliding properties:
\begin{align}
    (I\otimes\sigma_x)\ket{R_{xy}\sigma_x}\!\rangle &= (\sigma_y\otimes I)\ket{R_{xy}\sigma_x}\!\rangle, & (I\otimes\sigma_x)\ket{R_{xy}\sigma_y}\!\rangle &= -(\sigma_y\otimes I)\ket{R_{xy}\sigma_y}\!\rangle, \nonumber\\
    (I\otimes\sigma_y)\ket{R_{xy}\sigma_x}\!\rangle &= (\sigma_x\otimes I)\ket{R_{xy}\sigma_x}\!\rangle, & (I\otimes\sigma_y)\ket{R_{xy}\sigma_y}\!\rangle &= -(\sigma_x\otimes I)\ket{R_{xy}\sigma_y}\!\rangle, \nonumber\\
    (I\otimes\sigma_z)\ket{R_{xy}\sigma_x}\!\rangle &= (\sigma_z\otimes I)\ket{R_{xy}\sigma_x}\!\rangle, & (I\otimes\sigma_z)\ket{R_{xy}\sigma_y}\!\rangle &= (\sigma_z\otimes I)\ket{R_{xy}\sigma_y}\!\rangle.
\end{align}

\subsubsection{Spin-$1$}
\label{app:sec:dimex_spinone}

To extend the ladder construction for the $XY$ model to spin $1$, the states \eqref{eq:app:xydim} and \eqref{eq:app:xyrdim} must be modified accordingly. The operator $R_{xy}$ naturally extends to
\begin{equation}
    R_{xy} = \exp\left(-i\pi\mathbf{n}_{xy}\cdot\mathbf{S}\right),
\end{equation}
where $\mathbf{n}_{xy} = \frac{1}{\sqrt{2}}(\mathbf{e}_x + \mathbf{e}_y)$ as before, and $\mathbf{S}=\{S_x,\,S_y,\,S_z\}$ denotes the spin operator. This operator $R_{xy}$ is both unitary and Hermitian. The transformation properties of the spin operators under this rotation take the form:
\begin{equation}
    R_{xy}S_x = S_y R_{xy}, \quad R_{xy}S_y = S_x R_{xy}, \quad R_{xy}S_z = -S_z R_{xy}.
\end{equation}

To retain the commutation and anticommutation algebra in the spin-$1$ case, the Pauli operators in the vectorizations \eqref{eq:app:xydim} and \eqref{eq:app:xyrdim} are replaced with the parity operators $\mathcal{P}_\alpha = e^{i\pi S_\alpha}$, which were also used in Ref.~\cite{ZeroEig25}. One can show that the operator $\mathcal{P}_\alpha$ commutes with $S_\alpha$ but anticommutes with $S_\beta$ for $\beta \neq \alpha$, for arbitrary spin $S$.

The generalized dimers can now be expressed as vectorizations of the modified operators:
\begin{align}
    \ket{\phi}_X &= \ket{\mathcal{P}_x}\!\rangle,  & \ket{\phi}_Y &= \ket{\mathcal{P}_y}\!\rangle, \nonumber\\
    \ket{\phi}_{RX} &= \ket{R_{xy}\mathcal{P}_x}\!\rangle, & \ket{\phi}_{RY} &= \ket{R_{xy}\mathcal{P}_y}\!\rangle.
\end{align}

Similarly to the spin-$1/2$ case, we can formulate the sliding properties for these generalized spin-$1$ dimers. For the unrotated dimers, the sliding properties read:
\begin{align}\label{app:eq:s1xy}
    (I \otimes S_x)\ket{\mathcal{P}_x}\!\rangle &= (S_x \otimes I)\ket{\mathcal{P}_x}\!\rangle, & (I \otimes S_x)\ket{\mathcal{P}_y}\!\rangle &= -(S_x \otimes I)\ket{\mathcal{P}_y}\!\rangle, \nonumber\\
    (I \otimes S_y)\ket{\mathcal{P}_x}\!\rangle &= (S_y \otimes I)\ket{\mathcal{P}_x}\!\rangle, & (I \otimes S_y)\ket{\mathcal{P}_y}\!\rangle &= -(S_y \otimes I)\ket{\mathcal{P}_y}\!\rangle, \nonumber\\
    (I \otimes S_z)\ket{\mathcal{P}_x}\!\rangle &= -(S_z \otimes I)\ket{\mathcal{P}_x}\!\rangle, & (I \otimes S_z)\ket{\mathcal{P}_y}\!\rangle &= -(S_z \otimes I)\ket{\mathcal{P}_y}\!\rangle.
\end{align}
For the rotated dimers, combining the action of $R_{xy}$ with the algebraic properties of $\mathcal{P}_\alpha$ yields the following relations:
\begin{align}\label{app:eq:s1xyrot}
    (I\otimes S_x)\ket{R_{xy}\mathcal{P}_x}\!\rangle &= (S_y\otimes I)\ket{R_{xy}\mathcal{P}_x}\!\rangle, & (I\otimes S_x)\ket{R_{xy}\mathcal{P}_y}\!\rangle &= -(S_y\otimes I)\ket{R_{xy}\mathcal{P}_y}\!\rangle, \nonumber\\
    (I\otimes S_y)\ket{R_{xy}\mathcal{P}_x}\!\rangle &= (S_x\otimes I)\ket{R_{xy}\mathcal{P}_x}\!\rangle, & (I\otimes S_y)\ket{R_{xy}\mathcal{P}_y}\!\rangle &= -(S_x\otimes I)\ket{R_{xy}\mathcal{P}_y}\!\rangle, \nonumber\\
    (I\otimes S_z)\ket{R_{xy}\mathcal{P}_x}\!\rangle &= (S_z\otimes I)\ket{R_{xy}\mathcal{P}_x}\!\rangle, & (I\otimes S_z)\ket{R_{xy}\mathcal{P}_y}\!\rangle &= (S_z\otimes I)\ket{R_{xy}\mathcal{P}_y}\!\rangle.
\end{align}

\section{Mirrored dimers for the spin-$1$ $XY$ model}\label{app:sec:mirr}

By virtue of properties~\eqref{app:eq:s1xy}, the bulk alternating $\ket{\mathcal{P}_x}\!\rangle$ and $\ket{\mathcal{P}_y}\!\rangle$ dimers counterbalance the $XY$ Hamiltonian terms on the left and right legs of the ladder construction. The difficulty arises at the top and bottom boundaries, where the corresponding dimer states are no longer eigenstates of the $S_x \otimes S_x$ and $S_y \otimes S_y$ terms. Consequently, the ladder construction cannot be directly extended to the spin-1 case in its current form.

The authors of Ref.~\cite{SashMoudXY26} proposed a structure, originally for the isotropic uniform $XY$ chain ($J_x = J_y = J$), in which the top and bottom boundaries consist of a single site each. The remainder of the construction is completely analogous to the ladder layout, consisting of alternating dimer states that counterbalance the side terms of the Hamiltonian. Here, we provide an interpretation of how the boundary Hamiltonian terms interact with one another from the perspective of operator sliding.

\begin{figure*}[t]
\centering
\includegraphics[width=0.35\textwidth]{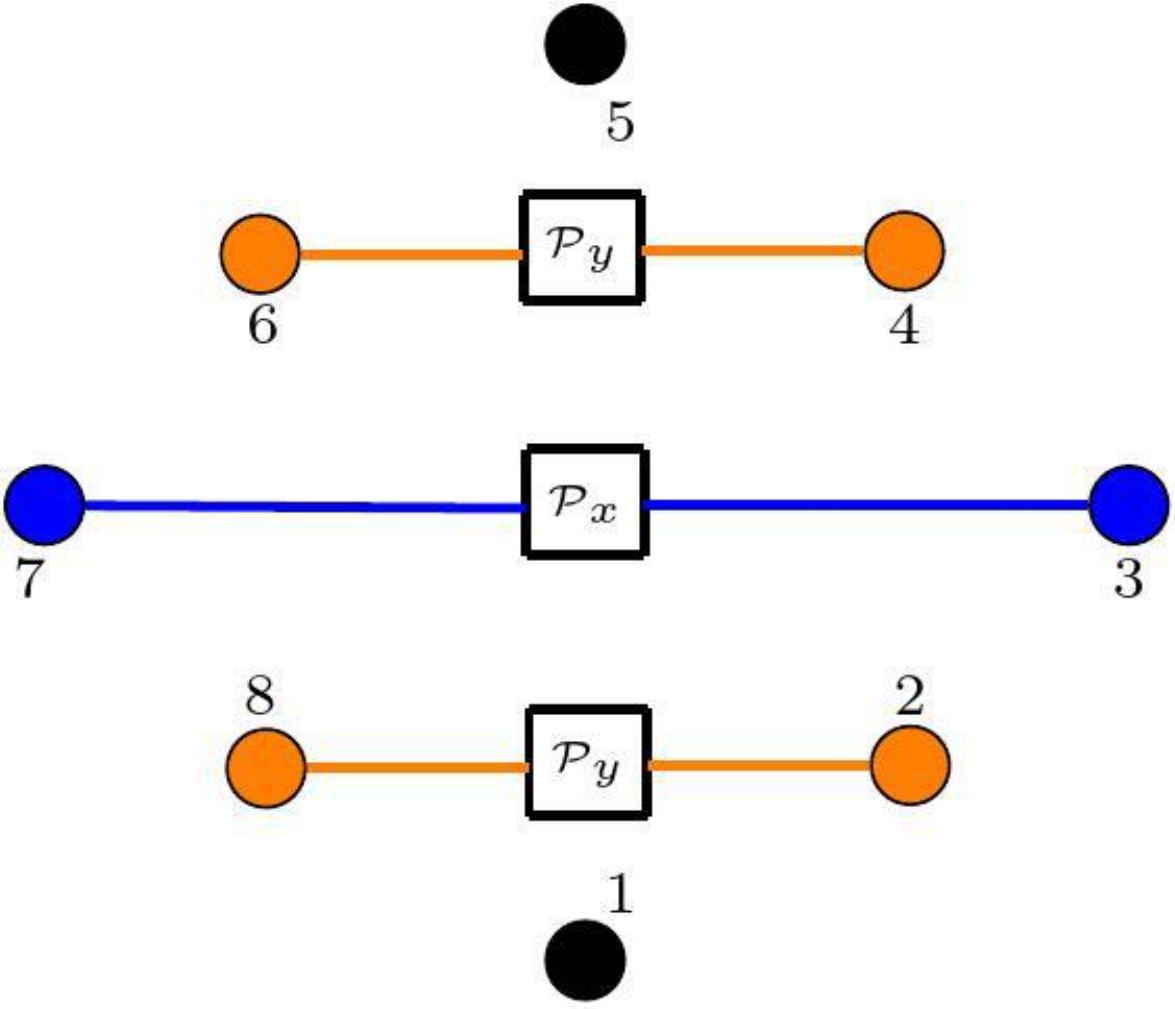}
\caption{Mirrored dimer construction for the $XY$ model on an $8$-site chain. The dimer states $\ket{\mathcal{P}_x}\!\rangle$ and $\ket{\mathcal{P}_y}\!\rangle$ are highlighted in blue and orange, respectively.}
\label{app:fig:mirr_dim_v}
\end{figure*}

The construction for an $8$-site spin chain in our formulation, which differs from that of Ref.~\cite{SashMoudXY26} in the choice of dimer states, is shown in Fig.~\ref{app:fig:mirr_dim_v}. Consider, for example, the terms supported on sites $1$, $2$, and $8$ at the bottom of the construction. Sites $2$ and $8$ form a joint dimer state $\ket{\mathcal{P}_y}\!\rangle_{8,2}$, while site $1$ is in a separate state $\ket{\phi}_1$ (or, alternatively, shares a pure bipartite state with the opposite boundary site $5$). The action of one such term, say $(S_x)_8 \otimes (S_x)_1$, on the composite state of this region can be evaluated using the sliding rule in Eq.~\eqref{app:eq:s1xy}:
\begin{equation}\label{app:eq:slxy1}
    (S_x)_8 \otimes (S_x)_1 \ket{\mathcal{P}_y}\!\rangle_{8,2} \otimes \ket{\phi}_1 = -(S_x)_2 \otimes (S_x)_1 \ket{\mathcal{P}_y}\!\rangle_{8,2} \otimes \ket{\phi}_1.
\end{equation}
That is, the sliding property applies exclusively to the $(S_x)_8$ operator of the Hamiltonian term: $(S_x)_8$ slides across the $\ket{\mathcal{P}_y}\!\rangle_{8,2}$ dimer to become $-(S_x)_2$, while $(S_x)_1$ remains unchanged on site $1$. Similarly, for the $S_y \otimes S_y$ term, we have:
\begin{equation}\label{app:eq:slxy2}
    (S_y)_8 \otimes (S_y)_1 \ket{\mathcal{P}_y}\!\rangle_{8,2} \otimes \ket{\phi}_1 = -(S_y)_2 \otimes (S_y)_1 \ket{\mathcal{P}_y}\!\rangle_{8,2} \otimes \ket{\phi}_1.
\end{equation}
As a result, the $S_x \otimes S_x$ terms supported on sites $(1,2)$ and $(1,8)$ cancel each other, and the same holds for the corresponding $S_y \otimes S_y$ terms. By Eq.~\eqref{app:eq:s1xy}, such cancellation is achieved only with the $\ket{\mathcal{P}_y}\!\rangle$ state. For this reason, $\ket{\mathcal{P}_y}\!\rangle$ must occupy both the top and bottom boundaries (Fig.~\ref{app:fig:mirr_dim_v}). 
Since the $2L$-site chain contains $L-1$ dimers
between the two unpaired boundary sites,
alternation with $\ket{\mathcal P_y}\!\rangle$
at both ends requires even $L$.
     The entire construction then, in the absence of an external magnetic field, yields a zero-energy eigenvalue. In the presence of such a field, the joint state of sites $1$ and $5$ must counterbalance the local $S_z$ operators supported on these sites, or, alternatively, each site must host an eigenstate of $S_z$. The overall construction extends directly to an anisotropic, non-uniform $XY$ chain, provided that the couplings $J_{x,i}$, $J_{y,i}$ and magnetic field parameters $h_i$ satisfy reflection symmetry conditions analogous to Eqs.~\eqref{eq:J_mirr} and~\eqref{eq:h_mirr}, respectively. Furthermore, the construction extends naturally to arbitrary spin $S$, as the (anti-)commutation relations of the parity operators $\mathcal{P}_x$, $\mathcal{P}_y$, $\mathcal P_z$ with the spin operators remain invariant.

To highlight the distinction between our approach and the construction of Ref.~\cite{SashMoudXY26}, we write the vectorizations of the spin-$1$ parity operators in the standard $S_z$ basis as follows:
\begin{align}
    \ket{\mathcal{P}_x}\!\rangle &= -\ket{+1}\ket{-1} - \ket{0}\ket{0} - \ket{-1}\ket{+1}, \nonumber\\[6pt]
    \ket{\mathcal{P}_y}\!\rangle &= +\ket{+1}\ket{-1} - \ket{0}\ket{0} + \ket{-1}\ket{+1}, \nonumber\\[6pt]
    \ket{\mathcal{P}_z}\!\rangle &= -\ket{+1}\ket{+1} + \ket{0}\ket{0} - \ket{-1}\ket{-1},
\end{align}
where the state $\ket{\mathcal{P}_y}\!\rangle$ is proportional to the isotropic singlet state $\ket{S_{\text{tot}}=0, M=0}$.

By contrast, the alternating dimers used in the construction of Ref.~\cite{SashMoudXY26} [Eqs.~(32) and~(34)] are
\begin{align}
    \ket{\phi^+} &= \ket{+1}\ket{+1} + \ket{0}\ket{0} + \ket{-1}\ket{-1} = \ket{I}\!\rangle, \nonumber\\[6pt]
    \ket{\phi^-} &= \ket{+1}\ket{+1} - \ket{0}\ket{0} + \ket{-1}\ket{-1} = -\ket{\mathcal{P}_z}\!\rangle. \label{app:eq:refst}
\end{align}
While the bulk dimers ensure the cancellation of side interaction terms along the legs, neither of the states $\ket{\phi^{\pm}}$, when placed at the boundaries, provides the required sliding behavior for \emph{both} the $S_x \otimes S_x$ and $S_y \otimes S_y$ Hamiltonian terms as in Eqs.~\eqref{app:eq:slxy1} and~\eqref{app:eq:slxy2}. Furthermore, in the presence of an external field, these states fail to allow cancellation of the corresponding $S_z$ terms. Consequently, it is unlikely that the choice of states in Eq.~\eqref{app:eq:refst} produces valid eigenstates in this scheme.

%\begin{figure*}[t]
%\centering
%\includegraphics[width=0.35\textwidth]{xyvert.pdf}
%\caption{Mirrored dimer construction for the $XY$ model on an $8$-site chain. The dimer states $\ket{\mathcal{P}_x}\!\rangle$ and $\ket{\mathcal{P}_y}\!\rangle$ are highlighted in blue and orange, respectively.}
%\label{app:fig:mirr_dim_v}
%\end{figure*}

\bibliography{bibliography}

@PREAMBLE{
 "\providecommand{\noopsort}[1]{}" 
 # "\providecommand{\singleletter}[1]{#1}" 
}

@article{DAlessio2016,
  author  = {D'Alessio, Luca and Kafri, Yariv and
             Polkovnikov, Anatoli and Rigol, Marcos},
  title   = {From quantum chaos and eigenstate thermalization
             to statistical mechanics and thermodynamics},
  journal = {Advances in Physics},
  volume  = {65},
  number  = {3},
  pages   = {239--362},
  year    = {2016},
  doi     = {10.1080/00018732.2016.1198134}
}

@article{Serbyn2021,
  author  = {Serbyn, Maksym and Abanin, Dmitry A. and
             Papi{\'c}, Zlatko},
  title   = {Quantum many-body scars and weak breaking of ergodicity},
  journal = {Nature Physics},
  volume  = {17},
  pages   = {675--685},
  year    = {2021},
  doi     = {10.1038/s41567-021-01230-2}
}

@article{Okatev2025,
  author  = {Okatev, V. S. and Sotnikov, Oleg M. and
             Mazurenko, V. V.},
  title   = {Exploring entanglement in finite-size quantum systems
             with degenerate ground state},
  journal = {Physical Review B},
  volume  = {111},
  pages   = {054443},
  year    = {2025},
  doi     = {10.1103/PhysRevB.111.054443}
}

@article{Karle2021,
  author  = {Karle, Volker and Serbyn, Maksym and
             Michailidis, Alexios A.},
  title   = {Area-Law Entangled Eigenstates from Nullspaces
             of Local Hamiltonians},
  journal = {Physical Review Letters},
  volume  = {127},
  pages   = {060602},
  year    = {2021},
  doi     = {10.1103/PhysRevLett.127.060602}
}

@article{Deutsch1991,
  author  = {Deutsch, J. M.},
  title   = {Quantum statistical mechanics in a closed system},
  journal = {Physical Review A},
  volume  = {43},
  pages   = {2046--2049},
  year    = {1991},
  doi     = {10.1103/PhysRevA.43.2046}
}

@article{Srednicki1994,
  author  = {Srednicki, Mark},
  title   = {Chaos and quantum thermalization},
  journal = {Physical Review E},
  volume  = {50},
  pages   = {888--901},
  year    = {1994},
  doi     = {10.1103/PhysRevE.50.888}
}

@article{Mori2018,
  author  = {Mori, Takashi and Ikeda, Tatsuhiko N. and
             Kaminishi, Eriko and Ueda, Masahito},
  title   = {Thermalization and prethermalization in isolated
             quantum systems: a theoretical overview},
  journal = {Journal of Physics B: Atomic, Molecular and
             Optical Physics},
  volume  = {51},
  number  = {11},
  pages   = {112001},
  year    = {2018},
  doi     = {10.1088/1361-6455/aabcdf}
}

@article{Turner2018,
  author  = {Turner, Christopher J. and Michailidis, Alexios A. and
             Abanin, Dmitry A. and Serbyn, Maksym and
             Papi{\'c}, Zlatko},
  title   = {Weak ergodicity breaking from quantum many-body scars},
  journal = {Nature Physics},
  volume  = {14},
  pages   = {745--749},
  year    = {2018},
  doi     = {10.1038/s41567-018-0137-5}
}

@article{Moudgalya2022,
  author  = {Moudgalya, Sanjay and Bernevig, B. Andrei and
             Regnault, Nicolas},
  title   = {Quantum many-body scars and {Hilbert} space
             fragmentation: a review of exact results},
  journal = {Reports on Progress in Physics},
  volume  = {85},
  number  = {8},
  pages   = {086501},
  year    = {2022},
  doi     = {10.1088/1361-6633/ac73a0}
}

@article{Ma2018,
  author  = {Ma, Han and Schmitz, A. T. and
             Parameswaran, S. A. and Hermele, Michael and
             Nandkishore, Rahul M.},
  title   = {Topological entanglement entropy of fracton
             stabilizer codes},
  journal = {Physical Review B},
  volume  = {97},
  pages   = {125101},
  year    = {2018},
  doi     = {10.1103/PhysRevB.97.125101}
}

@article{Sun2025,
  author  = {Sun, Jiace and Cheng, Lixue and Zhang, Shi-Xin},
  title   = {Stabilizer ground states for simulating quantum
             many-body physics: theory, algorithms, and applications},
  journal = {Quantum},
  volume  = {9},
  pages   = {1782},
  year    = {2025},
  doi     = {10.22331/q-2025-06-24-1782}
}

@article{Gu2024,
  author  = {Gu, Andi and Oliviero, Salvatore F. E. and
             Leone, Lorenzo},
  title   = {Doped stabilizer states in many-body physics
             and where to find them},
  journal = {Physical Review A},
  volume  = {110},
  pages   = {062427},
  year    = {2024},
  doi     = {10.1103/PhysRevA.110.062427}
}

@article{Dooley2026,
  author  = {Dooley, Shane},
  title   = {Parent {Hamiltonians} for Stabilizer
             Quantum Many-Body Scars},
  journal = {Physical Review Letters},
  volume  = {136},
  number  = {24},
  pages   = {240402},
  year    = {2026},
  doi     = {10.1103/3vk5-483c}
}

@article{Mesty_2025,
doi = {10.1088/1751-8121/ade380},
url = {https://doi.org/10.1088/1751-8121/ade380},
year = {2025},
month = {jun},
publisher = {IOP Publishing},
volume = {58},
number = {26},
pages = {265001},
author = {Mestyán, Márton and Pozsgay, Balázs},
title = {Crosscap states with tunable entanglement as exact eigenstates of local spin chain Hamiltonians},
journal = {Journal of Physics A: Mathematical and Theoretical}
}

@article{ZeroEig25,
  title = {Exact Volume-Law Entangled Zero-Energy Eigenstates in a Large Class of Spin Models},
  author = {Mohapatra, Sashikanta and Moudgalya, Sanjay and Balram, Ajit C.},
  journal = {Phys. Rev. Lett.},
  volume = {134},
  issue = {21},
  pages = {210403},
  numpages = {7},
  year = {2025},
  month = {May},
  publisher = {American Physical Society},
  doi = {10.1103/PhysRevLett.134.210403},
  url = {https://link.aps.org/doi/10.1103/PhysRevLett.134.210403}
}

@article{Mukherjee2026,
  title = {Symmetric Tensor Scars with Tunable Entanglement from Volume to Area Law},
  author = {Mukherjee, Bhaskar and Turner, Christopher J. and Szyniszewski, Marcin and Pal, Arijeet},
  journal = {Phys. Rev. Lett.},
  volume = {136},
  issue = {9},
  pages = {090401},
  numpages = {7},
  year = {2026},
  month = {Mar},
  publisher = {American Physical Society},
  doi = {10.1103/v8rf-p5zc},
  url = {https://link.aps.org/doi/10.1103/v8rf-p5zc}
}

@article{SashMoudXY26,
  title = {Additional quantum many-body scars of the spin-1 $XY$ model with Fock-space cages and commutant algebras},
  author = {Mohapatra, Sashikanta and Moudgalya, Sanjay and Balram, Ajit C.},
  journal = {Phys. Rev. B},
  volume = {113},
  issue = {5},
  pages = {054310},
  numpages = {29},
  year = {2026},
  month = {Feb},
  publisher = {American Physical Society},
  doi = {10.1103/4tv9-q7g7},
  url = {https://link.aps.org/doi/10.1103/4tv9-q7g7}
}

@phdthesis{Gottesman1997,
  author        = {Gottesman, Daniel},
  title         = {Stabilizer Codes and Quantum Error Correction},
  school        = {California Institute of Technology},
  year          = {1997},
  eprint        = {quant-ph/9705052},
  archiveprefix = {arXiv}
}

@article{AaronsonGottesman2004,
  title         = {Improved simulation of stabilizer circuits},
  author        = {Aaronson, Scott and Gottesman, Daniel},
  journal       = {Physical Review A},
  volume        = {70},
  number        = {5},
  pages         = {052328},
  year          = {2004},
  publisher     = {American Physical Society},
  doi           = {10.1103/PhysRevA.70.052328},
  eprint        = {quant-ph/0406196},
  archiveprefix = {arXiv}
}

@book{artin1988geometric,
  author    = {Artin, Emil},
  title     = {Geometric Algebra},
  publisher = {John Wiley \& Sons},
  series    = {Wiley Classics Library},
  year      = {1988},
  address   = {New York},
  isbn      = {978-0471608394}
}

@book{grove2002classical,
  author    = {Grove, Larry C.},
  title     = {Classical Groups and Geometric Algebra},
  publisher = {American Mathematical Society},
  series    = {Graduate Studies in Mathematics},
  volume    = {39},
  year      = {2002},
  address   = {Providence, Rhode Island},
  isbn      = {978-0821820193}
}

@misc{OllivierTillich2004,
      title={Quantum convolutional codes: fundamentals}, 
      author={H. Ollivier and J. -P. Tillich},
      year={2004},
      eprint={quant-ph/0401134},
      archivePrefix={arXiv},
      primaryClass={quant-ph},
      url={https://arxiv.org/abs/quant-ph/0401134}, 
}

@article{Schlingemann2008,
  author  = {Schlingemann, Dirk-M. and Vogts, Holger and
             Werner, Reinhard F.},
  title   = {On the structure of {Clifford} quantum cellular automata},
  journal = {Journal of Mathematical Physics},
  volume  = {49},
  pages   = {112104},
  year    = {2008},
  doi     = {10.1063/1.3005565}
}

@article{Haah2013,
  author  = {Haah, Jeongwan},
  title   = {Commuting {Pauli} Hamiltonians as Maps between Free Modules},
  journal = {Communications in Mathematical Physics},
  volume  = {324},
  pages   = {351--399},
  year    = {2013},
  doi     = {10.1007/s00220-013-1810-2}
}

@article{Haah2021,
  author  = {Haah, Jeongwan},
  title   = {Classification of translation invariant topological
             {Pauli} stabilizer codes for prime dimensional qudits
             on two-dimensional lattices},
  journal = {Journal of Mathematical Physics},
  volume  = {62},
  pages   = {012201},
  year    = {2021},
  doi     = {10.1063/5.0021068}
}

@misc{VanDeWetering2020,
  author        = {van de Wetering, John},
  title         = {{ZX}-calculus for the working quantum
                   computer scientist},
  year          = {2020},
  eprint        = {2012.13966},
  archivePrefix = {arXiv},
  primaryClass  = {quant-ph},
  url           = {https://arxiv.org/abs/2012.13966}
}

@article{Nicolau2026,
  author  = {Nicolau, Eloi and Ljubotina, Marko and Serbyn, Maksym},
  title   = {Fragmentation, Zero Modes, and Collective Bound States
             in Constrained Models},
  journal = {PRX Quantum},
  volume  = {7},
  pages   = {010352},
  year    = {2026},
  doi     = {10.1103/sl79-1xgb}
}

@book{CoeckeKissinger2017,
  author    = {Coecke, Bob and Kissinger, Aleks},
  title     = {Picturing Quantum Processes:
               A First Course in Quantum Theory and
               Diagrammatic Reasoning},
  publisher = {Cambridge University Press},
  year      = {2017},
  doi       = {10.1017/9781316219317}
}

@article{Biamonte2011,
  author  = {Biamonte, Jacob D. and Clark, Stephen R. and
             Jaksch, Dieter},
  title   = {Categorical tensor network states},
  journal = {AIP Advances},
  volume  = {1},
  number  = {4},
  pages   = {042172},
  year    = {2011},
  doi     = {10.1063/1.3672009}
}

@misc{BiamonteBergholm2017,
  author        = {Biamonte, Jacob and Bergholm, Ville},
  title         = {Tensor Networks in a Nutshell},
  year          = {2017},
  eprint        = {1708.00006},
  archivePrefix = {arXiv},
  primaryClass  = {quant-ph},
  url           = {https://arxiv.org/abs/1708.00006}
}
\bibliographystyle{apsrev4-2}

\end{document}